\documentclass[12pt, a4paper]{article}
\usepackage{geometry}
\usepackage{amsmath}
\usepackage{graphicx,psfrag,epsf}
\usepackage{enumerate}
\usepackage[hyphens]{url}
\usepackage{setspace}
\usepackage[svgnames]{xcolor}
\usepackage{xcolor}
\newcommand{\blind}{1}
\usepackage{float}
\date{\vspace{-5ex}}
\usepackage[utf8]{inputenc}
\usepackage[normalem]{ulem}
\usepackage{multirow}
\usepackage{multicol}
\usepackage{amsmath}
\usepackage{graphicx,psfrag,epsf}
\usepackage{enumerate}
\usepackage[hyphens]{url}
\usepackage{setspace}
\usepackage[section]{placeins}
\usepackage[utf8]{inputenc}
\usepackage{dcolumn}
\newcolumntype{d}[1]{D{.}{.}{#1}}
\usepackage{booktabs}
\usepackage{siunitx}
\usepackage{multirow}
\usepackage{makecell}
\usepackage{lscape}
\usepackage{ragged2e}
\usepackage{graphicx}
\usepackage{pdflscape}
\usepackage{graphicx,subcaption} 
\usepackage[demo]{rotating}
\usepackage{listings}
\usepackage{setspace}
\usepackage{titlesec}
\usepackage[textsize=small]{todonotes}
\titlelabel{\thetitle.\quad}
\usepackage[format=hang]{caption}
\usepackage{amssymb}
\usepackage[english]{babel}
\usepackage[autostyle, english = american]{csquotes}
\MakeOuterQuote{"}
\usepackage[toc,page]{appendix}
\usepackage{caption}
\usepackage{lipsum}
\usepackage{subcaption}

\usepackage[notes,backend=bibtex]{biblatex-chicago}

\newcommand{\CH}[1]{\textcolor{blue}{#1}}

\begin{document}

\if1\blind
{
    \title{\bf Dual system estimation using mixed effects loglinear models}
    \author{Ceejay Hammond\thanks{University of Southampton, Highfield, Southampton, SO17 1BJ, UK and Office for National Statistics, London, SW1P 4DF, UK. E-mail: cah1g17@soton.ac.uk},~
    Paul A. Smith\thanks{University of Southampton, Highfield, Southampton, SO17 1BJ, UK. E-mail: {p.a.smith@soton.ac.uk}}~
    and Peter G.M. van der Heijden\thanks{University of Southampton, Highfield, Southampton, SO17 1BJ, UK and Utrecht University, Padualaan 14, 3584 CH Utrecht, The Netherlands. E-mail: {p.g.m.vanderheijden@uu.nl}}}

\maketitle
} \fi

\if0\blind
{
  \begin{center}
    {\LARGE\bf Dual system estimation using mixed effects loglinear models}\\
\hfill \break 
\end{center}
} \fi

\linespread{2}
\newpage
\abstract
\label{abstract}
In official statistics, dual system estimation (DSE) is a well-known tool to estimate the size of a population. Two sources are linked, and the number of units that are missed by both sources is estimated. Often dual system estimation is carried out in each of the levels of a stratifying variable, such as region. DSE can be considered a loglinear independence model, and, with a stratifying variable, a loglinear conditional independence model. The standard approach is to estimate parameters for each level of the stratifying variable. Thus, when the number of levels of the stratifying variable is large, the number of parameters estimated is large as well. Mixed effects loglinear models, where sets of parameters involving the stratifying variable are replaced by a distribution parameterised by its mean and a variance, have also been proposed, and we investigate their properties through simulation. In our simulation studies the mixed effects loglinear model outperforms the fixed effects loglinear model although only to a small extent in terms of mean squared error. We show how mixed effects dual system estimation can be extended to multiple system estimation.


\newpage
\section{Introduction}
\label{introduction}


The dual system estimator is the most simple capture-recapture model and was originally used to estimate the size and density of wildlife populations (Seber, 1986)\nocite{seber1986review}. It is now commonly used across official statistics to estimate the population size for domains of interest. Two sources are linked and the unobserved number of units missed from both sources is estimated. In official statistics this estimator was originally used for coverage assessment in the 1950 decennial United States Census, to estimate coverage error (Wolter, 1986)\nocite{WolterKM1986}. This estimator relies on five key assumptions (International Working Group for Disease Monitoring and Forecasting, 1995)\nocite{IWGfDMaF1995} -- that the population is closed, there is perfect matching between the two lists, the capture/inclusion probability of elements in at least one of the lists is homogeneous (Seber, 1982;  van der Heijden \textit{et al}., 2012)\nocite{seber1982}\nocite{VanderHeijdenPGMWhittakerCruyffBakkerVanderVliet2012}, the inclusion in the lists is independent and there are no erroneous captures in any list (no overcoverage).
   
All of these key assumptions should be met, however, in practice it can be difficult to ensure this. One of the key assumptions that underpins dual system estimation is that the inclusion in the lists is independent. In practice independence can be violated for one of three reasons: (a) operationally, such as that the modes of data collection are the same for both lists, (b) conditional inclusion of units, where an individual's inclusion in one list is conditional on their inclusion in the other list, and (c) failure of the homogeneous inclusion probability assumption (Brown, Sexton, Abbott and Smith, 2019)\nocite{BrownJSextonCAbbottOSmithPA2019}. 


The simplest situation we consider for dual system estimation is by fitting a loglinear model with two dichotomous variables. To help to satisfy the homogeneous inclusion probability assumption, one can make use of one or more stratifying variables (Gerritse, Bakker and van der Heijden, 2015)\nocite{GerritseBakkervanderHeijden2015}. The dual system estimation is then carried out at each of the levels of the stratifying variables. The methodology we discuss in this paper can be extended to loglinear models of any size, in terms of both the number of variables, and the number of levels of the stratifying variables.

It suits our purposes to define loglinear models in terms of Poisson regression models with a log link (Nelder and Wedderburn, 1972)\nocite{Nelder1972}. One can make use of fixed effects loglinear models, which explain within-group variability for each of the stratifying variables. This leads to models with many parameters. These models are inefficient for large numbers of groups with small numbers of observations within groups 
(Townsend, Buckley, Harada and Scott, 2013\nocite{townsend2013choice}; Clarke, Crawford, Steele and Vignoles, 2015\nocite{clarke2015revisiting}). Therefore we investigate the use of mixed effects loglinear models, also known as multilevel, hierarchical or variance components models (Hox, Moerbeek and Van de Schoot, 2017)\nocite{hox2017multilevel}. These models replace the fixed parameters for the stratifying variables with normal distributions parameterised by their mean and variance (Berkhof and Snijders, 2001)\nocite{berkhof2001variance}.
Royle and Link (2002)\nocite{royle2002random} present the use of random effects to model heterogeneous capture probabilities and of shrinkage estimation to account for the variability introduced by using sample data. Agresti (2013)\nocite{AgrestiA2013} points out that mixed effects loglinear models can be used to model heterogeneous capture probabilities across groups by borrowing strength from the whole. The multilevel model can also be used as a bias-robust approach, because it borrows information across all of the groups, and therefore suffers less from small sample bias; it provides an alternative to the Chapman estimator. Therefore, we expect these models to improve population estimates in terms of mean squared error, compared to the corresponding fixed effects loglinear models.

We investigate the performance of mixed effects loglinear models through simulation, where population sizes are both small and large, and are made up of varying numbers of regions. For small sample sizes, bias can be a concern, as illustrated in recent contributions of work from Zult, van der Heijden and Bakker (2025)\nocite{zult2023bias}; Rivest and Yauck (2024)\nocite{rivest2024small}.

In mixed effects loglinear model parlance, we are interested in estimating the number of individuals (level 1) within each of the regions (level 2). We also show how the methodology discussed for dual system estimation is extended to multiple system estimation, which makes use of more than two lists to estimate the population for domains of interest.

The structure of the paper is as follows. Section \ref{Methodology} discusses the methodology for fixed effects loglinear models for the simplest situation involving tables without and with a stratifying variable. We discuss mixed effects loglinear models, including shrinkage estimation and model fitting. Section \ref{simulation Study} describes the set-up of the simulation study; we examine random effects which are normally distributed in line with the normality assumption, and also lognormally or Pareto distributed to evaluate the effect of departures from the assumption. Section~\ref{results} presents the results, where fixed and mixed effects loglinear models are compared by the mean absolute relative bias and the coefficient of variation for the total population and for the levels of the stratifying variable. Section~\ref{methods_mse} shows how the methods are extended to multiple system estimation. Section \ref{Discussion} makes a comparison, including a discussion of advantages and disadvantages.

\section{Methodology - Dual System Estimation}
\label{Methodology}

\subsection{Fixed effects loglinear models}
\label{Section 2.1}

\subsubsection{Models for two-way tables}
\label{m2way}

The simplest situation that we consider is the fixed effects loglinear model with two dichotomous variables $A$ and $B$. Let $A$ be indexed by $i = 0,1$ and $B$ by $j = 0,1$. We denote the expected count for cell $(i,j)$ of the two-way contingency table of variables $A$ and $B$ by $\mu_{ij}$. We denote the loglinear parameters by $\lambda$. Then the saturated loglinear model for two lists is

\begin{equation}
\label{satfixed}
\log \mu_{ij} = \lambda + \lambda_i^A + \lambda_j^B +\lambda_{ij}^{AB},
\end{equation}

\noindent where $\lambda$ is the intercept term, $\lambda_i^A$ and $\lambda_j^B$ are the respective parameters for variables A and B, and $\lambda_{ij}^{AB}$ is the two factor interaction parameter. The parameters are identified by setting them equal to 0 when an index is 0, i.e. $\lambda_0^A = \lambda_0^B = \lambda_{00}^{AB} = \lambda_{01}^{AB} = \lambda_{10}^{AB} = 0$, referred to as the corner point constraint parameterization. Thus the model has four parameters and four observed counts, and hence is called saturated.

In dual system estimation variable $A$ stands for "Being included in list $A$", where 0,1 stands for "missed", ``in'', and similarly for variable B. The size of the $(A=0,B=0)$ cell is not observed, so model (\ref{satfixed}) cannot be estimated. Therefore we proceed by assuming independence between $A$ and $B$ and fitting

\begin{equation}
\label{indep2}
\log \mu_{ij} = \lambda + \lambda_i^A + \lambda_j^B.
\end{equation}







One can estimate the population size $\hat{N}$ directly from the observed counts by making use of the Lincoln Peterson estimator 
(Bishop, Fienberg and Holland, 1975)\nocite{BishopYMMFienbergSEHollandPW1975}, that relies on the independence assumption,

\begin{equation}
    \hat{\mu}_{00} =  \frac{n_{10}n_{01}}{n_{11}},
    \label{LP}
\end{equation}

\noindent or one can also make use of the parameters of a loglinear model by

\begin{equation}   \hat{\mu}_{00} = \exp(\hat{\lambda}),
    \label{fixed}
\end{equation}

\noindent which makes use of the corner point constraints of the loglinear parameters. Thus the population size estimate $\hat{N} = n_{10} + n_{01} + n_{11} + \hat{\mu}_{00}$. $\hat{N}$ is a maximum likelihood estimate that is asymptotically unbiased.

\subsubsection{Models for two-way tables including a stratifying variable}
\label{Fixed}

The loglinear models discussed in section \ref{m2way} can be developed to include a stratifying variable, where parameters are estimated for each level of the variable. 
The situation that we consider here is the fixed effects loglinear model with two dichotomous variables $A$ and $B$ and a polytomous stratifying variable, denoted as $R$ (such as a regional level geography), 
that is associated with each of the lists $A$ and $B$.
The parameters for the levels of the stratifying parameter control for observed heterogeneity between groups. Stratification is often used because the assumption of homogeneous inclusion probabilities is more tenable within the strata.

Let the stratifying variable $R$ be indexed by $l = 1,...,r$.
We denote the expected count for cell $(i,j, l)$ of the two-way tables $A$, $B$ within strata $R$ by $\mu_{ijl}$. The saturated loglinear model for two lists with a stratifying variable is:

\begin{equation}
\label{sat2}
\log \mu_{ijl} = \lambda + \lambda_i^A + \lambda_j^B + \lambda_l^R +\lambda_{ij}^{AB}+ \lambda_{il}^{AR} +\lambda_{jl}^{BR} +\lambda_{ijl}^{ABR}
\end{equation}

\noindent where $\lambda$ is the intercept term, $\lambda_i^A$, $\lambda_j^B$ and $\lambda_l^R$ are the respective parameters for variables A, B and R, $\lambda_{ij}^{AB}$, $\lambda_{il}^{AR}$, $\lambda_{jl}^{BR}$ are the two factor interaction parameters, and  $\lambda_{ijl}^{ABR}$ 
is the three factor interaction parameter. 
The parameters are identified by corner point constraints, i.e. by setting the parameter equal to 0 when at least one of the indices is 0. Thus the model has $2\times2\times r$ independent parameters and $2\times2\times r$ observed counts, and hence is saturated.
When we fit model (\ref{sat2}) to observed counts $n_{ijl}$, the fitted values are equal to the observed counts, i.e. $\hat{\mu}_{ijl} = n_{ijl}$. This model is denoted by the variables constituting the highest-order fitted margins, i.e. [ABR]. 

In population size estimation, all $(A=0,B=0)$ cells are unobserved, so the most complex model that can be fitted is the conditional independence model denoted by [AR][BR], where lists $A$ and $B$ are assumed to be independent given the stratifying variable. In this 
model the number of observed counts is equal to the number of parameters, hence the number of degrees of freedom is zero. Following van der Heijden et al. (2022)\nocite{van2022multiple} we call this a {\em maximal} model, to distinguish it from the saturated model (\ref{sat2}). Fitting the conditional independence model requires leaving the interaction terms $\lambda_{ij}^{AB}$ and $\lambda_{ijl}^{ABR}$ out of equation (\ref{sat2})\CH{.} 


For population size estimation, we estimate the unobserved cell ${\mu}_{00l}$ directly from the observed counts, where the Lincoln Peterson estimator is applied to each level of the stratifying variable $R$,

\begin{equation}
    \hat{\mu}_{00l} =  \frac{n_{10l}n_{01l}}{n_{11l}}.
    \label{stratifying_LP}
\end{equation}

\noindent We can also estimate ${\mu}_{00l}$ using the loglinear parameters by,

\begin{equation}
    \hat{\mu}_{00l} = \exp(\hat{\lambda} + \hat{\lambda}_l^R).
    \label{strat_fixed}
\end{equation}

\noindent To estimate $\hat{N}$, for each level of $R$, $\hat{N_l}= n_{10l} + n_{01l} + n_{11l} + \hat{\mu}_{00l}$ and therefore, $\hat{N} = \sum_{l} \hat{N_l}$.

\subsection{Mixed effects loglinear models}
\label{mixed models}

\subsubsection{Models for two-way tables including a stratifying variable}
\label{m3waymixed}



We may develop the fixed effects loglinear models to include random terms. These methods solve problems which occur when fitting fixed effects loglinear models to data with small cell counts and/or a large number of parameters (Royle and Link, 2002)\nocite{royle2002random}. Snijders and Bosker (2011, pp. 46 -- 48)\nocite{bosker2011multilevel} suggest fixed effects loglinear models should be used in cases with a small number of groups, and random effects if the groups can be considered to be a sample from a population of groups, and if group sizes are small but there is a large number of groups (more than ten).

When sample sizes are large, the dual system estimator is applied separately to each stratum, which is equivalent to fitting the fixed effects loglinear model. However, when there are many strata and particularly when the sample sizes are small, this results in a biased estimate of $N$. To reduce this bias, one can use the Chapman estimator or the mixed effects loglinear model, which provides an alternative approach by borrowing strength across strata to improve the estimates. To contribute to understanding, we examine the properties of population size estimates derived from mixed effects models.



We can replace the fixed terms in model (\ref{sat2}), $\lambda_l^R$, $\lambda_{il}^{AR}$, $\lambda_{jl}^{BR}$ and $\lambda_{ijl}^{ABR}$, with a random intercept term for the stratifying variable R and random slope terms for the two and three factor interactions as follows, 

\begin{equation}
\label{mixedslopemaximal2}
\log \mu_{ijl} = (\lambda + u_{0l}) + \lambda_i^A + \lambda_j^B + \lambda_{ij}^{AB} + u_{1il} + u_{2jl} + u_{3ijl},
\end{equation}

\noindent where $u_{0l}$ is the random intercept parameter with $u_{0l} \sim N \left(0, \sigma^2_{u0}\right)$, $u_{1il}$ is the random slope parameter for list A with $u_{1il} \sim N \left(0, \sigma^2_{u1}\right)$, $u_{2jl}$ is the random slope parameter for list B with $u_{2jl} \sim N \left(0, \sigma^2_{u2}\right)$, and $u_{3ijl}$ is the random slope parameter for the two factor interaction between list A and B with $u_{3ijl} \sim N \left(0, \sigma^2_{u3}\right)$.
The parameters are identified by setting the sum of the fixed effects parameters $\lambda_i^A$, $\lambda_j^B$ and $\lambda_{ij}^{AB}$ equal to 0 i.e. $\Sigma_i\lambda_i^A = \Sigma_j\lambda_j^B = \Sigma_{i}\lambda_{ij}^{AB} = \Sigma_{j}\lambda_{ij}^{AB} =  0$, referred to as the sum-zero constraint parameterization. We prefer the sum-zero parameterisation to centre the random effects at zero. 
In practice, either constraint can be used and will produce equivalent estimates.

Population size estimation with the conditional independence model requires leaving the terms $\lambda_{ij}^{AB}$ and $u_{3ijl}$ out of equation (\ref{mixedslopemaximal2}). This allows us
to model between-region variability with mixed effects terms:
\begin{equation}
\label{mixedslopemaximal22}
\log \mu_{ijl} = (\lambda + u_{0l}) + \lambda_i^A + \lambda_j^B + u_{1il} + u_{2jl},
\end{equation}
\noindent where the random intercept term $u_{0l}$ captures variation in the population size by the stratifying variable $R$, and the random slope terms, $u_{1il}$ and $u_{2jl}$, capture variation in the inclusion probabilities of lists A and B respectively across the strata $R$.

Random effect estimators are often referred to as shrinkage estimators (Robinson, 1991)\nocite{robinson1991blup} or Best Unbiased Predictors (BUP) (Royle and Link, 2002)\nocite{royle2002random}. Royle and Link (2002)\nocite{royle2002random} pointed out that the use of the word `unbiased', in BUP, may be misleading as these estimators are (conditionally) biased, however they do have minimum variance.

`Shrinkage' refers to the property that the group-level estimate from the multilevel model is between the direct estimate and the overall mean -- the direct estimate is `shrunk' towards the overall mean. Post hoc, i.e. after the estimation of the model parameters, one can estimate these shrunk parameters. The level of shrinkage depends on the initial reliability of the parameter estimates, which is affected by group sample size and the distance between group-based and overall estimates (Hox, 2010)\nocite{hox2010multilevel}. This results in borrowing strength from groups across the whole dataset, 
and is a natural consequence of the use of mixed effects loglinear models. Shrinkage toward the mean is  useful to reduce the variability of estimates, especially when there are small sample sizes. It
is said to lead to better estimates in terms of the mean-squared error.

This is distinct from the use of shrinkage in regularisation (eg ridge regression, lasso) (Tibshirani, 1996)\nocite{tibshirani1996regression}, which we do not consider here.


For population size estimation one can also make use of the parameters of the mixed effects loglinear models to estimate $\hat{N}$. For this, the mixed effects conditional independence model (\ref{mixedslopemaximal22}) makes use of shrinkage estimators to estimate ${\mu}_{00l}$ as follows,

\begin{equation}
    \hat{\mu}_{00l} = \exp((\hat{\lambda} + \hat{u}_{0l}) + (\hat{\lambda}_i^A + \hat{u}_{1il}) + (\hat{\lambda}_j^B + \hat{u}_{2jl})),
    \label{stratifying_mixed}
\end{equation}

\noindent where the parameter $\lambda$ is estimated using sum zero constraints. In contrast with the fixed effects loglinear model the observed counts do not equal the fitted counts, i.e. $n_{10l} \neq \hat{\mu}_{10l}$, $ n_{01l} \neq \hat{\mu}_{01l}$ and $n_{11l} \neq \hat{\mu}_{11l}$. We use the observed counts when estimating $\hat{N_l}= n_{10l} + n_{01l} + n_{11l} + \hat{\mu}_{00l}$, as they are the best information we have for $n_{10l}$, $n_{01l}$ and $n_{11l}$, and estimate $\hat{N} = \sum_{l} \hat{N_l}$.

\subsubsection{Model Fitting}
\label{Model Fitting}

The models are estimated in \texttt{R} (R Core Team, 2015)\nocite{core2015team}
using the statistical package \texttt{lme4} (Bates, M{\"a}chler, Bolker and Walker, 2015)\nocite{lme4}.

One can choose the optimization algorithm to estimate the parameters for the specified model. Under the standard optimizer, small population sizes for the groups can result in convergence warnings.

We use only the \texttt{bobyqa} optimizer as it is reasonably fast and has fewer convergence warnings compared to the default glmer optimizer. In the cases where model fitting resulted in convergence warnings, the parameter estimates are comparable to the same models fitted with a different optimiser that have no warnings, so we deduce that the convergence problems appear close to the maximum of the likelihood and have only a minimal impact on the final result. 

\subsection{Bias reduction}
\label{Chapman}

We investigate the performance of the dual system estimator when sample counts are small, for example in regional sampling. In the fixed effects approach the Lincoln-Petersen estimator is asymptotically unbiased, but a small sample size results in a biased estimate of $N$ (Zult et al., 2025)\nocite{chapman1951some}\nocite{zult2023bias}. To reduce this bias, we make use of the Chapman estimator (Chapman, 1951)\nocite{chapman1951some} that is essentially unbiased. Recently this was illustrated in simulation studies by Zult et al. (2025)\nocite{zult2023bias}. 
We apply this approach to the fixed effects loglinear model only.
For the simplest dual system estimation situation we consider, the Chapman estimate $\hat{\mu}_{00}^{Chap}$ can be found directly from the observed counts by,

\begin{equation}
    \hat{\mu}_{00}^{Chap} = \frac{n_{10}n_{01}}{(n_{11}+1)}.
\end{equation}
\noindent Note that the estimate can be found simply by adding 1 to $n_{11}$ and then applying equation (\ref{LP}) or, similarly, (\ref{fixed}). Therefore, one can estimate the population size $\hat{N}$ directly from the observed counts.



Where the population includes a stratifying variable $R$, the Chapman estimator is applied to each level of $R$,

\begin{equation}
    \hat{\mu}_{00l}^{Chap} = \frac{n_{10l}n_{01l}}{(n_{11l}+1)}.
    \label{strat_chap}
\end{equation}\
\noindent where 1 is added to the (A, B) = (1, 1) cells for each level of the stratifying variable.



No adjustment of the $(A,B) = (1,1)$ cells is necessary in the mixed effects loglinear model. The Chapman estimator is not applied because of the use of shrinkage estimators, which provide an alternative approach to dealing with the bias in the dual system estimator.

\section{Set up of the simulation study}
\label{simulation Study}

\subsection{Simulated data}
\label{simulated data}


Simulations are used to illustrate the properties of dual system estimators that make use of fixed and mixed effects loglinear models. We present and compare the results where we fit loglinear models with and without the Chapman estimator.   
Mixed effects loglinear models assume normality for the random effects. Scenarios are investigated that are defined by the following factors:
\begin{itemize}
    \item The distribution of the random numbers in population generation: Normal, Pareto and Lognormal [3 options]
    \item The size of the population $N$: 500, 1,000, 1,500, 2,000, 2,500, 3,000, 30,000 and 300,000 [8 options]
    \item The number of regions in a population: 5, 10, 15, 20, 25 and 30 [6 options]
    \item The size of the capture/inclusion probabilities for list A ($\pi_A$) and list B ($\pi_B$): ($\pi_{A}$ = 0.8, $\pi_{B}$ = 0.7), which represent a census context where we mirror the capture/inclusion probabilities used by Brown, Abbott and Diamond (2006) and ($\pi_{A}$ = 0.4, $\pi_{B}$ = 0.2), which represent a non-census context where we mirror the capture/inclusion probabilities used by Baffour, Brown and Smith (2013)\nocite{BrownJAbbottODiamondI2006}\nocite{Baffour2013} [2 options]
    \item The fitted models: The fixed effects loglinear model without (\ref{strat_fixed}) or with (\ref{strat_chap}) the Chapman estimator, and the mixed effects loglinear model (\ref{mixedslopemaximal22}) [3 options]
    \item The chosen variances for $\sigma^2_{u0}$, $\sigma^2_{u1}$ and $\sigma^2_{u2}$, informed by the simulations discussed in supplementary material \ref{appendix A} [1 option]
\end{itemize}

This results in a total of 288 scenarios.
As an example we describe how we generate a single dataset for 10 regions, with the larger inclusion probabilities ($\pi_{A}$ = 0.8, $\pi_{B}$ = 0.7), where the mixed model normality assumption holds, as in model (\ref{mixedslopemaximal22}). This is discussed in detail in supplementary material \ref{Appendix A1}.

\begin{figure}[!htbp]
\includegraphics[height=10cm, width=18cm]{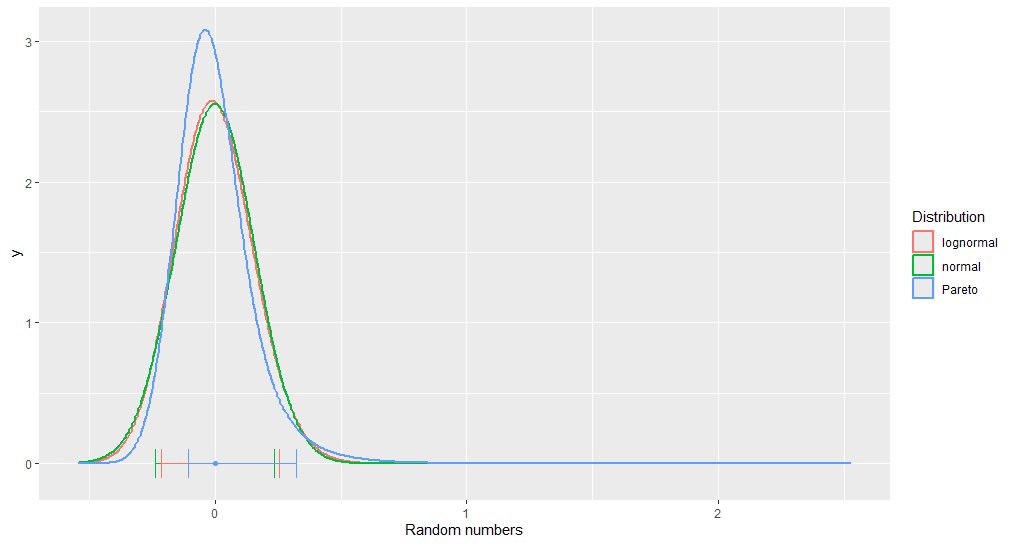}
\captionsetup{width=5.5in}
\caption{\small The distributions of the normally (in green), lognormally (in red) and Pareto (in blue) distributed random numbers.}
\label{Distributions}
\end{figure}

To assess the effect of deviations from the assumption of normally distributed random effects, we generate random numbers from the lognormal and Pareto distributions instead of normal distributions. This is discussed in detail in supplementary material \ref{Appendix D}.
The distributions of the normally, lognormally and Pareto distributed numbers for the intercept $\mu_{0l}$ are presented in Figure \ref{Distributions}. Although the distributions overlap, the vertical bars at the x-axis, providing the boundaries for the 95 \% densities, show that the normal distribution is symmetric, whereas the lognormal and Pareto are skewed to the right.

\subsection{Reported descriptive statistics}
\label{descriptive statistics}

In reporting the simulation studies we focus on the relative bias (\%) and coefficient of variation (\%) for the multinomial samples. In the simulations, there are two types of variation, population and multinomial sampling variation. To extract the variation from the multinomial samples we used the package \texttt{VCA} (Erhardt and Schuetzenmeister, 2022)\nocite{erhardt2022r} to perform variance component analysis by using ANOVA-type estimation via the \texttt{anovaVCA} function.

The mean absolute relative bias (marb) (\%), i.e. relative to $N_l$, is calculated by,

\begin{equation}
    \frac{\sum_p{\sum_m{\sum_l{(|{\hat{N}{_{pml}} - N_{pml}}|}/{N_{pml}})}}}{(100*100*l)}*100
\end{equation}

\noindent where the denominator presents the 10,000 samples drawn by generating 100 populations, and for each of these populations 100 multinomial samples are drawn, and $l$ denotes the number of regions. {In the numerator, $p$ denotes the populations, $m$ denotes the multinomial samples.
The coefficient of variation (cv) (\%), relative to $\hat{\mu_l}$, is calculated by $(\hat{\sigma_l}/{\hat{\mu_l}}) * 100$ of the region level estimates across all numbers of regions and population sizes $N$. We present the coefficient of variation to take care of the different sizes of estimates.


In section \ref{results} we present the marb and cv for 5, 15 and 30 regions. Occasionally, in particular in scenarios where counts are small, the count in the (1,1) cell of a region is zero and the fixed model estimate without the Chapman correction is infinite.
For this reason we do not present the results for the fixed effects model without the Chapman correction in figures but they can be found in the tables (\ref{table:1a}-
\ref{table:2e}) in the Supplementary material. 

Also in the Supplementary material, we present the marb and cv for 10, 20 and 25 regions. We report extensive tables for the regional estimates of the mean relative bias (mrb) (\%), the cv and the mean squared error (mse) for the fixed (with or without the Chapman estimator) and mixed effects loglinear models. We also present box plots for the regional estimates and relative bias (rb). We present an extensive set of results which include the rb and mrb to present the bias, whereas the marb and cv are used to assess the variability.
\section{Results}
\label{results}

\begin{sidewaysfigure}[p]
\begin{minipage}{0.9\textwidth}
\centering
\begin{subfigure}[b]{.3\linewidth}
         \includegraphics[width=7cm,height=6cm]{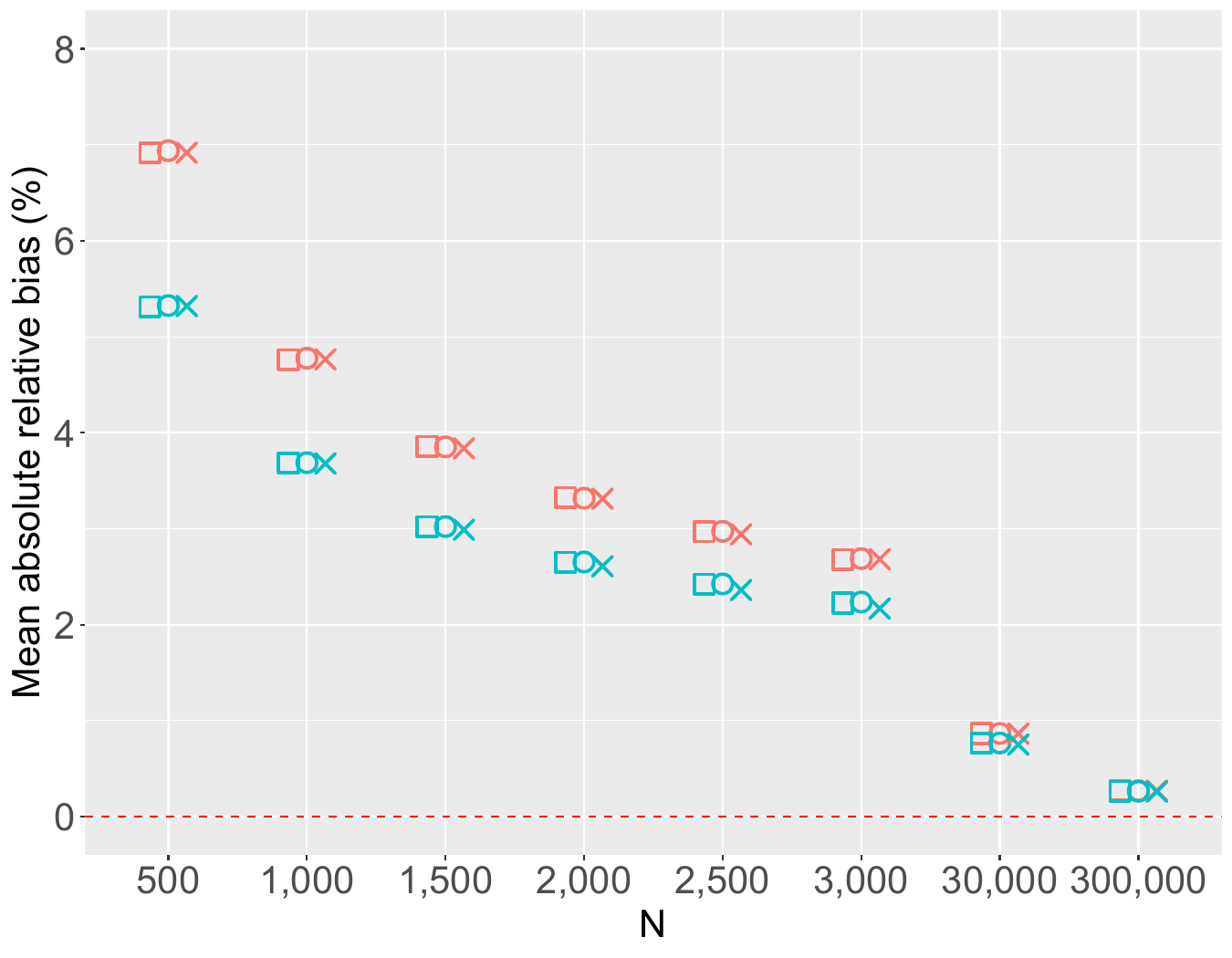}
         \caption{\small 30 Regions}
        \label{arobust_mean_bias_30_high_probs}
    \end{subfigure}\hspace*{\fill}
    \begin{subfigure}[b]{.3\linewidth}
         \includegraphics[width=7cm,height=6cm]{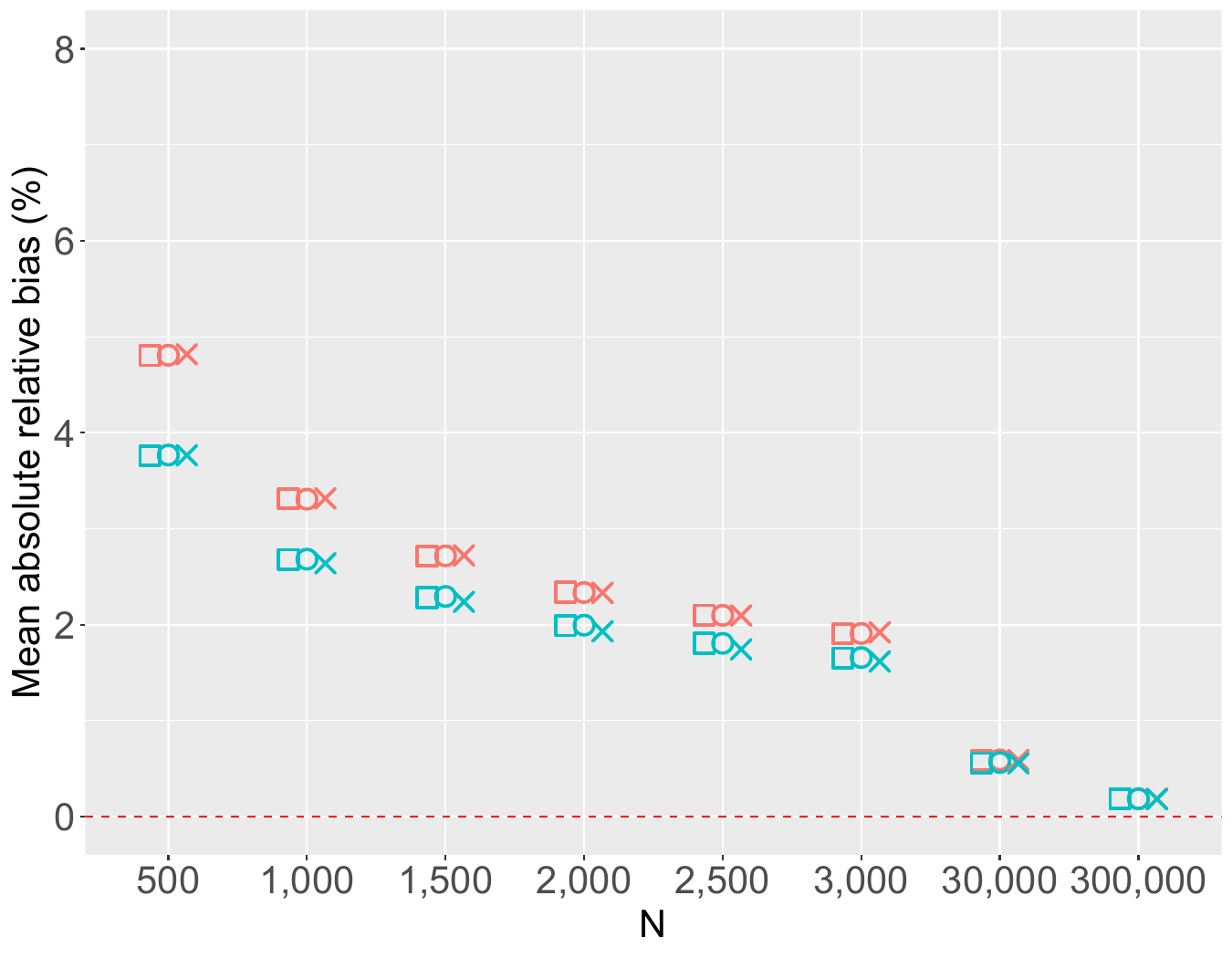}
        \caption{\small 15 Regions}
        \label{arobust_mean_bias_15_high_probs}
    \end{subfigure}\hspace*{\fill}
    \begin{subfigure}[b]{.3\linewidth}
         \includegraphics[width=7cm,height=6cm]{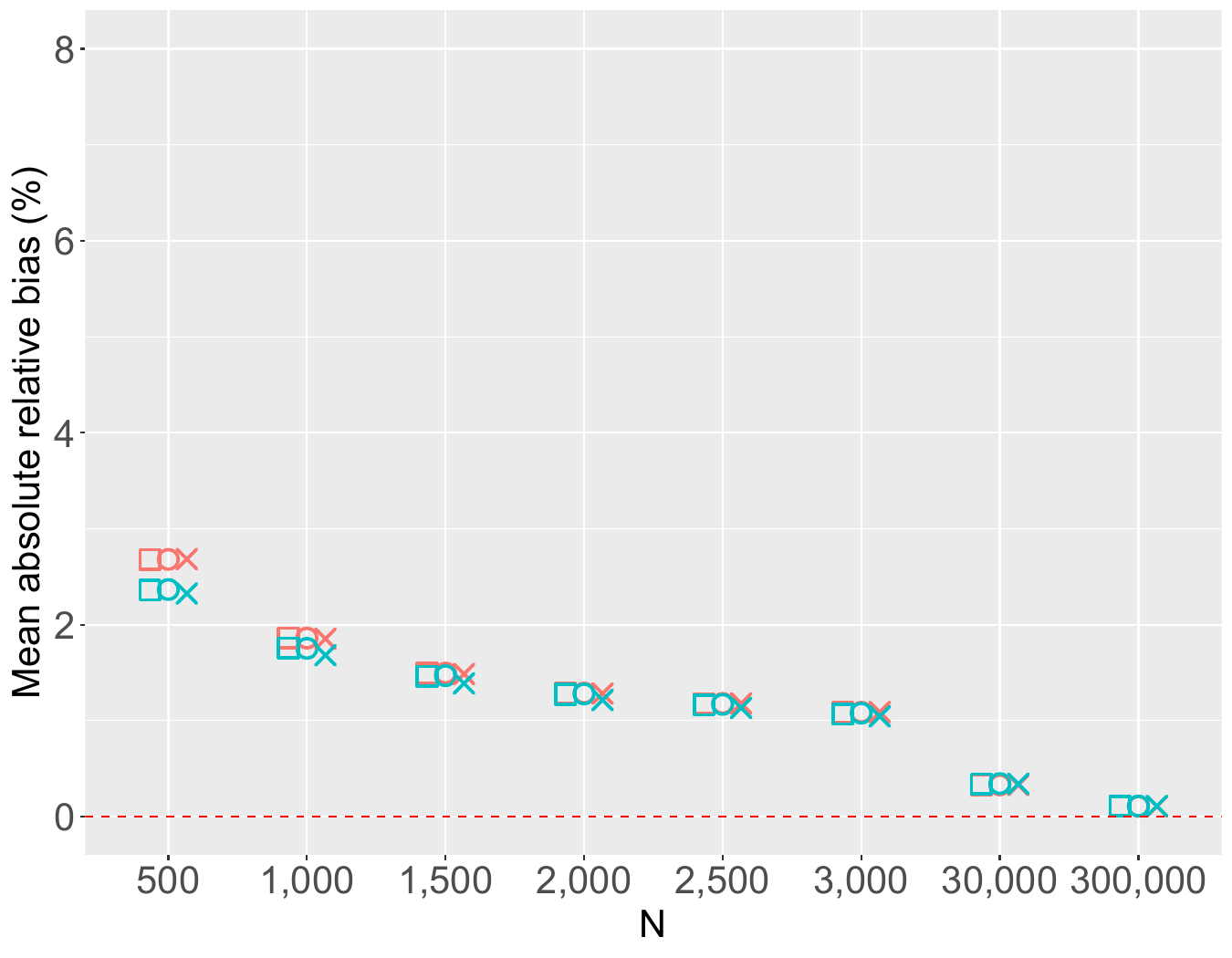}
         \caption{\small 5 Regions}
        \label{arobust_mean_bias_5_high_probs}
    \end{subfigure}\hspace*{\fill}

\caption{Mean absolute relative bias (\%) of estimates where initially $\pi_A = 0.8$ and $\pi_B = 0.7$ and the random effects follow a normal, lognormal or Pareto distribution.}\label{rb_high_probs}

\begin{subfigure}[b]{.3\linewidth}
         \includegraphics[width=7cm,height=6cm]{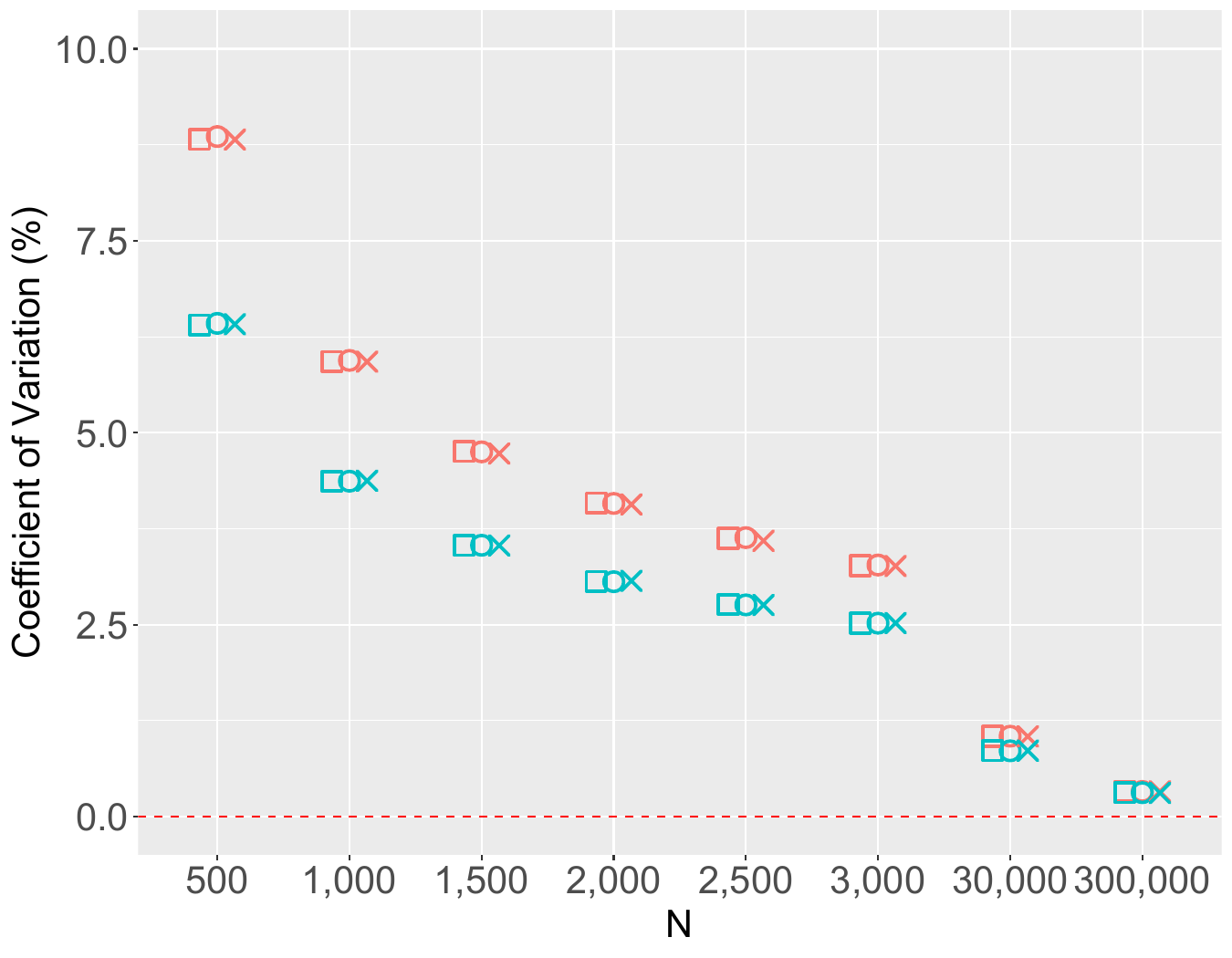}
         \caption{\small 30 Regions}
        \label{cv_30_high_probs}
    \end{subfigure}\hspace*{\fill}
    \begin{subfigure}[b]{.3\linewidth}
         \includegraphics[width=7cm,height=6cm]{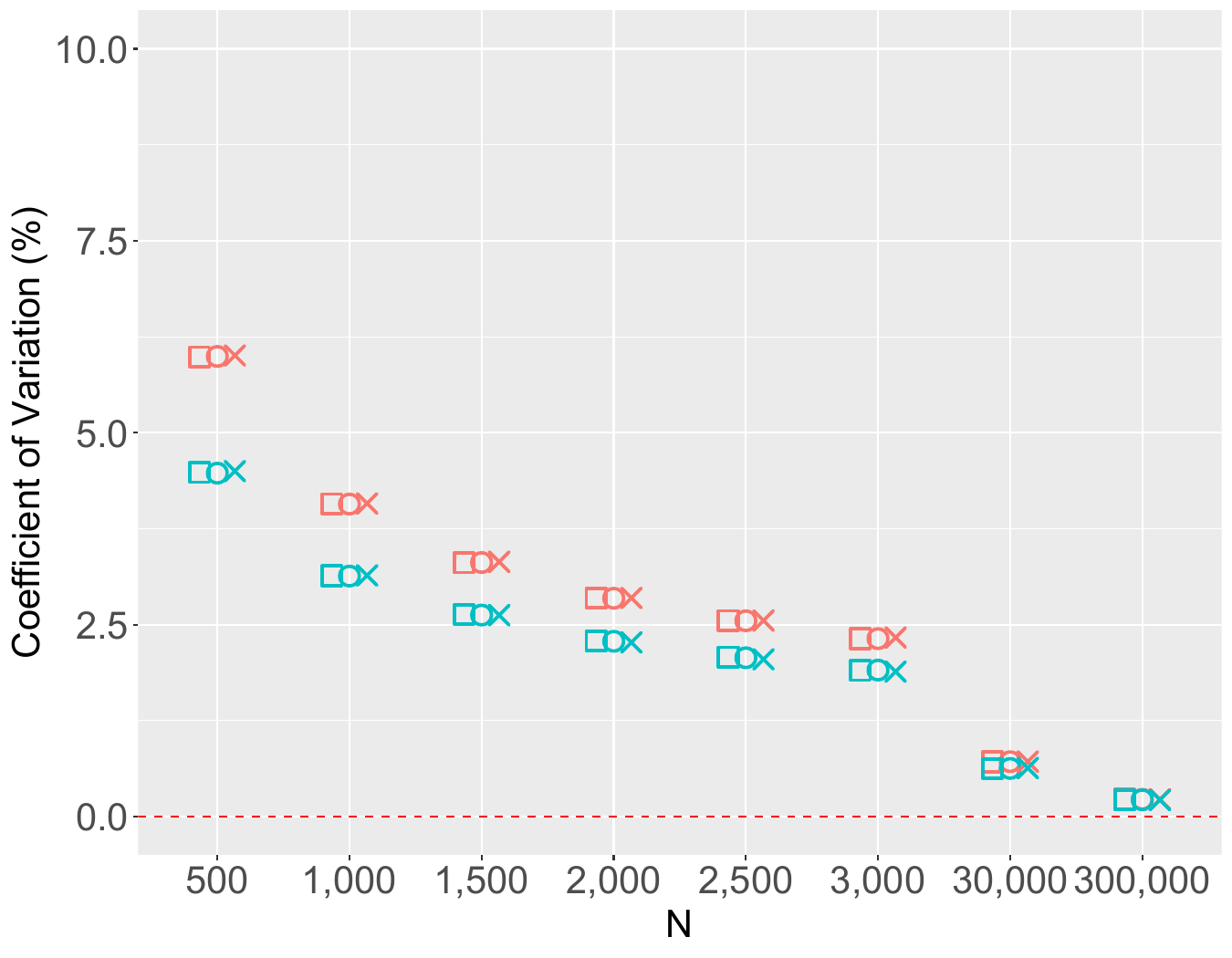}
        \caption{\small 15 Regions}
        \label{cv_15_high_probs}
    \end{subfigure}\hspace*{\fill}
    \begin{subfigure}[b]{.3\linewidth}
         \includegraphics[width=7cm,height=6cm]{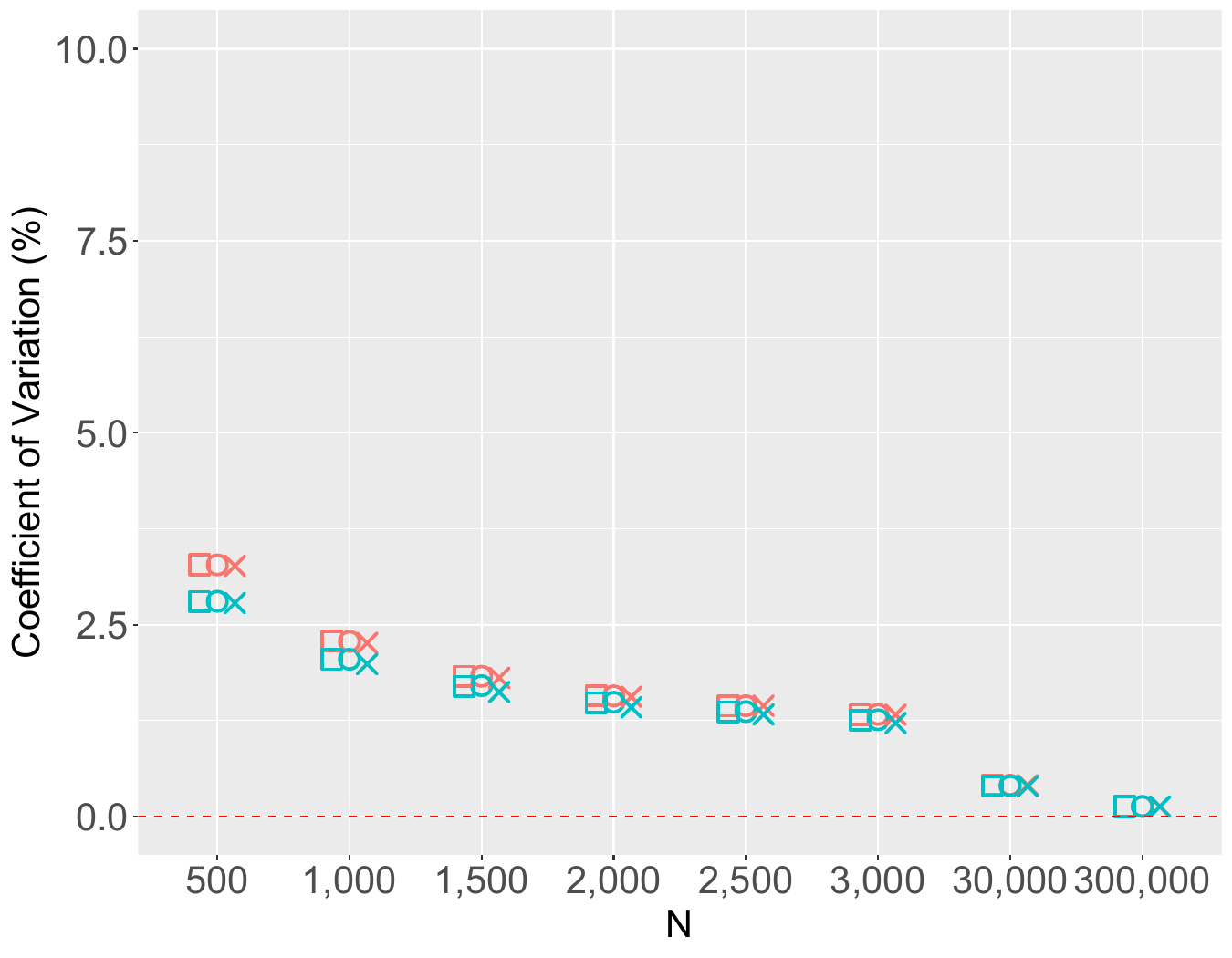}
         \caption{\small 5 Regions}
        \label{cv_5_high_probs}
    \end{subfigure}\hspace*{\fill}
    \end{minipage}
    \begin{minipage}[c]{0.1\textwidth}
        \includegraphics[width=4cm,height=4cm]{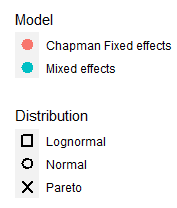}
    \end{minipage}

\caption{Coefficient of variation (\%) of estimates where initially $\pi_A = 0.8$ and $\pi_B = 0.7$ and the random effects follow a normal, lognormal or Pareto distribution.}\label{rv_high_probs}

\end{sidewaysfigure}

\begin{sidewaysfigure}[p]
\begin{minipage}{0.9\textwidth}
\centering
\begin{subfigure}[b]{.3\linewidth}
         \includegraphics[width=7cm,height=6cm]{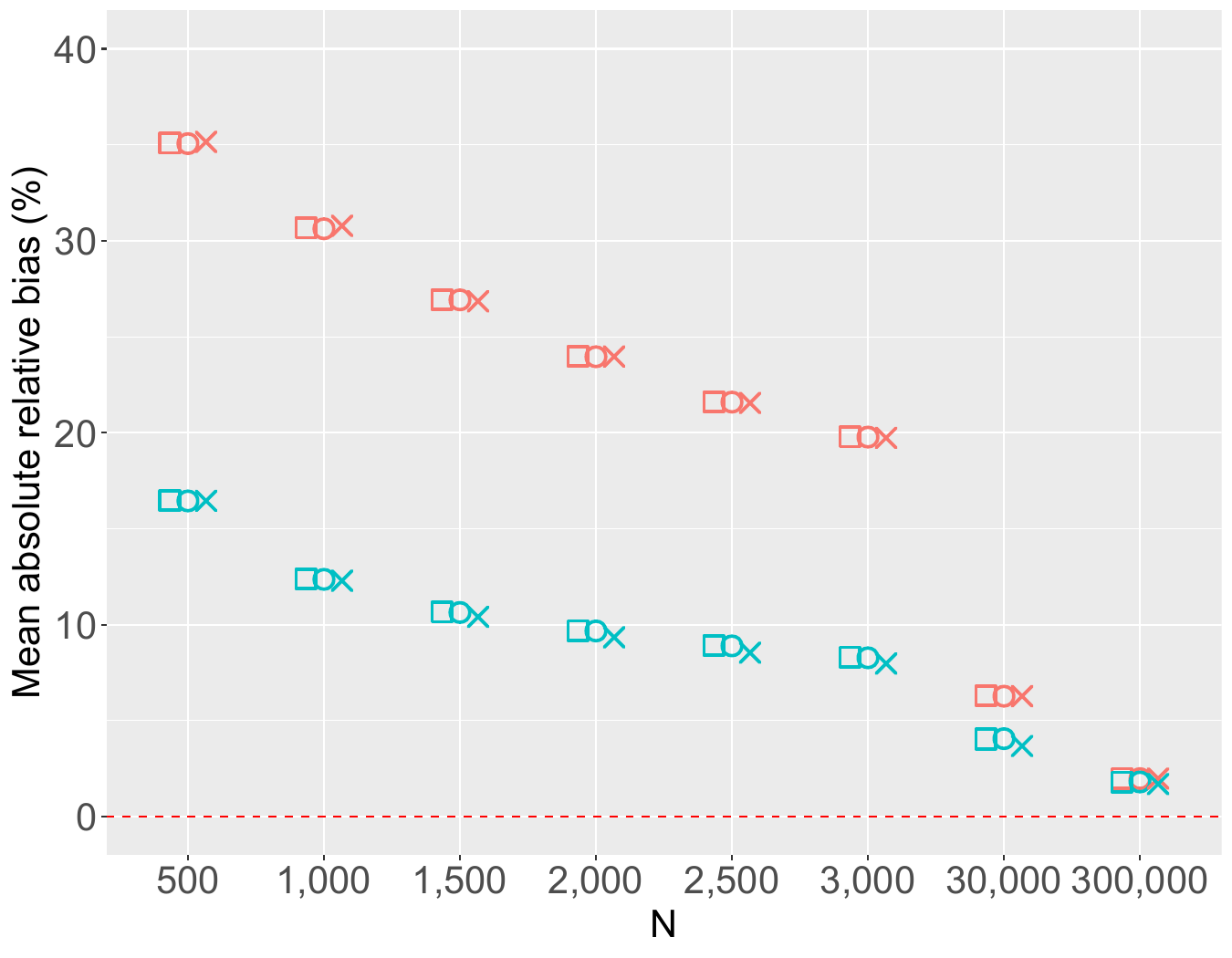}
         \caption{\small 30 Regions}
        \label{arobust_mean_bias_30_low_probs}
    \end{subfigure}\hspace*{\fill}
    \begin{subfigure}[b]{.3\linewidth}
         \includegraphics[width=7cm,height=6cm]{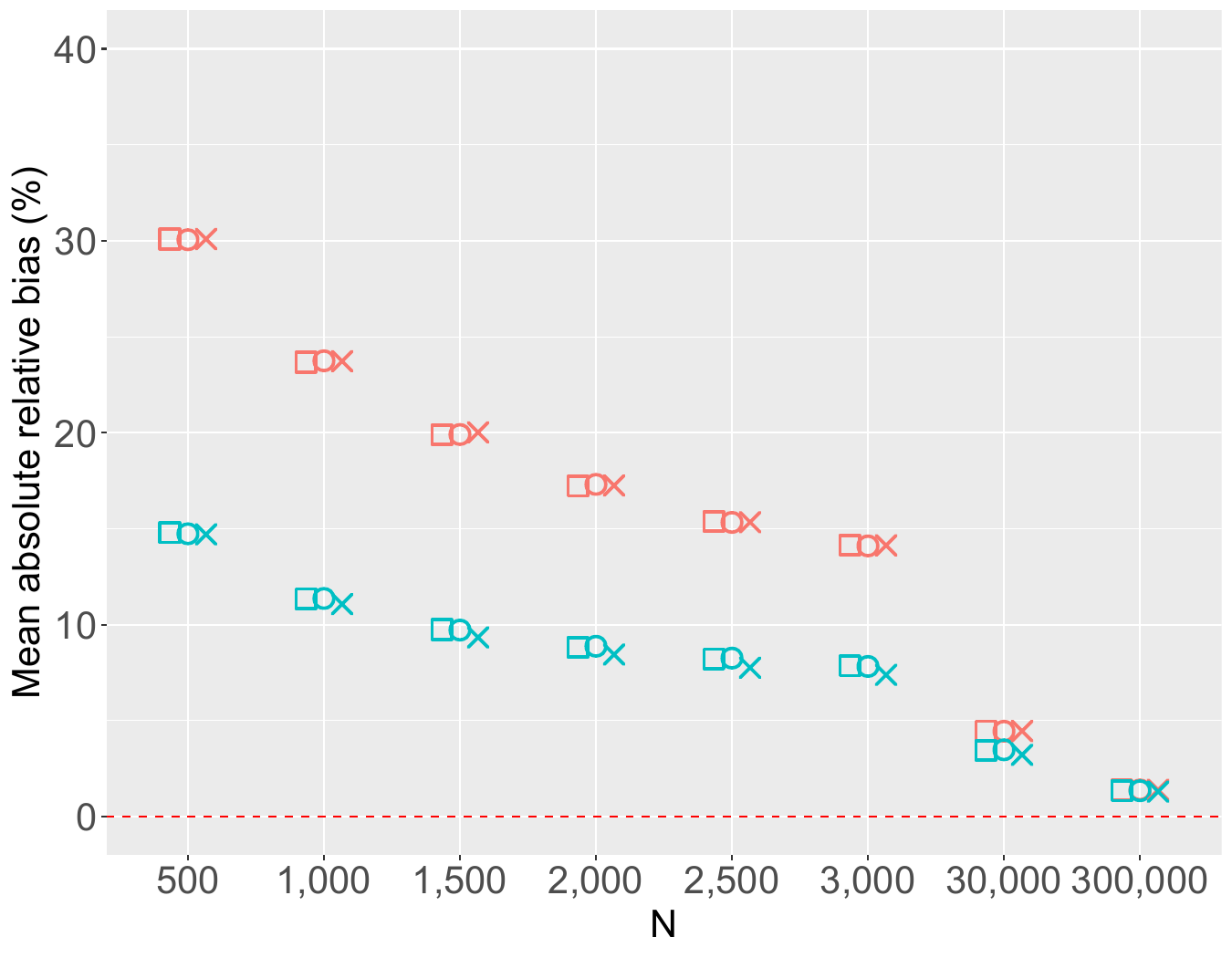}
        \caption{\small 15 Regions}
        \label{arobust_mean_bias_15_low_probs}
    \end{subfigure}\hspace*{\fill}
    \begin{subfigure}[b]{.3\linewidth}
         \includegraphics[width=7cm,height=6cm]{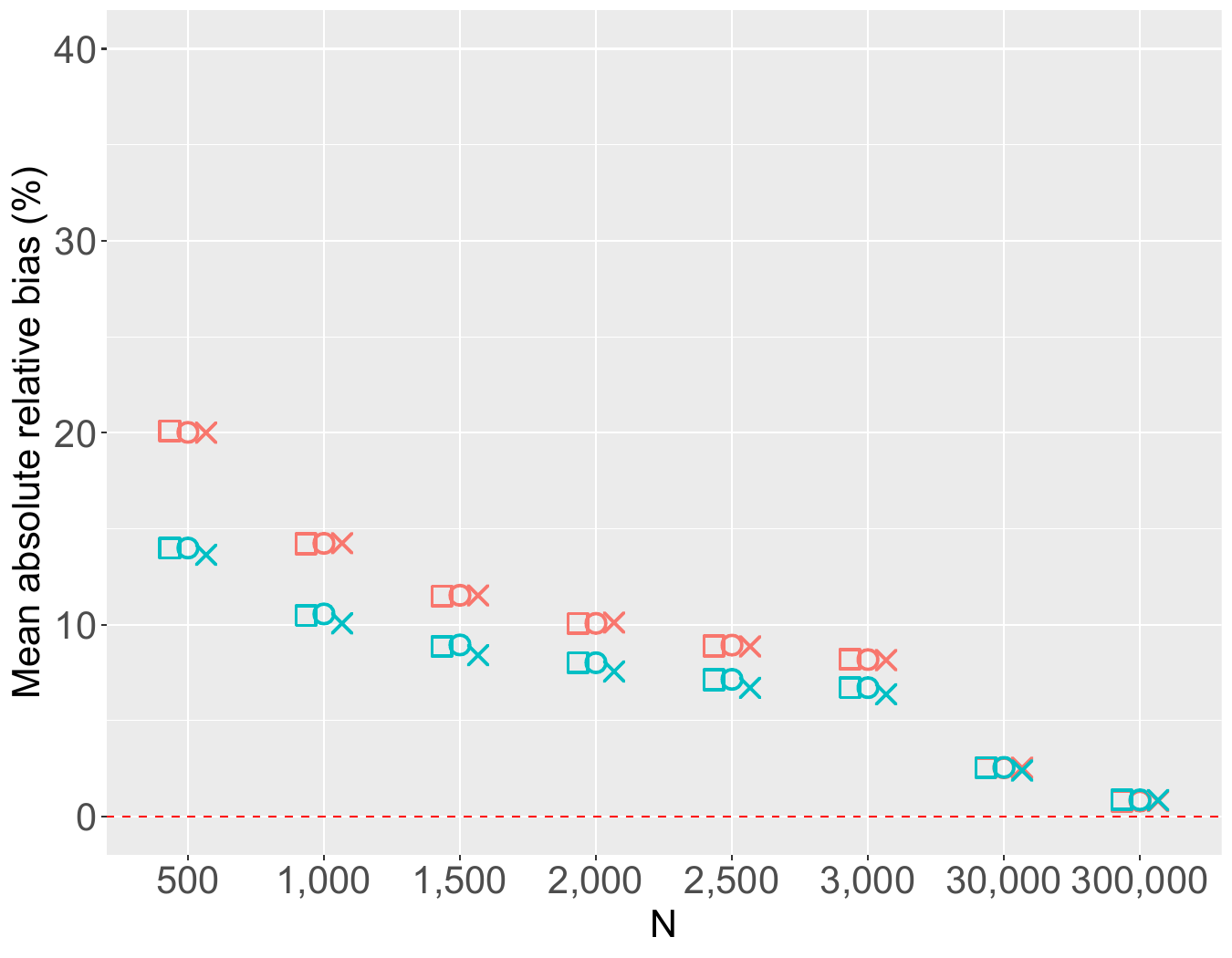}
         \caption{\small 5 Regions}
        \label{arobust_mean_bias_5_low_probs}
    \end{subfigure}\hspace*{\fill}

\caption{Mean absolute relative bias (\%) of estimates where initially $\pi_A = 0.4$ and $\pi_B = 0.2$ and the random effects follow a normal, lognormal or Pareto distribution.}\label{rb_low_probs}

\begin{subfigure}[b]{.3\linewidth}
         \includegraphics[width=7cm,height=6cm]{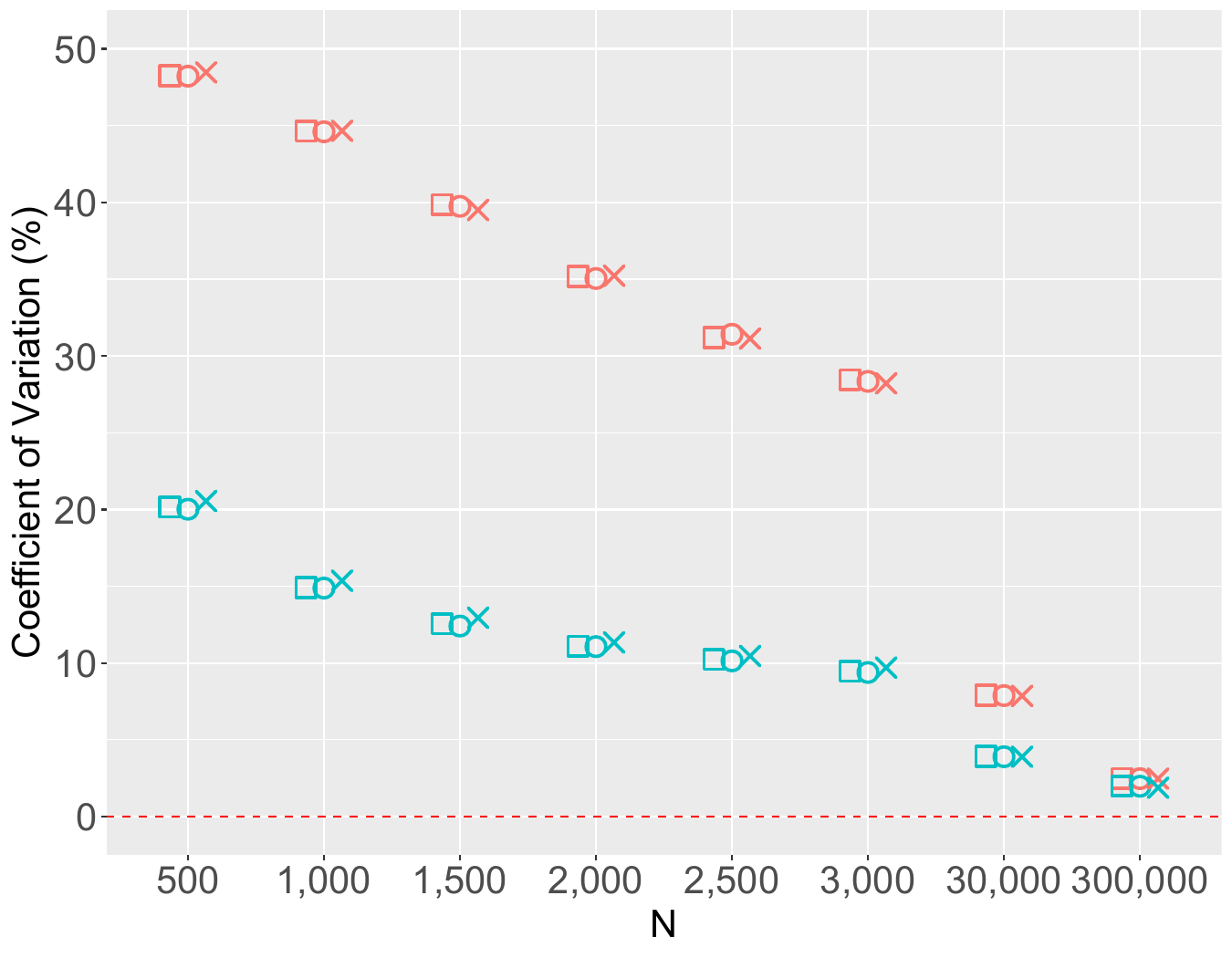}
         \caption{\small 30 Regions}
        \label{cv_30_low_probs}
    \end{subfigure}\hspace*{\fill}
    \begin{subfigure}[b]{.3\linewidth}
         \includegraphics[width=7cm,height=6cm]{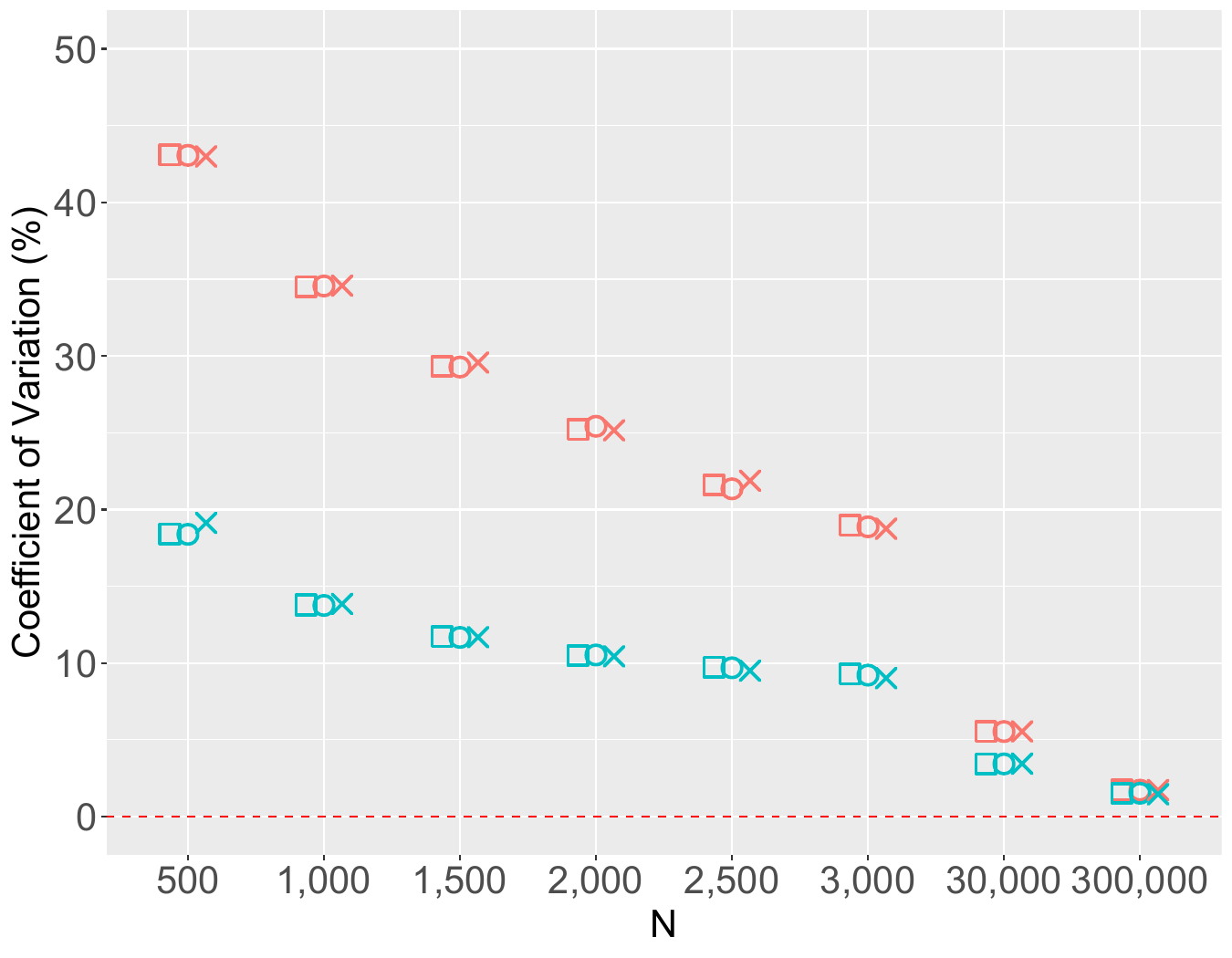}
        \caption{\small 15 Regions}
        \label{cv_15_low_probs}
    \end{subfigure}\hspace*{\fill}
    \begin{subfigure}[b]{.3\linewidth}
         \includegraphics[width=7cm,height=6cm]{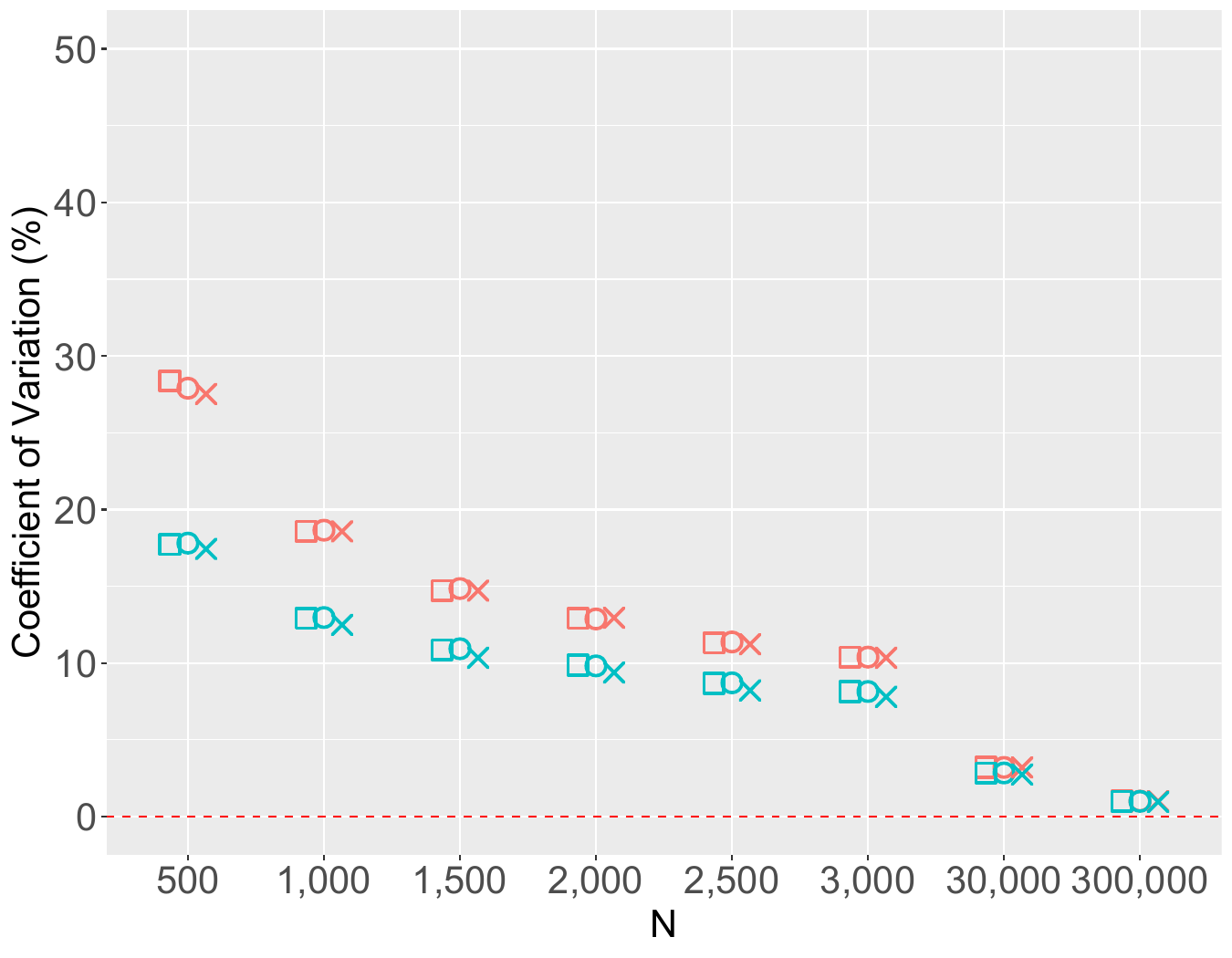}
         \caption{\small 5 Regions}
        \label{cv_5_low_probs}
    \end{subfigure}\hspace*{\fill}
    \end{minipage}
    \begin{minipage}[c]{0.1\textwidth}
        \includegraphics[width=4cm,height=4cm]{legend.png}
    \end{minipage}

\caption{Coefficient of variation (\%) of estimates where initially $\pi_A = 0.4$ and $\pi_B = 0.2$ and the random effects follow a normal, lognormal or Pareto distribution.}\label{rv_low_probs}

\end{sidewaysfigure}

\noindent Figure \ref{rb_high_probs} (and Figure \ref{rb_high_probs2} in Supplementary materials) displays the results for marb, with initial probabilities of $\pi_A = 0.8$ and $\pi_B = 0.7$. The results show that, as the population size $N$ increases, the marb decreases, which is found within all scenarios. This is also found in Figure \ref{rb_high_probs_region} in the supplementary materials, that 
presents box-plots for the regional level rb for all 10,000 samples across all scenarios. Across the scenarios, as the number of regions increases, the marb increases, which is due to decreasing sizes of simulated counts per region. The mixed effects loglinear model either outperforms or performs very similarly to the fixed effects loglinear model with the Chapman estimator; the largest differences are found when $N$ is small and the number of regions is large, a situation where the simulated counts are smallest. This is due to the fixed effects loglinear model with the Chapman estimator overestimating the size of the regions, which is illustrated in Figure \ref{rb_high_probs_region}.
Across all scenarios, when $N = 300,000$, fitting either model results in very small rb. For the mixed effects model, the largest marb is found across all scenarios when $N$ is small, particularly when $N = 500$. Remarkably, the results show that violating the normality assumption by choosing the lognormal or Pareto distribution, results in a marb that is comparable to when the normality assumption is satisfied.

Figure \ref{rb_low_probs} (and Figure \ref{rb_low_probs2} in Supplementary materials) displays the results for marb and Figure \ref{rb_low_probs_region} displays the rb for initial probabilities of $\pi_A = 0.4$ and $\pi_B = 0.2$. Overall the results mirror those in Figure \ref{rb_high_probs} (and Figures \ref{rb_high_probs2} and \ref{rb_high_probs_region} in Supplementary materials). The impact of making capture/inclusion probabilities lower can be seen from the range of the marb. In Figure \ref{rb_high_probs} (and Figure \ref{rb_high_probs2} in Supplementary materials) the marb ranges from 0.11\% to 7.43\%, whereas in Figure \ref{rb_low_probs} (and Figure \ref{rb_high_probs2} in Supplementary materials) the marb ranges from 0.81\% to 35.1\%.

Figure \ref{rv_high_probs} (and Figure \ref{rv_high_probs2} in Supplementary materials) displays the results for the cv, with initial probabilities of $\pi_A = 0.8$ and $\pi_B = 0.7$. As the population size $N$ increases and therefore the simulated counts increase, the cv decreases, within all scenarios. Across the scenarios, as the number of regions increases, the cv increases, due to decreasing simulated counts. The mixed effects loglinear model is again at least as good as the fixed effects loglinear model with the Chapman estimator, with the largest differences where the simulated counts are smallest. For these scenarios, results are comparable when violating the normality assumption by choosing the lognormal or Pareto distribution, compared to when the assumption is satisfied.

Figure \ref{rv_low_probs} (and Figure \ref{rv_low_probs2} in Supplementary materials) displays the results for cv, with initial probabilities of $\pi_A = 0.4$ and $\pi_B = 0.2$. Overall the results mirror those in Figure \ref{rv_high_probs}, however the impact of the lower capture/inclusion probabilities can be seen from the coefficients of variation (\%).

The full mrb, cv and mse results are shown in tables \ref{table:1a}-\ref{table:2e} in the supplementary material.

\section{Models for three-way tables including a stratifying variable}
\label{methods_mse}

\subsection{Fixed effects loglinear models}
\label{fixed_mse_models}

To show how our results can be generalised for more than two lists, we develop the loglinear models discussed in section (\ref{Fixed}) to include a third list, C. For population size estimation, this is referred to as the triple system estimator.
We consider the fixed effects loglinear model, with three dichotomous variables $A$, $B$ and $C$ and a polytomous stratifying variable, such as a  regional level geography, that is available for each of the variables, denoted as $R$. Let $A$ be indexed by $i = 0,1$, $B$ by $j = 0,1$, $C$ by $k = 0,1$ and $R$ by $l = 0,...,r$.
We denote the expected count for cell $(i,j, k, l)$ of the four-way 
contingency table including a stratifying variable by $\mu_{ijkl}$. The saturated loglinear model for three lists is:

\begin{equation}
\begin{split}
\log \mu_{ijkl} = &\lambda + \lambda_i^A + \lambda_j^B + \lambda_k^C + \lambda_{l}^{R} +\lambda_{il}^{AR}+\lambda_{jl}^{BR}+\lambda_{kl}^{CR}+\lambda_{ij}^{AB}+\lambda_{ik}^{AC}+\lambda_{jk}^{BC}+\\
&\lambda_{ijl}^{ABR}+\lambda_{ikl}^{ACR}+\lambda_{jkl}^{BCR}+\lambda_{ijk}^{ABC}+\lambda_{ijkl}^{ABCR}
\end{split}
\label{fixed-3way}
\end{equation}

\noindent where $\lambda$ is the intercept term, $\lambda_i^A$ $\lambda_j^B$, $\lambda^C_{k}$ and $\lambda_l^R$ are respectively the parameters for variables A, B, C and R, $\lambda_{il}^{AR}$, $\lambda_{jl}^{BR}$,  $\lambda_{kl}^{CR}$, $\lambda_{ij}^{AB}$, $\lambda_{ik}^{AC}$ and $\lambda_{jk}^{BC}$ are the two factor interaction parameters, $\lambda_{ijl}^{ABR}$, $\lambda_{ikl}^{ACR}$,  $\lambda_{jkl}^{BCR}$ and $\lambda_{ijk}^{ABC}$ are the three factor interaction parameters and $\lambda_{ijkl}^{ABCR}$ is the four factor interaction parameter. In margin notation this model is [ABCR].

The parameters can be identified by corner point constraints The model is saturated, with the fitted values equal to the observed counts.

In population size estimation, all $(A=0,B=0,C=0)$ cells are unobserved, so the maximal
model is [ABR][ACR][BCR], which omits the interaction terms $\lambda_{ij}^{ABC}$ and $\lambda_{ijl}^{ABCR}$ from equation (\ref{fixed-3way}).

For triple system estimation where the population includes a stratifying variable, one can estimate the population size $\hat{N}$ directly from the observed counts making use of the estimator described by Fienberg (1972)\nocite{FienbergS1972}, 

\begin{equation}
    \hat{\mu}_{000l} = \frac{n_{111l}n_{001l}n_{100l}n_{010l}}{n_{101l}n_{011l}n_{110l}}.
    \label{SF}
\end{equation}

\noindent Or one can also make use of the parameters of a loglinear model as follows,

\begin{equation}
    \hat{\mu}_{000l} = \exp(\hat{\lambda} + \hat{\lambda}_l^R).
    \label{stratifying_fixed}
\end{equation}


\noindent Therefore, $\hat{N_l} = n_{111l} + n_{100l} + n_{010l} + n_{001l} + n_{110l} + n_{011l} + n_{101l} + \hat{\mu}_{000l}$ and $\hat{N} = \sum_{l} \hat{N_l}$.


\subsection{Mixed effects loglinear models}


We can replace the fixed terms in the saturated model (\ref{fixed-3way}), $\lambda_l^R$, $\lambda_{il}^{AR}$, $\lambda_{jl}^{BR}$, $\lambda_{kl}^{CR}$, $\lambda_{ijl}^{ABR}$, $\lambda_{ikl}^{ACR}$, $\lambda_{jkl}^{BCR}$ and $\lambda_{ijkl}^{ABCR}$ with a random intercept term for the stratifying variable R and random slope terms for the two, three and four factor interactions as follows, 

\begin{equation}
\label{covsatmixedall2}
\begin{split}
\log \mu_{ijkl} =&(\lambda + u_{0l}) + \lambda_i^A + \lambda_j^B + \lambda_k^C +\lambda_{ij}^{AB}+\lambda_{ik}^{AC}+\lambda_{jk}^{BC}+
\lambda_{ijk}^{ABC}\\
&+ u_{1il} + u_{2jl} + u_{3kl} + u_{4ijl} + u_{5ikl} + u_{6jkl} + u_{7ijkl}.
\end{split}
\end{equation}


\noindent Here $u_{0l}$ is the random intercept parameter with $u_{0l} \sim N \left(0, \sigma^2_{u0}\right)$, $u_{1il}$ is the random slope parameter for list A with $u_{1il} \sim N \left(0, \sigma^2_{u1}\right)$, $u_{2jl}$ is the random slope parameter for list B with $u_{2jl} \sim N \left(0, \sigma^2_{u2}\right)$, and $u_{3kl}$ is the random slope parameter for list C with $u_{3kl} \sim N \left(0, \sigma^2_{u3}\right)$, $u_{4ijl}$ is the random slope parameter for the two factor interaction between list A and B with $u_{4ijl} \sim N \left(0, \sigma^2_{u4}\right)$, $u_{5ikl}$ is the random slope parameter for the two factor interaction between list A and C with $u_{5ikl} \sim N \left(0, \sigma^2_{u5}\right)$, $u_{6jkl}$ is the random slope parameter for the two factor interaction between list B and C with $u_{6jkl} \sim N \left(0, \sigma^2_{u6}\right)$ and $u_{7ijkl}$ is the random slope parameter for the three factor interaction between list A, B and C with $u_{7ijkl} \sim N \left(0, \sigma^2_{u7}\right)$.



For population size estimation, $n_{000l}$ is not observed and therefore, $\lambda_{ijk}^{ABC}$ and $u_{7ijkl}$ cannot be estimated. The maximal model is

\begin{equation}
\label{covsatmixedall3}
\begin{split}
\log \mu_{ijkl} =&(\lambda + u_{0l}) + \lambda_i^A + \lambda_j^B + \lambda_k^C +\lambda_{ij}^{AB}+\lambda_{ik}^{AC}+\lambda_{jk}^{BC}+ u_{1il} + u_{2jl}\\
& + u_{3kl} +u_{4ijl} + u_{5ikl} + u_{6jkl}.
\end{split}
\end{equation}

By making use of the parameters of the mixed effects loglinear model, we can estimate the unobserved cell ${n}_{000l}$ for each level of the stratifying variable:

\begin{equation}
\begin{split}
\hat{\mu}_{000l} = &\exp((\hat{\lambda} + \hat{u}_{0l}) + (\hat{\lambda}_i^A + \hat{u}_{1il}) + (\hat{\lambda}_j^B + \hat{u}_{2jl}) + (\hat{\lambda}_k^C + \hat{u}_{3kl}) +(\hat{\lambda}_{ij}^{AB} + \hat{u}_{4ijl})\\ 
&  +(\hat{\lambda}_{ik}^{AC} + \hat{u}_{5ikl})+ (\hat{\lambda}_{jk}^{BC} + \hat{u}_{6jkl})).
\end{split}
\label{stratifying_mixed2}
\end{equation}

\noindent Once again, $\hat{N_l} = n_{111l} + n_{100l} + n_{010l} + n_{001l} + n_{110l} + n_{011l} + n_{101l} + \hat{\mu}_{000l}$ and $\hat{N} = \sum_{l} \hat{N_l}$.

\section{Discussion and conclusion}
\label{Discussion}

For dual system estimation, the mixed effects loglinear model outperforms the fixed effects (with or without the Chapman estimator) in terms of the mean absolute relative bias, the coefficient of variation and the mean squared error. The fixed effects loglinear model without the Chapman estimator regularly results in infinite estimates, particularly for small sample sizes. The mixed effects model has better properties when the sample sizes are small -- either when the overall population is small, when a fixed population is split over many regions, or when the coverage probability in one or more sources is low.

Under the high and low inclusion probability scenarios, one can relax the normality assumption, as fitting a mixed effects loglinear model where this assumption is satisfied or violated produces comparable results. However, if the variance of the distributions is substantially larger than those tested in the simulation study, there is still a risk that violating the normality assumption will have a more substantial impact on the estimates.

The simulation data are generated under the independence model. We also introduced dependence between the lists A and B for each Region, specifically when the odds ratio = 2, using the method proposed by Hammond, van der Heijden and Smith (2024)\nocite{hammond2024generating} (results not presented). The results of the mixed effects loglinear model (\ref{mixedslopemaximal22}) looked as expected (Gerritse, van der Heijden and Bakker, 2015)\nocite{GerritsevanderHeijdenBakker2015}, where positive odds ratios result in population size estimates being negatively biased. But the relative performance of the mixed and fixed effects models were qualitatively the same, with the mixed effects model performing better.

In summary the well-known and much-used Chapman estimator performs well in our scenarios, but the mixed effects model has better properties when there are many regions, although the differences are small. The mixed effects models' performance is robust to the assumption that the random effects are normally distributed, at least over the range of variation considered here.

\clearpage

\begingroup
\setstretch{1.0}
\printbibliography
\endgroup

\clearpage

\section*{Supplementary Material}
\label{Supplementary Material}
\setcounter{figure}{0}
\renewcommand{\thefigure}{S.\arabic{figure}}

\renewcommand{\thesubsection}{\Alph{subsection}} 
\setcounter{subsection}{0}

\setcounter{table}{0}
\renewcommand{\thetable}{S.\arabic{table}}

\subsection{Simulated data}
\label{Appendix A1}

We describe how we generate a single dataset for 10 regions, with the larger inclusion probabilities ($\pi_{A}$ = 0.8, $\pi_{B}$ = 0.7), where the mixed model normality assumption holds, as in model (\ref{mixedslopemaximal22}).

\begin{enumerate}

    \item{For a population of 10 regions we initially assume homogeneous region sizes $N_l$ and for each region the conditional inclusion probabilities for the lists are $\pi_{A|l} = 0.8$ and $\pi_{B|l} = 0.7$. Therefore, at this step, for each region, the joint probabilities conditional on $l$ are ($\pi_{11|l}, \pi_{10|l}, \pi_{01|l}, \pi_{00|l}$) = (0.56, 0.24, 0.14, 0.06). As there are 10 regions of identical size $N_l$, the unconditional probabilities in each region $l$ are
    (0.056, 0.024, 0.014, 0.006). For the moment the probabilities are multiplied by some arbitrary population size, for example, 10,000, so that the data can be analysed.} 
    \item{The fixed effects independence model is fitted to the data. The loglinear parameter estimates 
    $\hat{\lambda}$, $\hat{\lambda}_1^A$, $\hat{\lambda}_1^B$, $\hat{\lambda}_l^R$, $\hat{\lambda}_{1l}^{AR}$ and  $\hat{\lambda}_{1l}^{BR}$ are stored.
    As all regions have the same initial $N_l$ and inclusion probabilities, the loglinear parameters are the same across all regions, so that $\hat{\lambda}_l^R =  \hat{\lambda}_{1l}^{AR} =  \hat{\lambda}_{1l}^{BR} = 0$.}
    
    \item{To introduce heterogeneous region sizes where the data follows the normality assumption (where the random effects are normally distributed), and there are 10 regions, we generate 10 random numbers from a normal distribution with mean $\mu = 0$ and variance $\sigma^2_{u0}$. A simulation study used to inform the chosen variance for $\sigma^2_{u0}$ is discussed in supplementary material \ref{appendix A}. We use a variance of 0.0143.
    One of the 10 generated numbers is added to each $\hat{\lambda}_l^R$, $l=1,...,10$.} 

    \item{To introduce heterogeneous 
    inclusion probabilities for each list across the 10 regions, where the random effects are normally distributed, we follow the same idea as for the random slope terms by generating two sets of ten normally distributed random numbers (one each for A and B) with means 0 and variances  $\sigma^2_{u1}$ and $\sigma^2_{u2}$ respectively. The simulation study used to inform the chosen variances for $\sigma^2_{u1}$ and $\sigma^2_{u2}$ is also discussed in supplementary material \ref{appendix A}. We use variances of 0.0253 and 0.0232. 
    These numbers are added to the fixed loglinear parameters $\hat{\lambda}_{1l}^{AR}$ and  $\hat{\lambda}_{1l}^{BR}$, $l = 1,...10$.}

\end{enumerate}

This results in a population with 10 regional population sizes, and 10 inclusion probabilities for lists $A$ and $B$ that are variable across regions. Note that we started with 
conditional inclusion probabilities  $\pi_{A|l} = 0.8$ and $\pi_{B|l} = 0.7$ for each list, and the loglinear parameter estimates are modified symmetrically by normally distributed numbers. However, from the inverse link function, the regional probabilities are no longer symmetric and this results in mean coverage probabilities that deviate slightly from 0.8 and 0.7 respectively (even in expectation).
   
This process generates a single population for 10 regions. In the simulation the values in the $10 \times 2 \times 2$ table are converted to a single set of probabilities. These probabilities are used to draw 100 multinomial samples of size $N$. 
If we draw an example sample for population size $N = 1,000$, the simulated region sizes $N_l$ range from 75.41 to 128.64, the conditional inclusion probabilities $\pi_{A|l}$ range from 0.74 to 0.84, and $\pi_{B|l}$ range from 0.63 to 0.73. 
Steps 3 and 4 are repeated 100 times to generate 100 populations, each with new random numbers from the normal distributions. For each of the 100 populations, 100 multinomial samples are drawn, and we are then able to assess the impacts of the variation in region properties between populations and of the sampling process within populations. 
Because of the way the populations are generated, the true region sizes $N_l$ are decimal numbers. However, to generate multinomial samples requires the regional counts to be integers. The procedure we use to solve for this while keeping the region sizes fixed is explained in Appendix \ref{Appendix B}.

\subsection{Simulation study for chosen variances}
\label{appendix A}

To inform the chosen variances for $\sigma^2_{u0}$, $\sigma^2_{u1}$ and $\sigma^2_{u2}$, in the simulated data (section \ref{simulated data}) a separate simulation study with the following steps is used. Contrary to the main simulation study where 100 populations are generated and within each population 100 multinomial samples are drawn, in this simulation study one population is generated and 1000 multinomial samples are drawn. The purpose of this simulation study is to come up with variances that are large, yet result in the smallest number of convergence warnings from the mixed effects loglinear model, which is subject to convergence warnings particularly for small sample sizes.
\begin{enumerate}[i)]
    \item {Initially the parameters, $N = 500$, 30 regions, $\pi_A = 0.8$ and $\pi_B = 0.7$ were chosen. 
    When $N = 500$ and the number of regions is 30, the average number of observations per cell is lowest over all conditions, and this results in the largest number of convergence warnings from the mixed effects loglinear model across all of the scenarios.}
     \item{To introduce heterogeneous inclusion probabilities for each list across the 30 regions, uniform random numbers are generated, where $ 0.7 \leq \pi_A \leq 0.9$ and $ 0.6 \leq \pi_B \leq 0.8$. Therefore, this results in varying conditional inclusion probabilities for each list, $\pi_{A|l}$ ranges from 0.705 to 0.896 and $\pi_{B|l}$ ranges from 0.602 to 0.794.}
     \item{To introduce heterogeneous region sizes, Poisson distributed numbers are generated with rate $\lambda$. For example, for a population with 30 regions, 30 Poisson distributed numbers are generated, where $\lambda = \frac{N}{30}$. This results in $N_l$ ranging from 10 to 25 and $N =$ 526.}
    \item{The previous steps generate a single population for 30 regions. In the simulation the values in the $30 \times 2 \times 2$ table are converted to a single set of probabilities. These probabilities are used to draw 1000 multinomial samples of region sizes $N_l$.}
    \item{Model (\ref{mixedslopemaximal22}) is fitted to each of the 1000 multinomial samples, and for models where there were no convergence warnings, the mean of the estimated random effects for $\sigma^2_{u0}$, $\sigma^2_{u1}$ and $\sigma^2_{u2}$ are calculated, giving $\sigma^2_{u0}$ = 0.0143, $\sigma^2_{u1}$ = 0.0253 and $\sigma^2_{u2}$ = 0.0232.}
\end{enumerate}
These values of $\sigma^2_{u0}$, $\sigma^2_{u1}$ and $\sigma^2_{u2}$ are used across all scenarios. We present the results for one option for the chosen variances informed by the simulation study described in supplementary material \ref{appendix A}. We investigated the impacts of different variances for $\sigma^2_{u0}$, $\sigma^2_{u1}$ and $\sigma^2_{u2}$ by changing the the initial parameters in Step 1 and also choosing the maximum value of the estimated random effects at which no convergence issues occurred. For example, when $N$ = 500, 30 regions, $\pi_A = 0.4$ and $\pi_B = 0.2$, which resulted in $\sigma^2_{u0} = 0.1583$, $\sigma^2_{u1} = 0.3515$ and $\sigma^2_{u2} = 0.2859$. The simulation study involves generating 100 populations and 100 multinomial samples, but the increased variance mainly affects the generation of populations. For example, an increase in $\sigma^2_{u0}$, $\sigma^2_{u1}$ and $\sigma^2_{u2}$ results in an increase in the variability across the populations, whereas the results for the multinomial samples are comparable with those presented in section \ref{results} and supplementary material \ref{Supplementary Material Results}.

\subsection{How to resolve decimal numbers for $N_l$}
\label{Appendix B}

The simulated data discussed in section (\ref{simulated data}) is generated in such a way that the true region sizes $N_l$ in a generated population are decimal numbers. However, to generate multinomial samples for each region, using \texttt{rmultinom()} in \texttt{R}, requires integer region counts. The way this is resolved, is explained in the following steps, which are followed afresh for each multinomial sample from the generated population:

\begin{enumerate}
    \item The true region counts $N_l$ are calculated by multiplying $\pi_{11l}$, $\pi_{10l}$, $\pi_{01l}$ and $\pi_{00l}$ by $N$, which results in fractional numbers.
    \item The fractional part from each $N_l$ is stored.
    \item For each region, \texttt{rbinom()} is used to draw a binary outcome, where the fractional parts from each $N_l$ are used as the probabilities.
    \item If the generated binary outcome is 1, the region count $N_l$ is rounded up to the nearest integer, if the outcome is 0, the region count $N_l$ is rounded down to the nearest integer. Label the resulting integer value as $N_l^*$.
    \item Once $N_l^*$ is obtained for each region, calculate the total population size $N_S=\sum_l N_l^*$.
    \item If $N = N_S$, the multinomial samples are drawn for each region, which results in one of the 100 multinomial samples from the generated population.
    \item \begin{enumerate}
        \item If $N > N_S$, $N_S$ will need to be increased by $N - N_S$. Therefore, the differences between the true (fractional) $N_l$ and rounded $N_{l}^*$ are calculated. 
        The regions ($l$) with the largest negative difference are where the remaining counts should preferentially be added. 
        \item This is done by multiplying all the differences between the true $N_l$ and rounded $N_{l}^*$ by $-1$. Negative differences become positive and positive differences become negative. The range remains the same. 
        \item Then take the most negative difference value and add this value onto all the differences so the minimum will be zero. 
        \item Draw one multinomial sample of size $N - N_S$, using the adjusted differences between the true $N_l$ and rounded $N_{l}^*$ for the probabilities.
        \item Add the generated sample to the $N_{l}^*$. Now $N = N_S$ and a multinomial sample is drawn for each region, which results in one multinomial sample for the given population.
        \end{enumerate}
    \item \begin{enumerate}
        \item If $N < N_S$ the difference between $N$ and $N_S$ will be removed from $N_S$. Therefore, the differences between the true (fractional) $N_l$ and rounded $N_{l}^*$ are calculated. The regions ($l$) with the largest positive difference are where the remaining counts should preferentially be removed. 
        \item This is done by taking the minimum difference between the true $N_l$ and rounded $N_{l}^*$ and then adding this value onto all the differences. The minimum will be zero where maximum will be the range value. 
        \item ($\dagger$) Draw one multinomial samples of size $N_S-N$ and the adjusted difference between the true $N_l$ and rounded $N_{l}^*$ for the probabilities. 
        \item Subtract the generated sample from the $N_{l}^*$. If any resulting region size is $<0$, return to ($\dagger$) and redraw the multinomial sample. 
        \item Now, $N = N_S$ and a multinomial sample is drawn for each region, which results in one multinomial sample for the given population.
        \end{enumerate}
\end{enumerate}

\subsection{Assessing the effect of deviations from the normality assumption}
\label{Appendix D}

To assess the effect of deviations from the assumption of normally distributed random effects, we generate random numbers from the lognormal and Pareto distributions instead of a normal distribution.
The parameters of the lognormal and Pareto distributions are fitted by the method of moments to match the normal; the expected value of the lognormal and Pareto ($\hat{\mu} = 1$) is subtracted from the generated values so that they are still centered at 0.
For the lognormal distribution, if we draw a sample for 10 regions and population size $N = 1,000$, for this sample the simulated region sizes $N_l$ range from 80.69 to 127.09 and the conditional inclusion probabilities $\pi_{A|l}$ range from 0.75 to 0.84, and $\pi_{B|l}$ range from 0.64 to 0.73.
For the Pareto distribution, if we draw a sample for 10 regions and population size $N = 1,000$, for this sample the simulated region sizes $N_l$ range from 80.17 to 166.64 and the conditional inclusion probabilities $\pi_{A|l}$ range from 0.78 to 0.89, and $\pi_{B|l}$ range from 0.67 to 0.75.

\clearpage
\begin{sidewaysfigure}[p]
\subsection{Results}
\label{Supplementary Material Results}
\begin{minipage}{0.9\textwidth}
\centering
\begin{subfigure}[b]{.3\linewidth}
         \includegraphics[width=7cm,height=6cm]{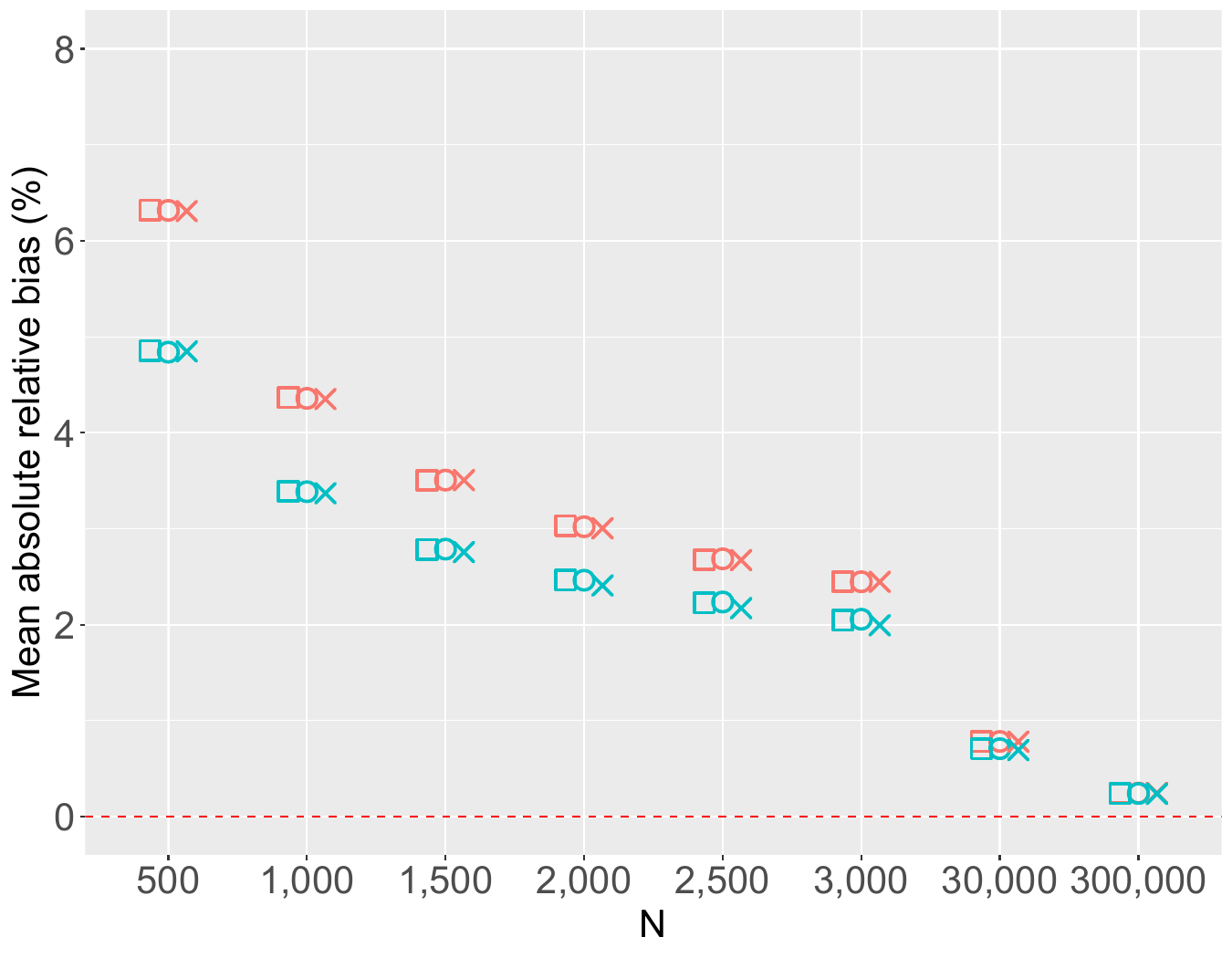}
         \caption{\small 25 Regions}
        \label{arobust_mean_bias_25_high_probs}
    \end{subfigure}\hspace*{\fill}
    \begin{subfigure}[b]{.3\linewidth}
         \includegraphics[width=7cm,height=6cm]{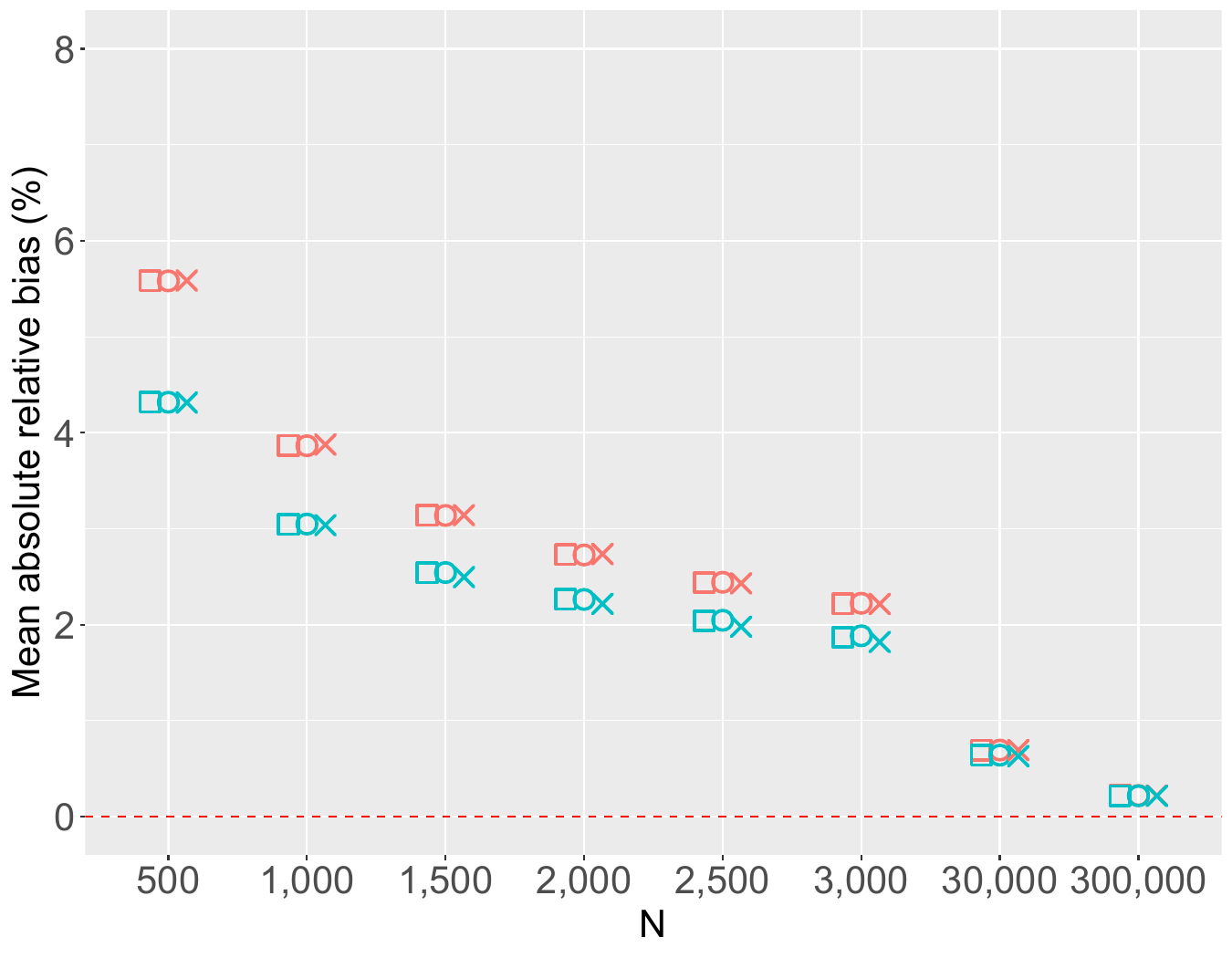}
        \caption{\small 20 Regions}
        \label{arobust_mean_bias_20_high_probs}
    \end{subfigure}\hspace*{\fill}
    \begin{subfigure}[b]{.3\linewidth}
         \includegraphics[width=7cm,height=6cm]{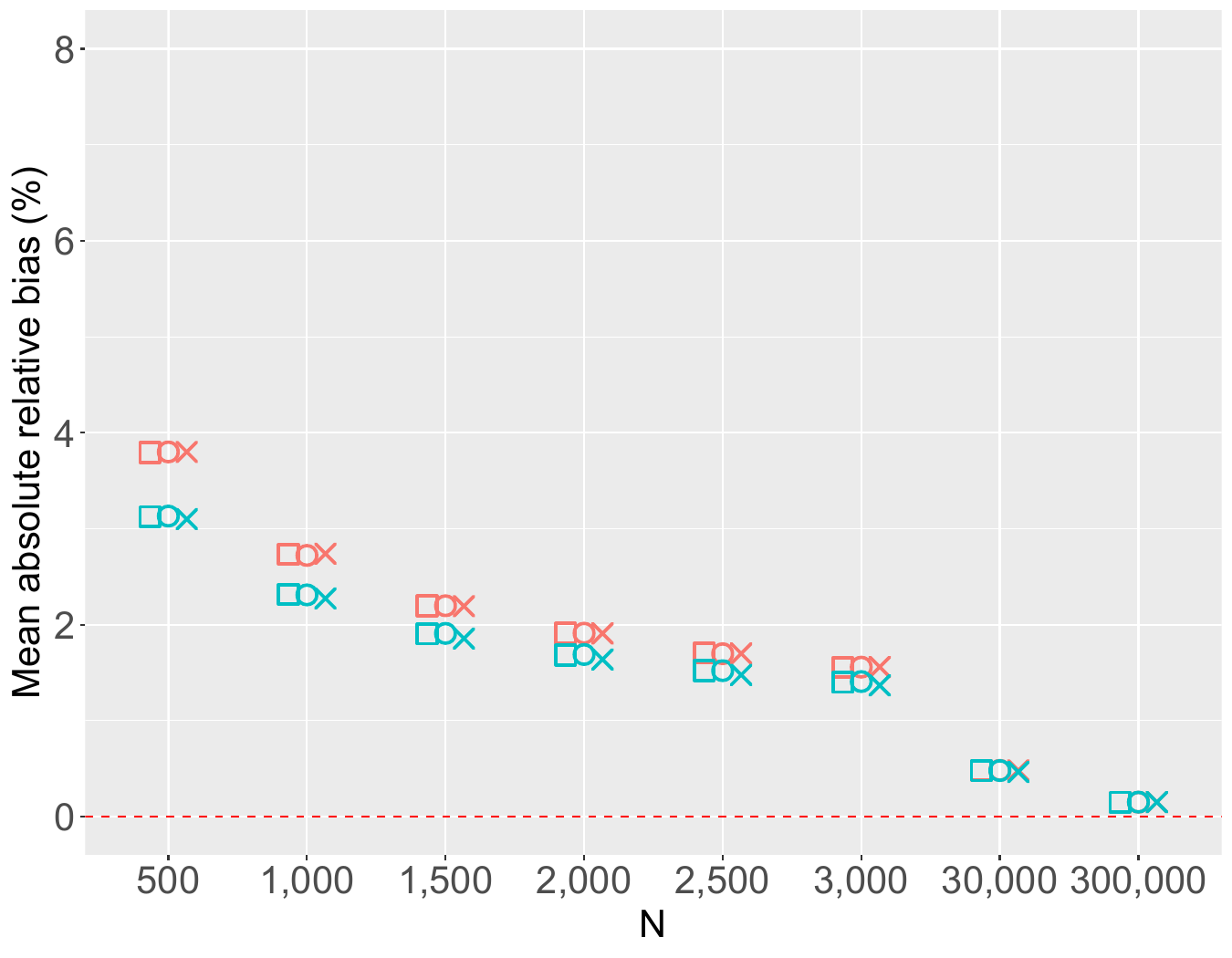}
         \caption{\small 10 Regions}
        \label{arobust_mean_bias_10_high_probs}
    \end{subfigure}\hspace*{\fill}

\caption{Mean absolute relative bias (\%) of estimates where initially $\pi_A = 0.8$ and $\pi_B = 0.7$ and the random effects follow a normal, lognormal or Pareto distribution.}\label{rb_high_probs2}

\begin{subfigure}[b]{.3\linewidth}
         \includegraphics[width=7cm,height=6cm]{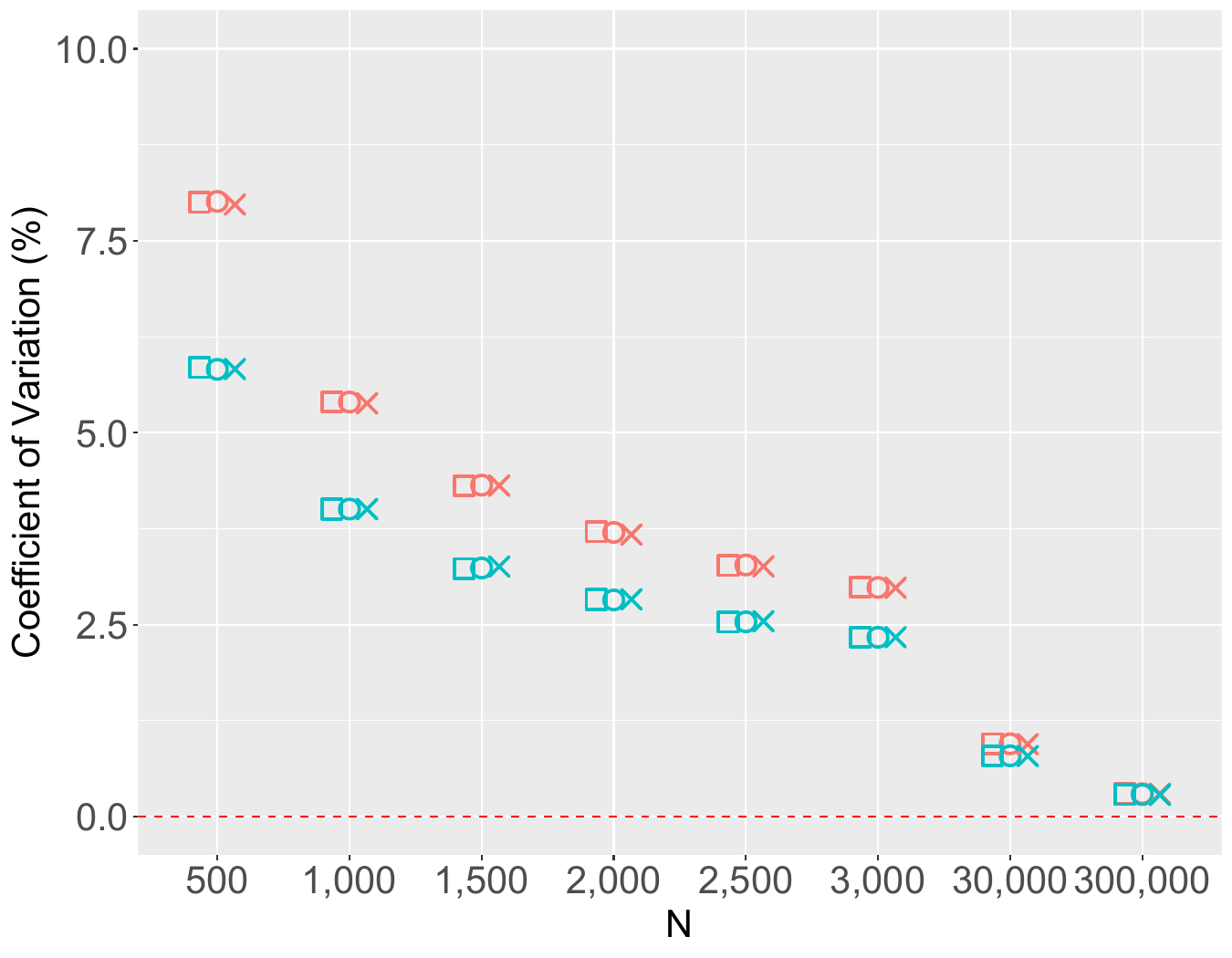}
         \caption{\small 25 Regions}
        \label{cv_25_high_probs}
    \end{subfigure}\hspace*{\fill}
    \begin{subfigure}[b]{.3\linewidth}
         \includegraphics[width=7cm,height=6cm]{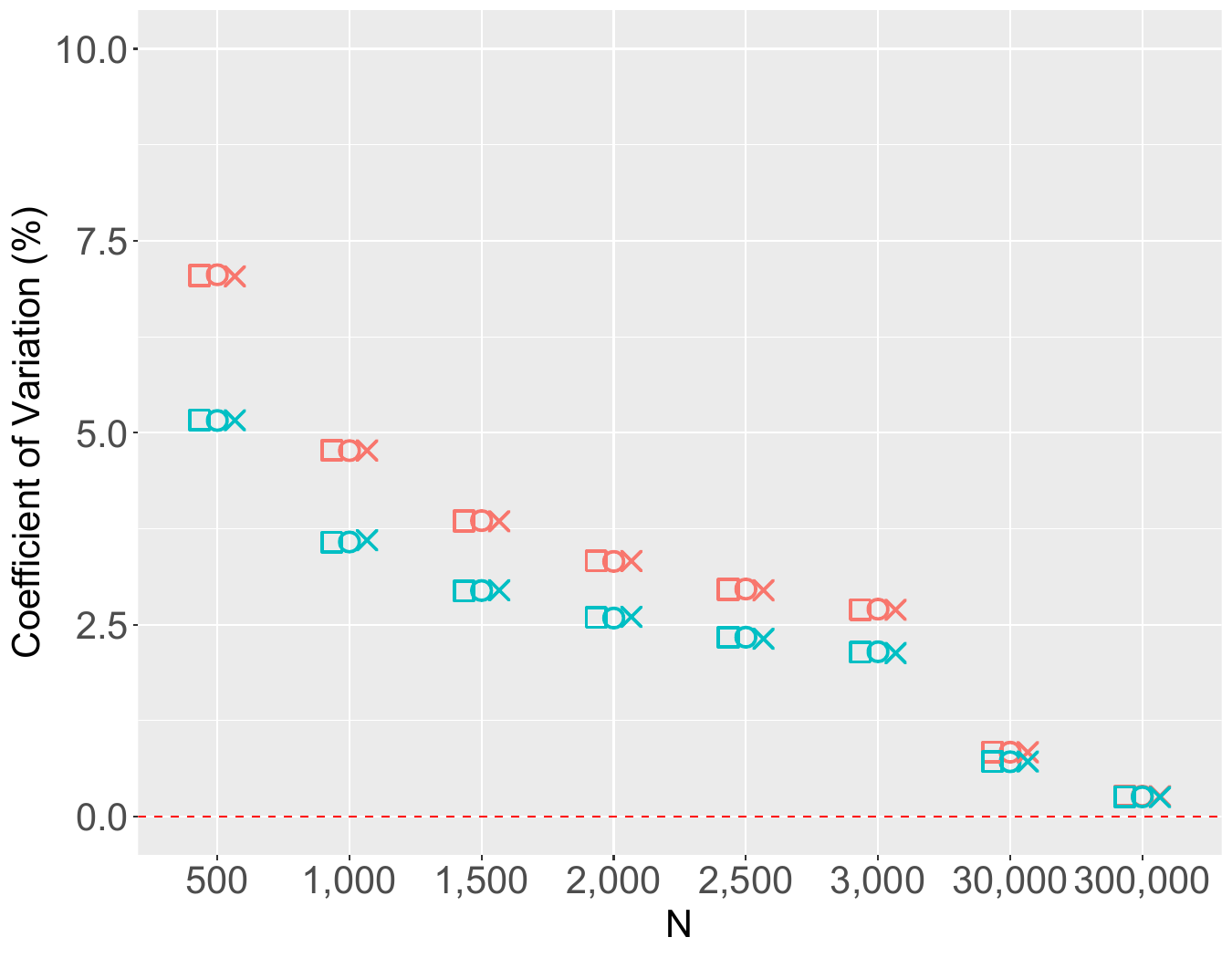}
        \caption{\small 20 Regions}
        \label{cv_20_high_probs}
    \end{subfigure}\hspace*{\fill}
    \begin{subfigure}[b]{.3\linewidth}
         \includegraphics[width=7cm,height=6cm]{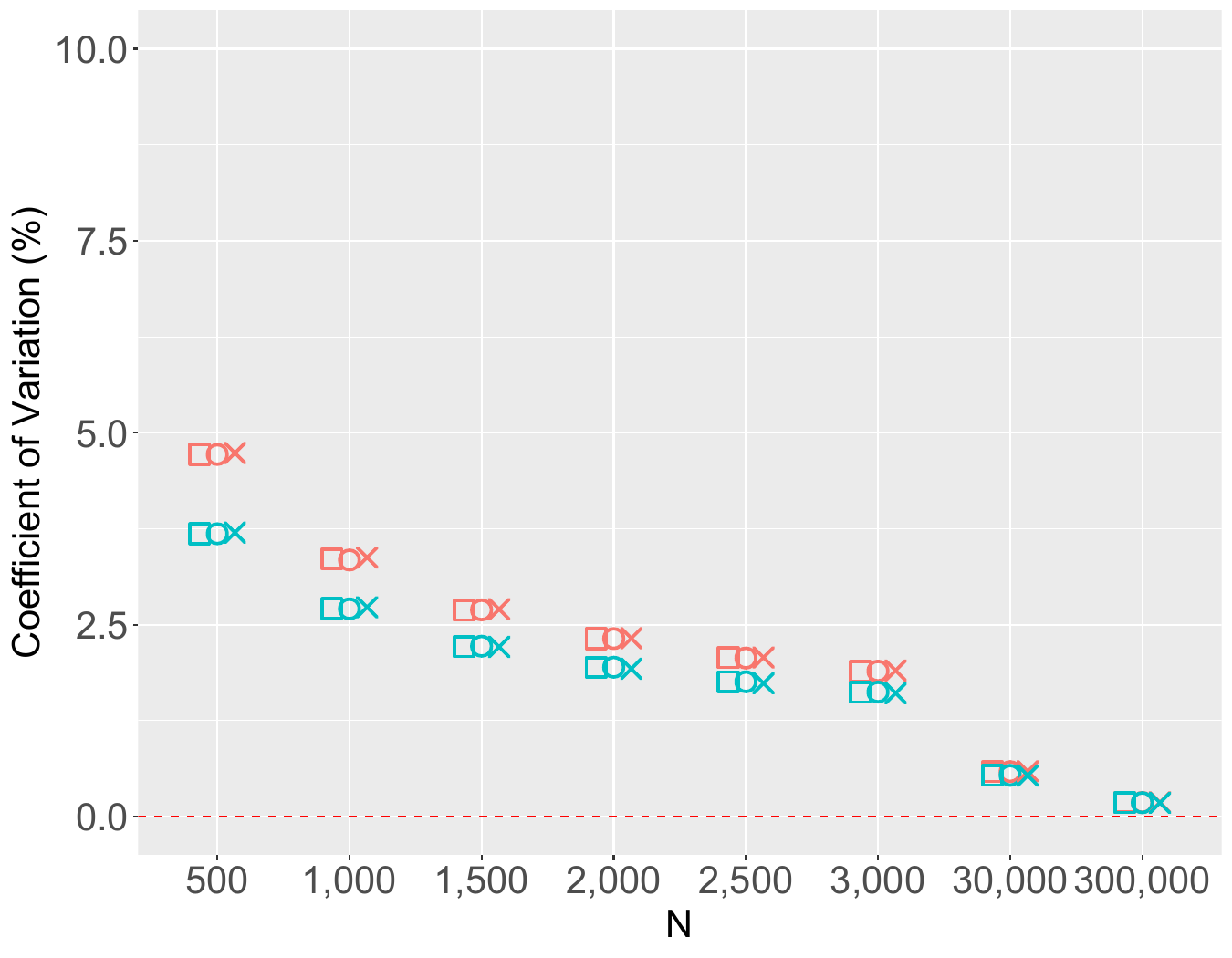}
         \caption{\small 10 Regions}
        \label{cv_10_high_probs}
    \end{subfigure}\hspace*{\fill}
    \end{minipage}
    \begin{minipage}[c]{0.1\textwidth}
        \includegraphics[width=4cm,height=4cm]{legend.png}
    \end{minipage}

\caption{Coefficient of variation (\%) of estimates where initially $\pi_A = 0.8$ and $\pi_B = 0.7$ and the random effects follow a normal, lognormal or Pareto distribution.}\label{rv_high_probs2}

\end{sidewaysfigure}

\clearpage

\begin{sidewaysfigure}[p]
\begin{minipage}{0.9\textwidth}
\centering
\begin{subfigure}[b]{.3\linewidth}
         \includegraphics[width=7cm,height=6cm]{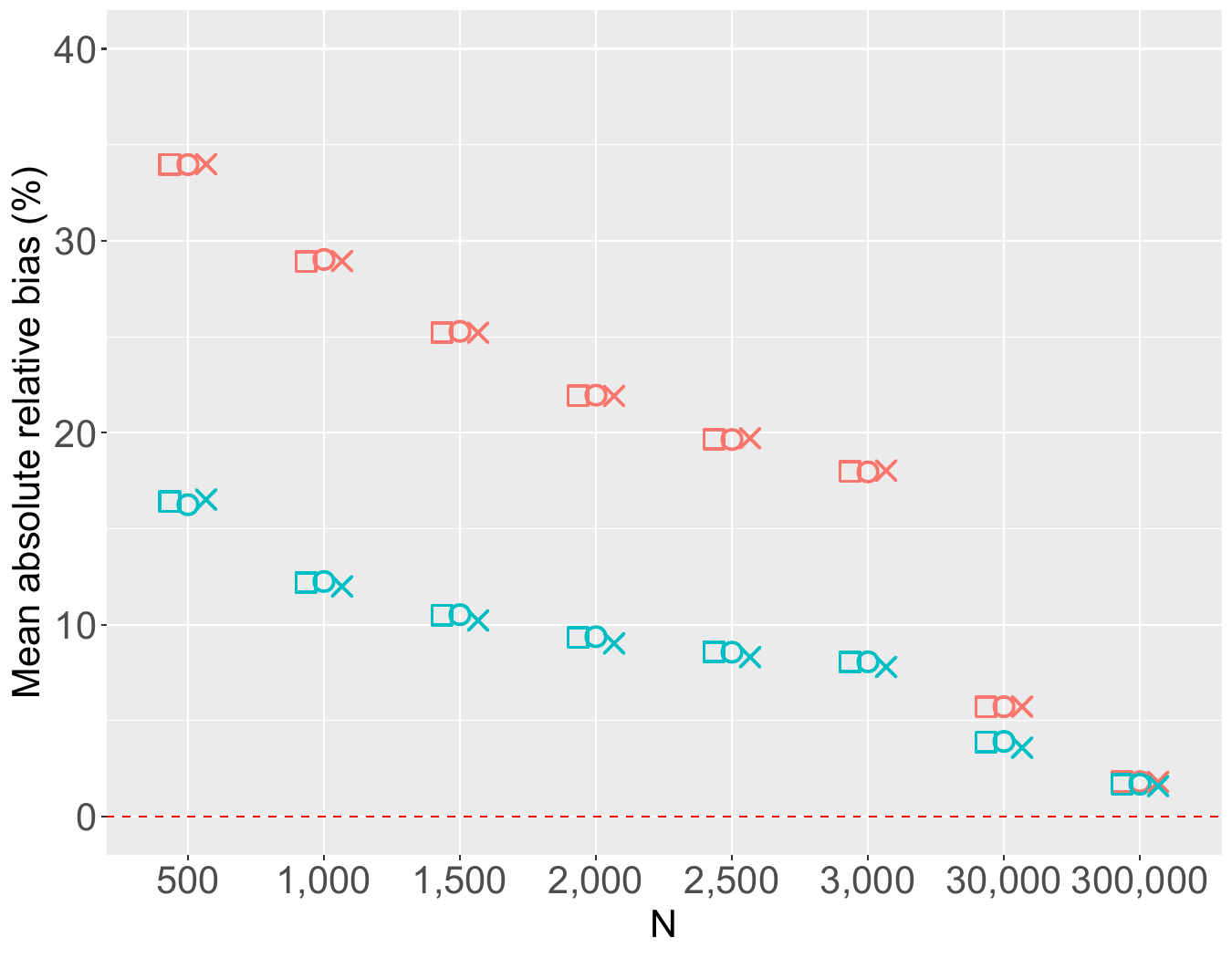}
         \caption{\small 25 Regions}
        \label{arobust_mean_bias_25_low_probs}
    \end{subfigure}\hspace*{\fill}
    \begin{subfigure}[b]{.3\linewidth}
         \includegraphics[width=7cm,height=6cm]{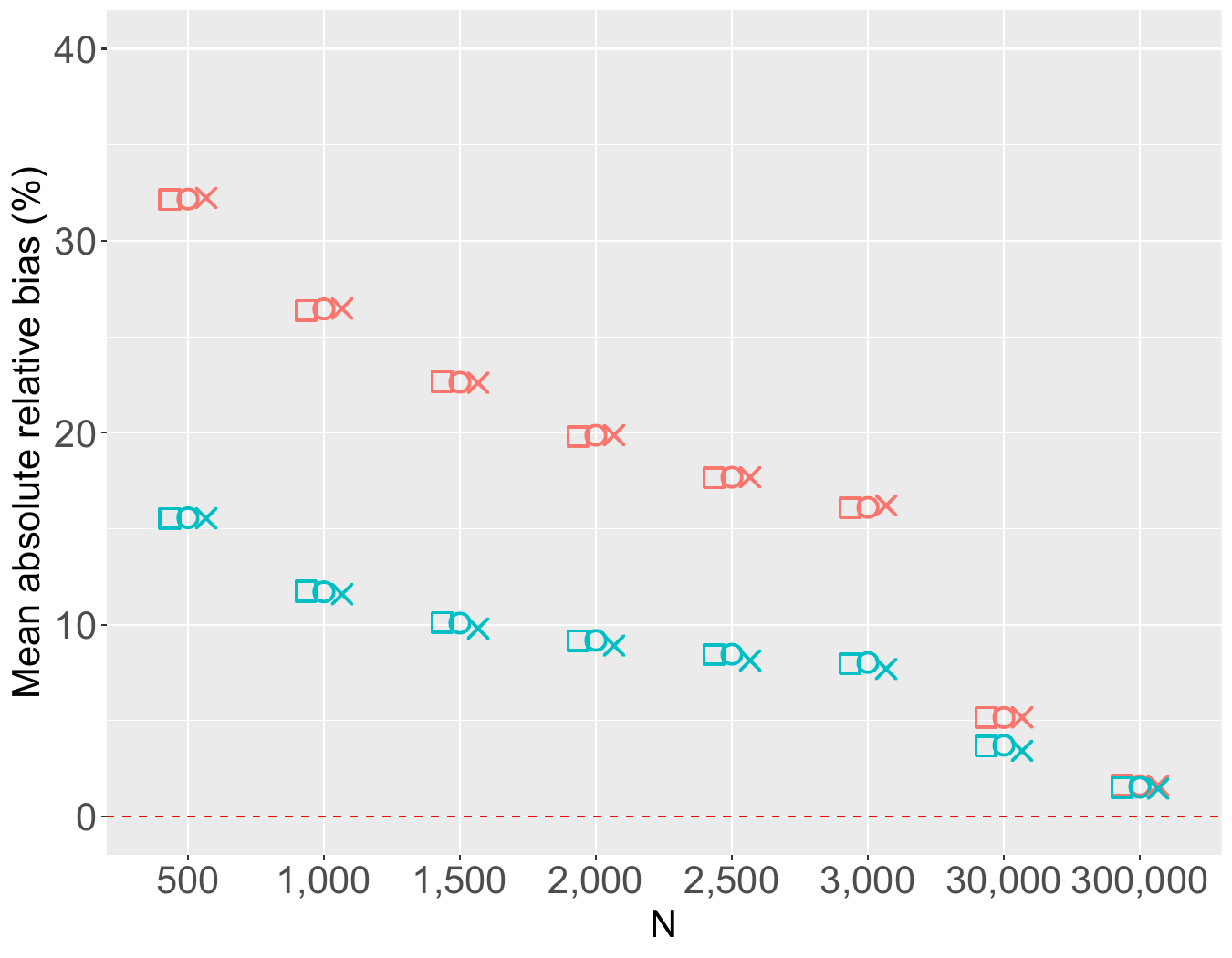}
        \caption{\small 20 Regions}
        \label{arobust_mean_bias_20_low_probs}
    \end{subfigure}\hspace*{\fill}
    \begin{subfigure}[b]{.3\linewidth}
         \includegraphics[width=7cm,height=6cm]{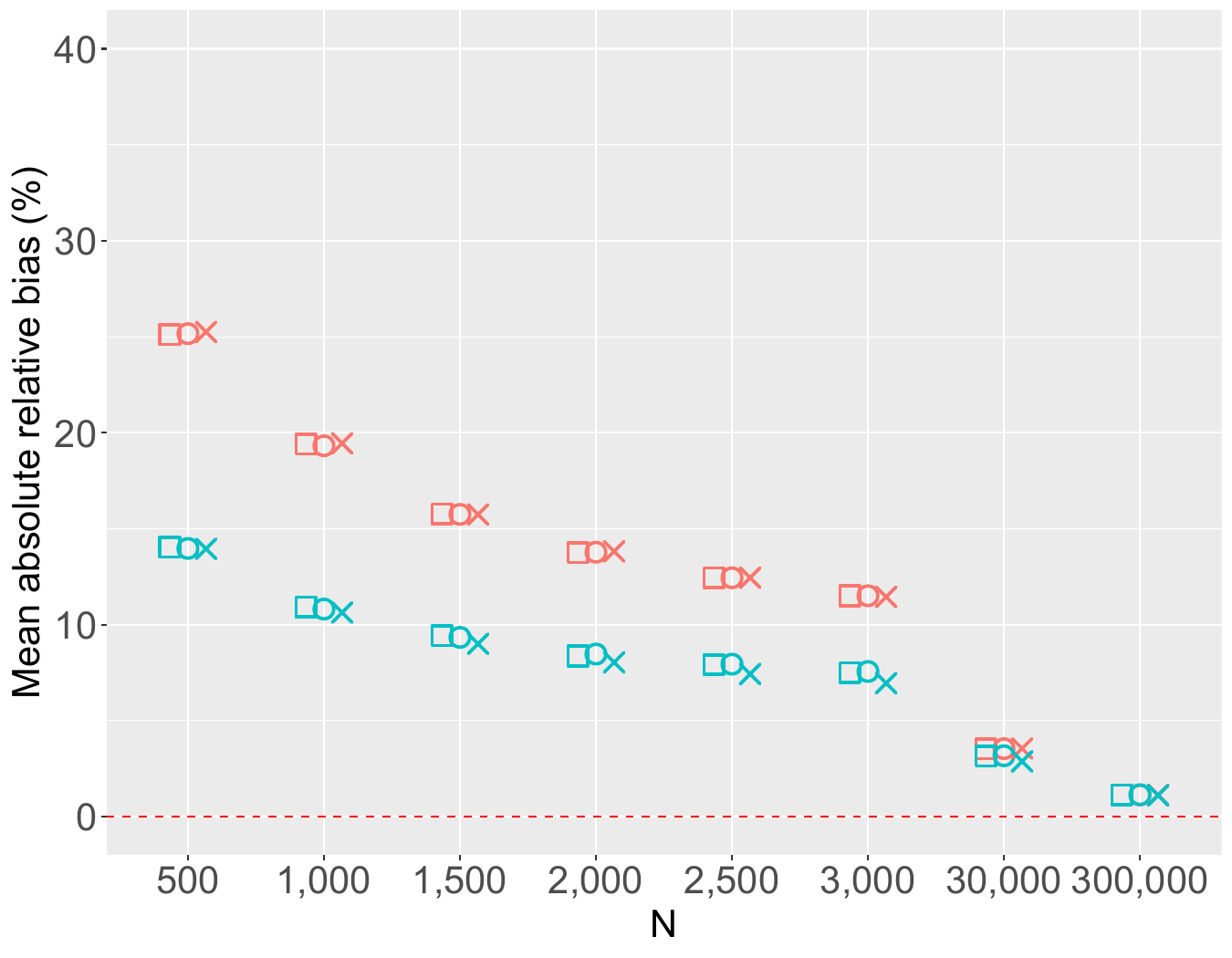}
         \caption{\small 10 Regions}
        \label{arobust_mean_bias_10_low_probs}
    \end{subfigure}\hspace*{\fill}

\caption{Mean absolute relative bias (\%) of estimates where initially $\pi_A = 0.4$ and $\pi_B = 0.2$ and the random effects follow a normal, lognormal or Pareto distribution.}\label{rb_low_probs2}

\begin{subfigure}[b]{.3\linewidth}
         \includegraphics[width=7cm,height=6cm]{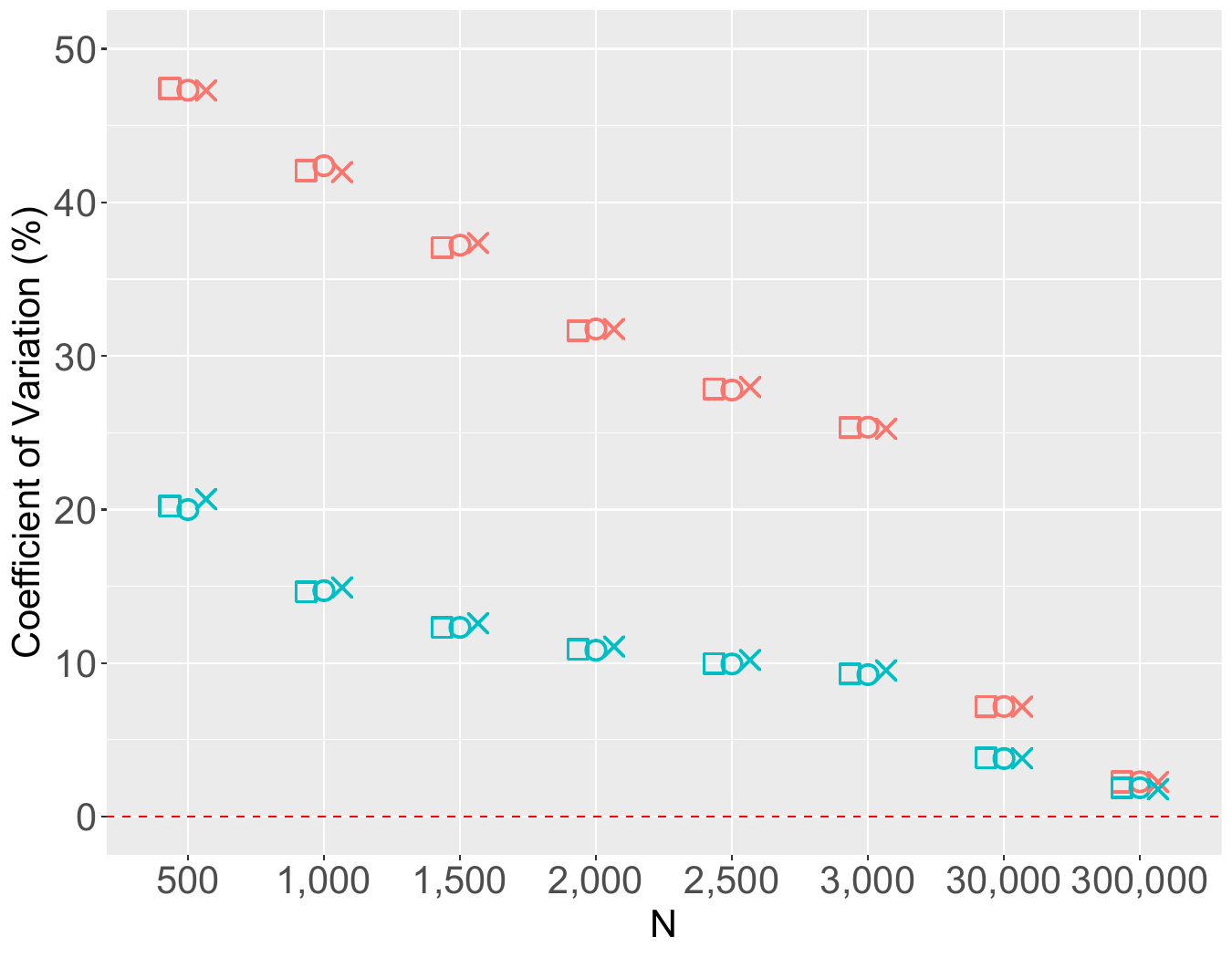}
         \caption{\small 25 Regions}
        \label{cv_25_low_probs}
    \end{subfigure}\hspace*{\fill}
    \begin{subfigure}[b]{.3\linewidth}
         \includegraphics[width=7cm,height=6cm]{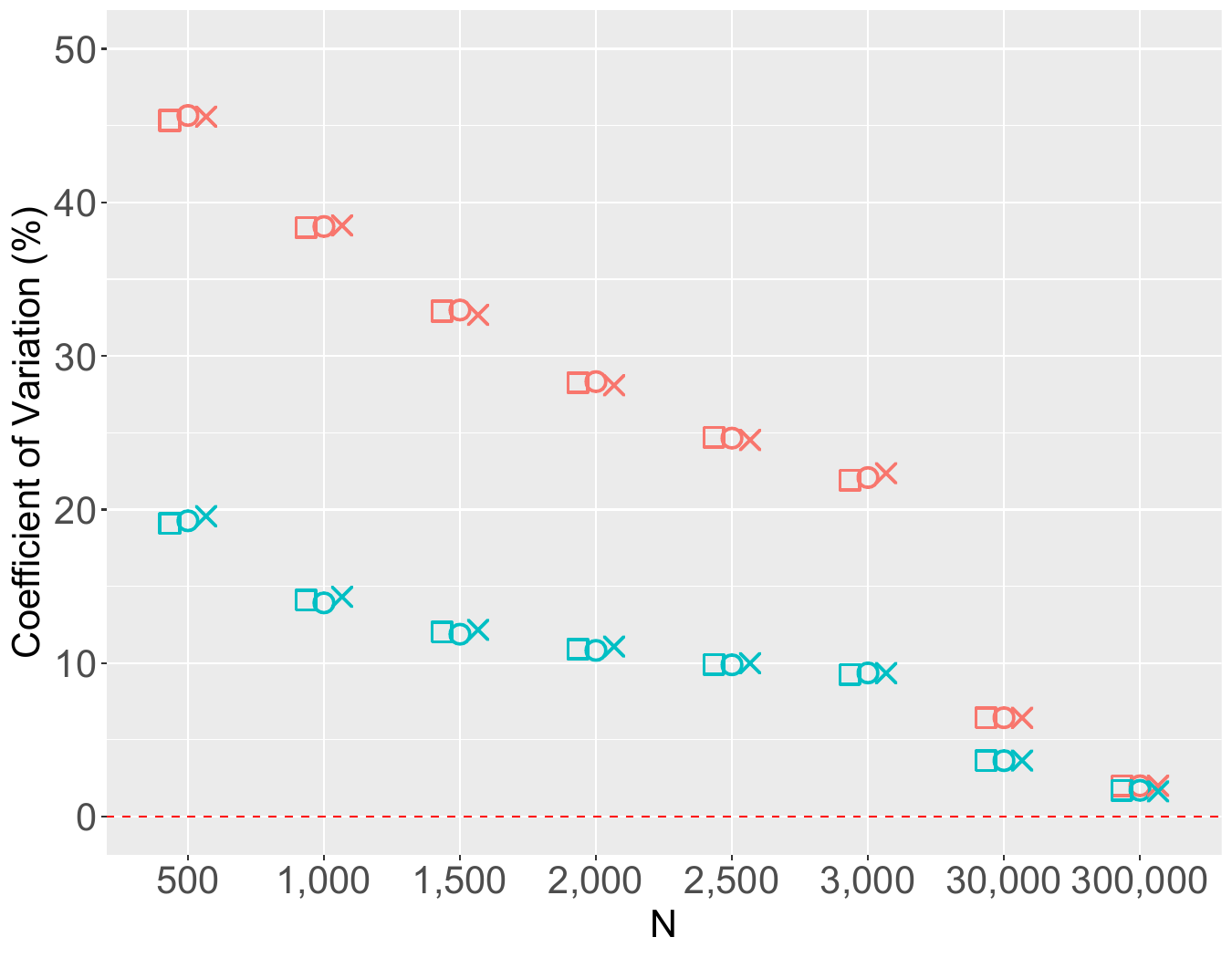}
        \caption{\small 20 Regions}
        \label{cv_20_low_probs}
    \end{subfigure}\hspace*{\fill}
    \begin{subfigure}[b]{.3\linewidth}
         \includegraphics[width=7cm,height=6cm]{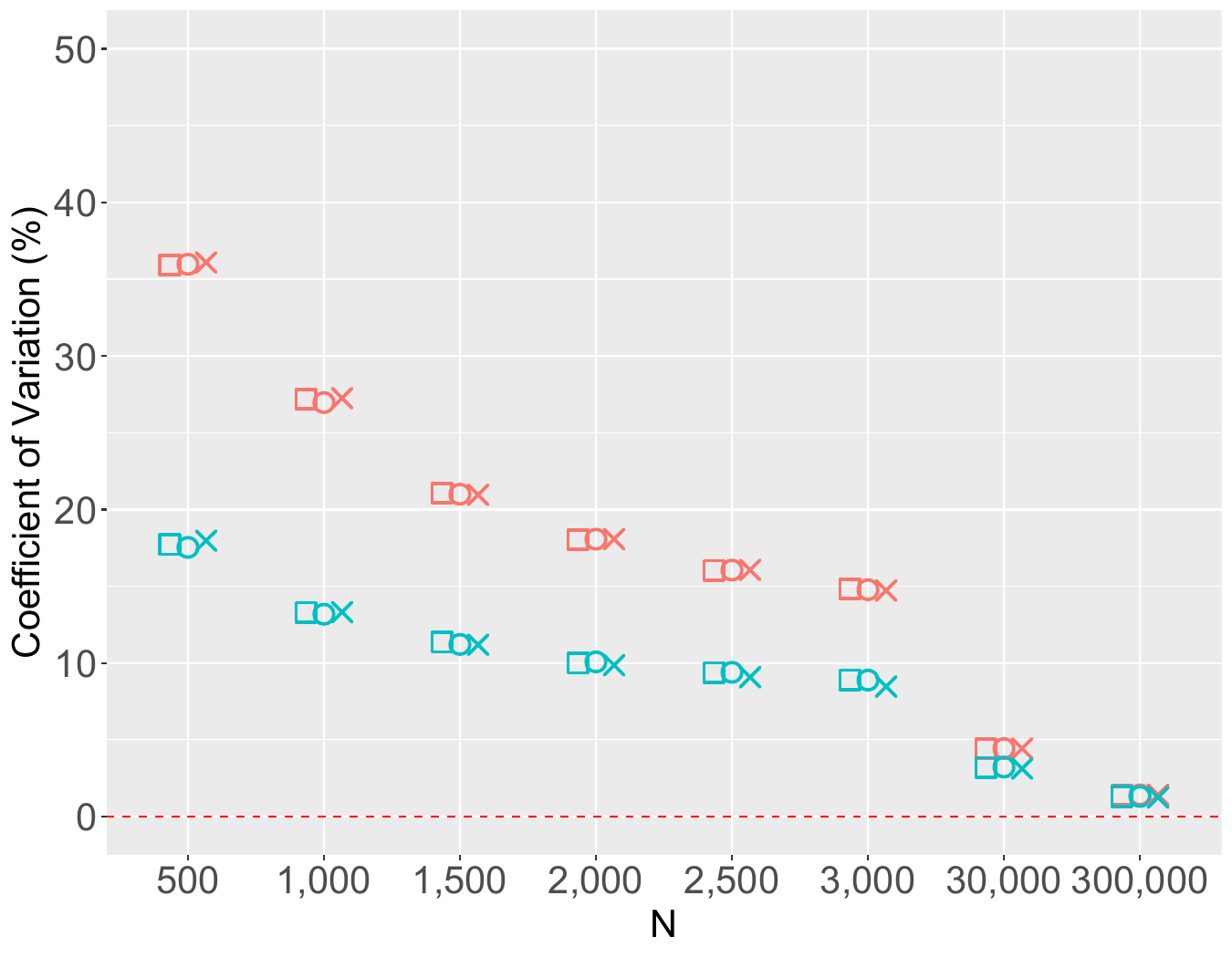}
         \caption{\small 10 Regions}
        \label{cv_10_low_probs}
    \end{subfigure}\hspace*{\fill}
    \end{minipage}
    \begin{minipage}[c]{0.1\textwidth}
        \includegraphics[width=4cm,height=4cm]{legend.png}
    \end{minipage}

\caption{Coefficient of variation (\%) of estimates where initially $\pi_A = 0.4$ and $\pi_B = 0.2$ and the random effects follow a normal, lognormal or Pareto distribution.}\label{rv_low_probs2}

\end{sidewaysfigure}

\clearpage

\begin{table}[hbt!]\footnotesize
\centering
\captionsetup{width=5.6in}
\caption{\small Simulation results where $\pi_A = 0.8$ and $\pi_B = 0.7$, the random effects follow a normal distribution, with total population size N = 500, 1,000, 1,500, 2,000, 2,500, 3,000, 30,000 and 300,000, the mean relative bias (mrb)\%, the coefficient of variation (cv)\% and the mean squared error (mse) of the estimates for the fitted fixed effects (F), Chapman corrected fixed effects (CF) and mixed effects (M) models. $\infty$ indicates that, among the 10,000 samples drawn for each combination of $N$ and Number of regions, at least one estimated population size is infinite.}
\scalebox{0.9}{
\begin{tabular}{rrrrrrrrrrr*{1}{S[table-format=2]}} 
 \hline
\multirow{-2.5}{*}{}&\multicolumn{1}{c}{No.}&\multicolumn{1}{c}{F}&\multicolumn{1}{c}{CF}&\multicolumn{1}{c}{M}&\multicolumn{1}{c}{F}&\multicolumn{1}{c}{CF}&\multicolumn{1}{c}{M}&\multicolumn{1}{c}{F}&\multicolumn{1}{c}{CF}&\multicolumn{1}{c}{M}\\
\multicolumn{1}{c}{N}&\multicolumn{1}{c}{regions}&\multicolumn{1}{c}{mrb\%}&\multicolumn{1}{c}{mrb\%}&\multicolumn{1}{c}{mrb\%}&\multicolumn{1}{c}{cv\%}&\multicolumn{1}{c}{cv\%}&\multicolumn{1}{c}{cv\%}&\multicolumn{1}{c}{mse}&\multicolumn{1}{c}{mse}&\multicolumn{1}{c}{mse}\\
\hline
500&30&$\infty$&-0.041&0.034&$\infty$&8.857&6.424&$\infty$&2.2&1.1\\
500&25&$\infty$&-0.085&0.007&$\infty$&8.012&5.825&$\infty$&2.6&1.4\\
500&20&0.610&0.025&0.150&7.458&7.057&5.159&3.5&3.1&1.7\\
500&15&0.468&0.057&0.217&6.204&5.994&4.473&4.3&4.0&2.2\\
500&10&0.515&0.256&0.442&4.816&4.716&3.684&5.9&5.6&3.4\\
500&5&0.308&0.185&0.324&3.309&3.278&2.804&11.1&10.8&8.0\\
\hline
1000&30&0.408&0.000&0.166&6.144&5.941&4.368&4.2&3.9&2.1\\
1000&25&0.285&-0.046&0.126&5.547&5.399&4.005&5.0&4.7&2.6\\
1000&20&0.249&-0.011&0.184&4.868&4.767&3.576&6.0&5.7&3.2\\
1000&15&0.219&0.030&0.234&4.131&4.070&3.133&7.6&7.4&4.4\\
1000&10&0.189&0.065&0.231&3.373&3.342&2.705&11.4&11.2&7.3\\
1000&5&0.155&0.095&0.234&2.292&2.282&2.049&21.1&20.9&16.9\\
\hline
1500&30&0.251&-0.008&0.195&4.847&4.75&3.534&5.9&5.6&3.1\\
1500&25&0.234&0.021&0.228&4.389&4.317&3.238&7.0&6.7&3.8\\
1500&20&0.193&0.026&0.238&3.904&3.855&2.944&8.6&8.4&4.9\\
1500&15&0.221&0.096&0.295&3.340&3.309&2.625&11.2&11.0&6.9\\
1500&10&0.116&0.034&0.180&2.708&2.692&2.219&16.6&16.3&11.1\\
1500&5&0.107&0.067&0.189&1.833&1.828&1.705&30.4&30.1&26.4\\
\hline
2000&30&0.225&0.034&0.262&4.139&4.079&3.059&7.7&7.4&4.2\\
2000&25&0.201&0.044&0.263&3.744&3.701&2.822&9.0&8.8&5.1\\
2000&20&0.158&0.034&0.241&3.354&3.323&2.586&11.3&11.1&6.7\\
2000&15&0.187&0.094&0.258&2.863&2.844&2.283&14.7&14.4&9.3\\
2000&10&0.097&0.036&0.167&2.329&2.318&1.946&21.8&21.5&15.2\\
2000&5&0.083&0.053&0.154&1.575&1.572&1.489&39.9&39.6&35.7\\
\hline
2500&30&0.187&0.037&0.270&3.675&3.633&2.757&9.4&9.2&5.3\\
2500&25&0.157&0.033&0.256&3.307&3.277&2.537&11.0&10.7&6.5\\
2500&20&0.134&0.035&0.220&2.985&2.964&2.336&14.0&13.7&8.6\\
2500&15&0.157&0.083&0.227&2.564&2.550&2.069&18.4&18.1&12.0\\
2500&10&0.048&0.000&0.116&2.072&2.065&1.756&26.9&26.6&19.3\\
2500&5&0.071&0.047&0.130&1.446&1.443&1.367&52.5&52.2&47.0\\
\hline
3000&30&0.163&0.039&0.274&3.307&3.277&2.519&11.0&10.7&6.4\\
3000&25&0.136&0.034&0.233&3.006&2.984&2.335&13.1&12.8&7.9\\
3000&20&0.138&0.056&0.220&2.718&2.702&2.147&16.7&16.4&10.4\\
3000&15&0.107&0.047&0.175&2.328&2.318&1.908&21.8&21.5&14.6\\
3000&10&0.062&0.022&0.122&1.900&1.895&1.622&32.6&32.3&23.7\\
3000&5&0.058&0.038&0.107&1.333&1.331&1.261&64.2&63.9&57.5\\
\hline
30000&30&0.007&-0.005&-0.001&1.048&1.047&0.857&109.8&109.6&73.4\\
30000&25&-0.004&-0.014&-0.009&0.949&0.948&0.792&129.6&129.4&90.3\\
30000&20&0.007&-0.001&0.003&0.839&0.839&0.714&158.5&158.3&114.8\\
30000&15&0.009&0.003&0.005&0.715&0.715&0.627&204.8&204.6&157.2\\
30000&10&0.016&0.012&0.015&0.585&0.585&0.537&308.8&308.5&259.8\\
30000&5&0.005&0.003&0.005&0.410&0.410&0.399&606.4&606.1&573.0\\
\hline
300000&30&0.004&0.003&0.003&0.328&0.328&0.311&1074.2&1073.9&970.1\\
300000&25&0.003&0.002&0.002&0.299&0.299&0.286&1288.1&1287.8&1180.7\\
300000&20&-0.001&-0.002&-0.002&0.266&0.266&0.257&1596.5&1596.3&1486.9\\
300000&15&0.003&0.002&0.002&0.225&0.225&0.219&2023.7&2023.4&1912.6\\
300000&10&0.004&0.003&0.003&0.184&0.184&0.181&3035.7&3035.3&2939.7\\
300000&5&0.005&0.005&0.005&0.132&0.132&0.132&6263.4&6262.5&6244.6\\
\hline
\end{tabular}}
\label{table:1a}
\end{table}

\clearpage

\begin{table}[hbt!]\footnotesize
\centering
\captionsetup{width=5.6in}
\caption{\small Simulation results where $\pi_A = 0.8$ and $\pi_B = 0.7$, the random effects follow a lognormal distribution, with total population size N = 500, 1,000, 1,500, 2,000, 2,500, 3,000, 30,000 and 300,000, the mean relative bias (mrb)\%, the coefficient of variation (cv)\% and the mean squared error (mse) of the estimates for the fitted fixed effects (F), Chapman corrected fixed effects (CF) and mixed effects (M) models. $\infty$ indicates that, among the 10,000 samples drawn for each combination of $N$ and Number of regions, at least one estimated population size is infinite.}
\scalebox{0.9}{
\begin{tabular}{rrrrrrrrrrr*{1}{S[table-format=2]}} 
 \hline
\multirow{-2.5}{*}{}&\multicolumn{1}{c}{No.}&\multicolumn{1}{c}{F}&\multicolumn{1}{c}{CF}&\multicolumn{1}{c}{M}&\multicolumn{1}{c}{F}&\multicolumn{1}{c}{CF}&\multicolumn{1}{c}{M}&\multicolumn{1}{c}{F}&\multicolumn{1}{c}{CF}&\multicolumn{1}{c}{M}\\
\multicolumn{1}{c}{N}&\multicolumn{1}{c}{regions}&\multicolumn{1}{c}{mrb\%}&\multicolumn{1}{c}{mrb\%}&\multicolumn{1}{c}{mrb\%}&\multicolumn{1}{c}{cv\%}&\multicolumn{1}{c}{cv\%}&\multicolumn{1}{c}{cv\%}&\multicolumn{1}{c}{mse}&\multicolumn{1}{c}{mse}&\multicolumn{1}{c}{mse}\\
\hline
500&30&$\infty$&-0.070&0.016&$\infty$&8.819&6.403&$\infty$&2.2&1.1\\
500&25&$\infty$&-0.120&-0.023&$\infty$&8.000&5.849&$\infty$&2.6&1.4\\
500&20&0.594&0.013&0.141&7.441&7.048&5.165&3.5&3.1&1.7\\
500&15&0.442&0.034&0.195&6.192&5.986&4.482&4.3&4.0&2.2\\
500&10&0.489&0.232&0.422&4.817&4.719&3.683&5.9&5.6&3.4\\
500&5&0.308&0.185&0.328&3.310&3.279&2.802&11.1&10.8&7.9\\
\hline
1000&30&0.396&-0.009&0.164&6.124&5.926&4.367&4.2&3.9&2.1\\
1000&25&0.280&-0.049&0.128&5.546&5.401&4.006&5.0&4.7&2.6\\
1000&20&0.251&-0.008&0.189&4.871&4.770&3.573&6.0&5.7&3.2\\
1000&15&0.216&0.027&0.228&4.134&4.074&3.137&7.6&7.4&4.4\\
1000&10&0.200&0.076&0.243&3.387&3.356&2.709&11.6&11.3&7.4\\
1000&5&0.173&0.113&0.248&2.295&2.285&2.048&21.2&21.0&16.9\\
\hline
1500&30&0.271&0.012&0.214&4.851&4.755&3.532&5.9&5.7&3.1\\
1500&25&0.233&0.020&0.227&4.380&4.308&3.230&7.0&6.7&3.8\\
1500&20&0.210&0.043&0.250&3.900&3.851&2.938&8.6&8.3&4.9\\
1500&15&0.229&0.104&0.291&3.337&3.307&2.632&11.2&11.0&7.0\\
1500&10&0.122&0.041&0.186&2.703&2.687&2.213&16.5&16.3&11.1\\
1500&5&0.098&0.059&0.178&1.833&1.828&1.698&30.4&30.1&26.1\\
\hline
2000&30&0.227&0.036&0.258&4.144&4.084&3.064&7.7&7.4&4.2\\
2000&25&0.210&0.053&0.268&3.753&3.709&2.828&9.1&8.8&5.1\\
2000&20&0.166&0.042&0.240&3.360&3.330&2.594&11.4&11.1&6.8\\
2000&15&0.195&0.102&0.261&2.865&2.845&2.288&14.7&14.4&9.4\\
2000&10&0.097&0.036&0.161&2.328&2.318&1.944&21.8&21.5&15.2\\
2000&5&0.075&0.045&0.144&1.577&1.574&1.482&39.9&39.7&35.3\\
\hline
2500&30&0.183&0.033&0.256&3.666&3.625&2.763&9.4&9.1&5.3\\
2500&25&0.152&0.028&0.243&3.301&3.271&2.535&10.9&10.7&6.4\\
2500&20&0.134&0.036&0.213&2.981&2.960&2.335&13.9&13.7&8.5\\
2500&15&0.142&0.068&0.209&2.567&2.554&2.076&18.4&18.2&12.0\\
2500&10&0.057&0.009&0.117&2.079&2.071&1.756&27.0&26.8&19.3\\
2500&5&0.070&0.047&0.127&1.445&1.442&1.364&52.4&52.1&46.7\\
\hline
3000&30&0.167&0.043&0.269&3.297&3.268&2.520&10.9&10.7&6.4\\
3000&25&0.134&0.031&0.225&3.010&2.988&2.334&13.1&12.9&7.9\\
3000&20&0.130&0.048&0.199&2.711&2.695&2.143&16.6&16.4&10.4\\
3000&15&0.101&0.041&0.159&2.327&2.317&1.903&21.7&21.5&14.5\\
3000&10&0.063&0.023&0.119&1.897&1.891&1.618&32.5&32.2&23.6\\
3000&5&0.053&0.034&0.099&1.331&1.329&1.254&63.9&63.7&56.8\\
\hline
30000&30&0.008&-0.004&0.001&1.046&1.045&0.860&109.5&109.3&73.9\\
30000&25&-0.009&-0.019&-0.014&0.947&0.946&0.791&129.1&129.0&90.0\\
30000&20&0.007&-0.001&0.002&0.840&0.840&0.717&158.8&158.6&115.7\\
30000&15&0.006&0.000&0.002&0.714&0.714&0.625&204.1&203.9&156.3\\
30000&10&0.015&0.011&0.013&0.587&0.587&0.538&310.2&309.9&260.4\\
30000&5&0.005&0.003&0.006&0.410&0.410&0.398&605.5&605.3&569.2\\
\hline
300000&30&0.004&0.003&0.003&0.328&0.328&0.312&1077.9&1077.6&973.7\\
300000&25&0.002&0.001&0.001&0.299&0.299&0.286&1289.2&1288.9&1181.6\\
300000&20&-0.002&-0.003&-0.003&0.266&0.266&0.257&1597.2&1597.1&1487.0\\
300000&15&0.002&0.001&0.001&0.225&0.225&0.219&2027.3&2027.1&1915.5\\
300000&10&0.004&0.004&0.004&0.184&0.184&0.181&3051.8&3051.3&2951.3\\
300000&5&0.004&0.004&0.004&0.131&0.131&0.131&6208.4&6207.6&6200.1\\
\hline
\end{tabular}}
\label{table:1c}
\end{table}

\clearpage

\begin{table}[hbt!]\footnotesize
\centering
\captionsetup{width=5.6in}
\caption{\small Simulation results where $\pi_A = 0.8$ and $\pi_B = 0.7$, the random effects follow a Pareto distribution, with total population size N = 500, 1,000, 1,500, 2,000, 2,500, 3,000, 30,000 and 300,000, the mean relative bias \%, the coefficient of variation (cv) \% and the mean squared error (mse) of the estimates for the fitted fixed effects (F), Chapman corrected fixed effects (CF) and mixed effects (M) models. $\infty$ indicates that, among the 10,000 samples drawn for each combination of $N$ and Number of regions, at least one estimated population size is infinite.}
\scalebox{0.9}{
\begin{tabular}{rrrrrrrrrrr*{1}{S[table-format=2]}} 
 \hline
\multirow{-2.5}{*}{}&\multicolumn{1}{c}{No.}&\multicolumn{1}{c}{F}&\multicolumn{1}{c}{CF}&\multicolumn{1}{c}{M}&\multicolumn{1}{c}{F}&\multicolumn{1}{c}{CF}&\multicolumn{1}{c}{M}&\multicolumn{1}{c}{F}&\multicolumn{1}{c}{CF}&\multicolumn{1}{c}{M}\\
\multicolumn{1}{c}{N}&\multicolumn{1}{c}{regions}&\multicolumn{1}{c}{mrb\%}&\multicolumn{1}{c}{mrb\%}&\multicolumn{1}{c}{mrb\%}&\multicolumn{1}{c}{cv\%}&\multicolumn{1}{c}{cv\%}&\multicolumn{1}{c}{cv\%}&\multicolumn{1}{c}{mse}&\multicolumn{1}{c}{mse}&\multicolumn{1}{c}{mse}\\
\hline
500&30&0.886&-0.056&0.043&9.701&8.818&6.412&2.7&2.2&1.1\\
500&25&0.669&-0.072&0.038&8.533&7.972&5.829&3.0&2.5&1.4\\
500&20&0.584&0.013&0.140&7.413&7.036&5.161&3.5&3.1&1.7\\
500&15&0.457&0.053&0.197&6.209&6.006&4.500&4.3&4.0&2.3\\
500&10&0.471&0.216&0.369&4.832&4.735&3.698&5.9&5.6&3.5\\
500&5&0.282&0.161&0.276&3.296&3.265&2.781&11.0&10.7&7.8\\
\hline
1000&30&0.398&-0.002&0.165&6.119&5.925&4.374&4.2&3.9&2.1\\
1000&25&0.286&-0.040&0.123&5.524&5.382&4.004&4.9&4.6&2.6\\
1000&20&0.234&-0.022&0.139&4.866&4.768&3.599&6.0&5.7&3.2\\
1000&15&0.202&0.015&0.155&4.137&4.078&3.141&7.7&7.4&4.4\\
1000&10&0.167&0.045&0.171&3.404&3.374&2.726&11.7&11.4&7.5\\
1000&5&0.171&0.111&0.212&2.271&2.261&1.988&20.8&20.5&15.9\\
\hline
1500&30&0.274&0.019&0.200&4.823&4.728&3.531&5.9&5.6&3.1\\
1500&25&0.214&0.004&0.166&4.382&4.311&3.257&7.0&6.7&3.8\\
1500&20&0.210&0.045&0.192&3.895&3.846&2.948&8.6&8.3&4.9\\
1500&15&0.236&0.112&0.241&3.349&3.319&2.624&11.3&11.1&6.9\\
1500&10&0.118&0.037&0.147&2.717&2.701&2.211&16.7&16.4&11.0\\
1500&5&0.095&0.056&0.143&1.813&1.808&1.623&29.7&29.5&23.8\\
\hline
2000&30&0.230&0.042&0.216&4.124&4.065&3.070&7.6&7.4&4.2\\
2000&25&0.190&0.035&0.187&3.716&3.674&2.829&8.9&8.6&5.1\\
2000&20&0.158&0.035&0.171&3.359&3.329&2.602&11.3&11.1&6.8\\
2000&15&0.177&0.085&0.192&2.867&2.848&2.268&14.7&14.5&9.2\\
2000&10&0.092&0.032&0.134&2.333&2.323&1.925&21.8&21.6&14.9\\
2000&5&0.080&0.051&0.129&1.564&1.561&1.423&39.3&39.1&32.6\\
\hline
2500&30&0.192&0.045&0.211&3.631&3.592&2.756&9.2&9.0&5.3\\
2500&25&0.155&0.032&0.182&3.288&3.259&2.544&10.9&10.6&6.5\\
2500&20&0.144&0.046&0.168&2.968&2.947&2.315&13.8&13.6&8.4\\
2500&15&0.146&0.074&0.168&2.566&2.552&2.049&18.4&18.1&11.7\\
2500&10&0.044&-0.003&0.086&2.080&2.073&1.737&27.1&26.9&18.9\\
2500&5&0.071&0.048&0.116&1.446&1.444&1.330&52.5&52.2&44.4\\
\hline
3000&30&0.174&0.052&0.218&3.293&3.264&2.521&10.9&10.7&6.4\\
3000&25&0.135&0.034&0.166&3.001&2.979&2.336&13.0&12.8&7.9\\
3000&20&0.133&0.052&0.159&2.708&2.693&2.130&16.6&16.3&10.2\\
3000&15&0.105&0.045&0.135&2.342&2.332&1.888&22.0&21.8&14.3\\
3000&10&0.047&0.007&0.090&1.909&1.904&1.607&32.9&32.6&23.3\\
3000&5&0.052&0.033&0.091&1.329&1.327&1.218&63.7&63.5&53.6\\
\hline
30000&30&0.008&-0.004&0.002&1.045&1.044&0.858&109.2&108.9&73.7\\
30000&25&-0.005&-0.015&-0.007&0.944&0.943&0.788&128.3&128.1&89.5\\
30000&20&0.008&0.001&0.006&0.838&0.838&0.717&158.2&158.0&115.6\\
30000&15&0.007&0.001&0.007&0.716&0.716&0.633&205.2&205.0&160.1\\
30000&10&0.014&0.010&0.014&0.589&0.589&0.534&312.5&312.2&257.2\\
30000&5&0.005&0.003&0.008&0.408&0.408&0.395&600.7&600.4&562.8\\
\hline
300000&30&0.004&0.002&0.002&0.328&0.328&0.307&1075.8&1075.5&944.2\\
300000&25&0.002&0.001&0.001&0.298&0.298&0.282&1281.2&1281.0&1144.0\\
300000&20&-0.002&-0.003&-0.003&0.266&0.266&0.254&1593.7&1593.6&1449.1\\
300000&15&0.002&0.002&0.002&0.225&0.225&0.216&2018.4&2018.1&1864.8\\
300000&10&0.005&0.005&0.005&0.185&0.185&0.179&3069.1&3068.6&2900.4\\
300000&5&0.005&0.004&0.004&0.131&0.131&0.130&6180.7&6179.9&6073.7\\
\hline
\end{tabular}}
\label{table:1e}
\end{table}

\begin{sidewaysfigure}[p]
\begin{minipage}{0.9\textwidth}
\centering
\begin{subfigure}[b]{.5\linewidth}
         \includegraphics[width=11cm,height=6cm]{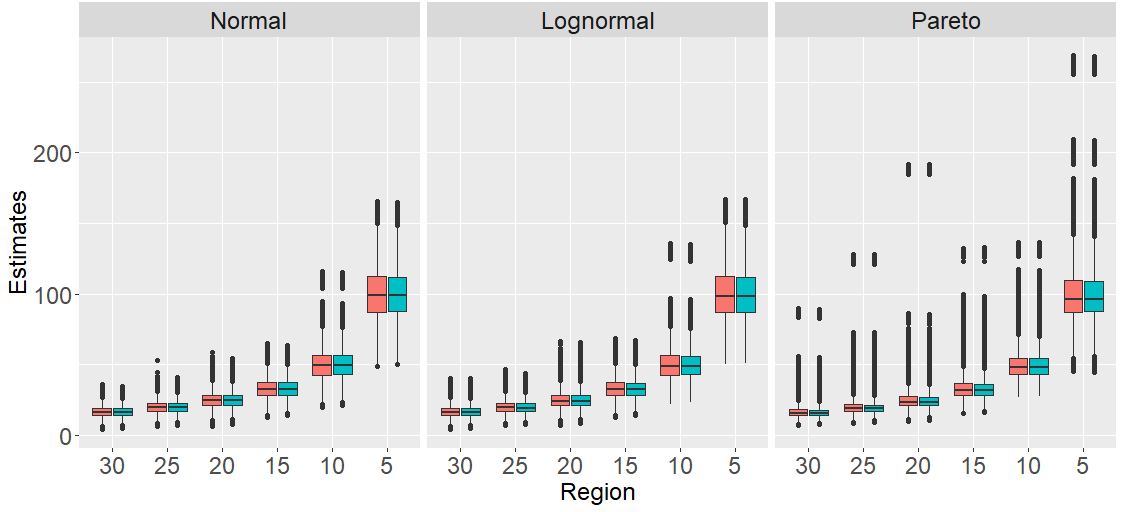}
         \caption{\small N = 500}
        \label{500_box_high_probs}
    \end{subfigure}\hspace*{\fill}
    \begin{subfigure}[b]{.5\linewidth}
         \includegraphics[width=11cm,height=6cm]{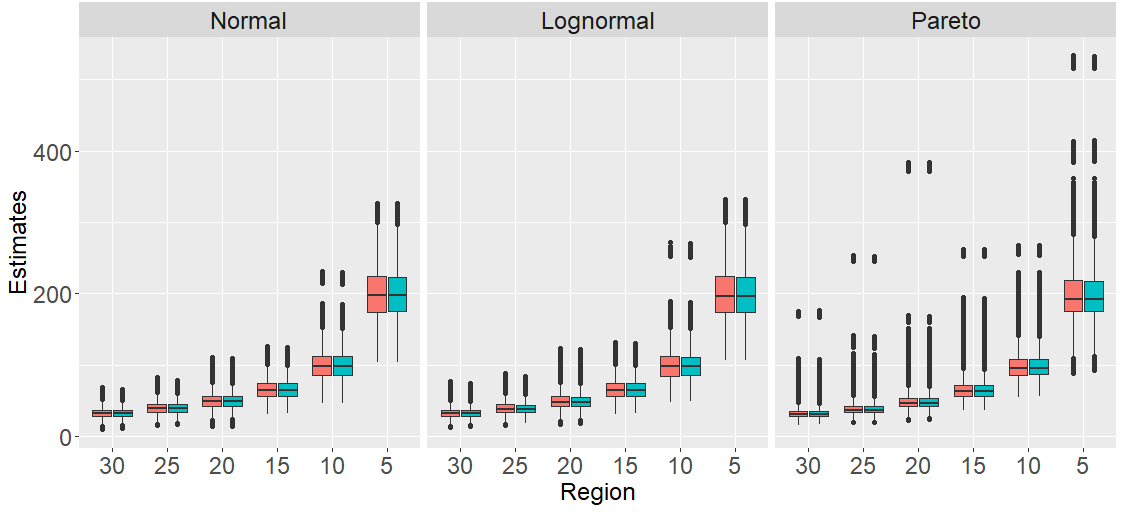}
        \caption{\small N = 1,000}
        \label{1000_box_high_probs}
    \end{subfigure}\hspace*{\fill}

    \begin{subfigure}[b]{.5\linewidth}
         \includegraphics[width=11cm,height=6cm]{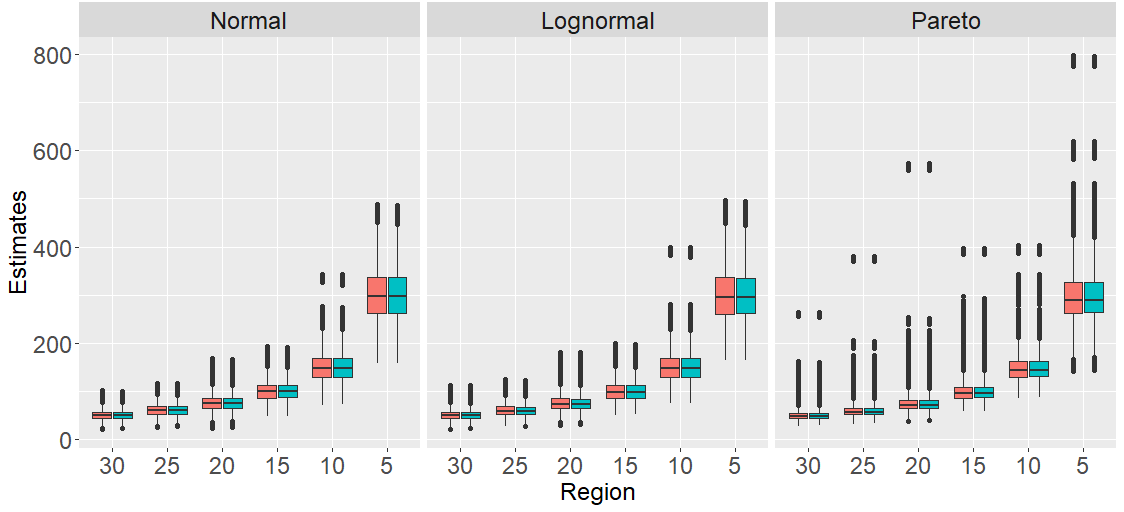}
         \caption{\small N = 1,500}
        \label{1500_box_high_probs}
    \end{subfigure}\hspace*{\fill}
    \begin{subfigure}[b]{.5\linewidth}
         \includegraphics[width=11cm,height=6cm]{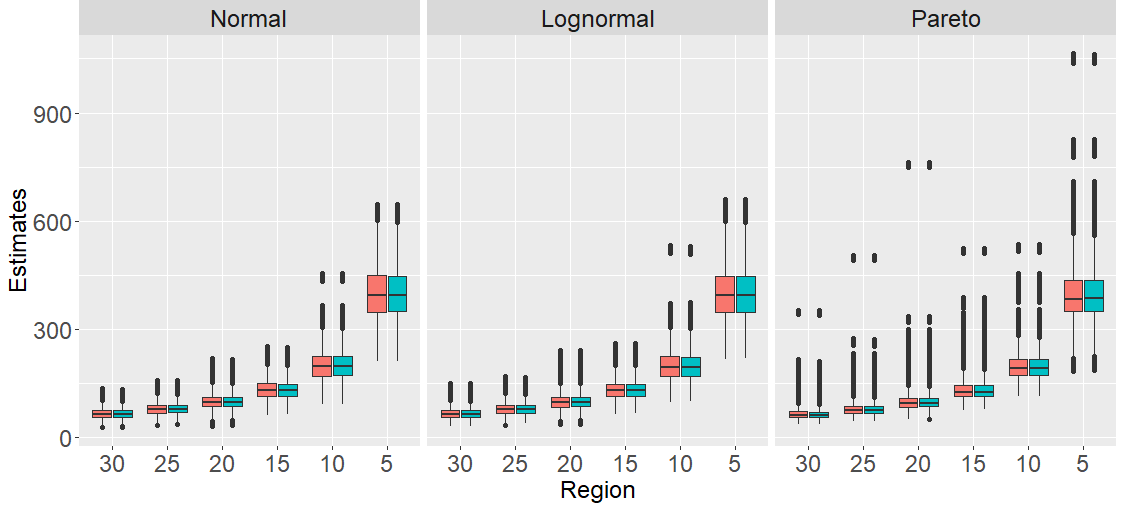}
         \caption{\small N = 2,000}
        \label{2000_box_high_probs}
    \end{subfigure}\hspace*{\fill}
    \end{minipage}
    \begin{minipage}[c]{0.1\textwidth}
        \includegraphics[width=3cm,height=2cm]{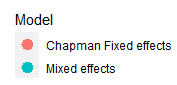}
    \end{minipage}

\caption{Regional level estimates where initially $\pi_A = 0.8$ and $\pi_B = 0.7$ and the random effects follow a normal, lognormal\\ or Pareto distribution. Note that the y-axis scale varies across $N$.}\label{box_high_probs_region}

\end{sidewaysfigure}

\begin{sidewaysfigure}[p]\ContinuedFloat
\begin{minipage}{0.9\textwidth}
\centering
\begin{subfigure}[b]{.5\linewidth}
         \includegraphics[width=11cm,height=6cm]{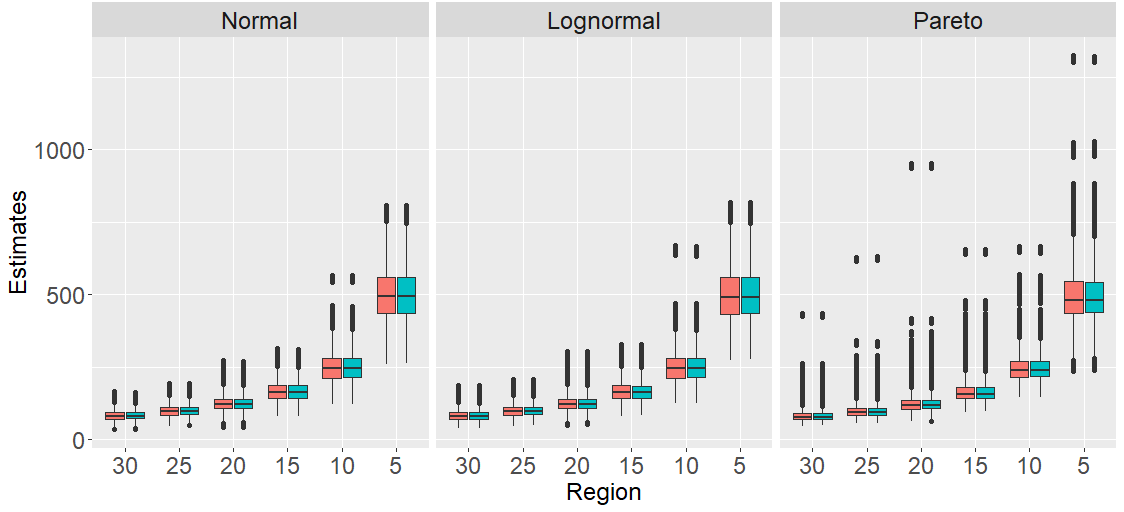}
         \caption{\small N = 2,500}
        \label{2500_box_high_probs}
    \end{subfigure}\hspace*{\fill}
    \begin{subfigure}[b]{.5\linewidth}
         \includegraphics[width=11cm,height=6cm]{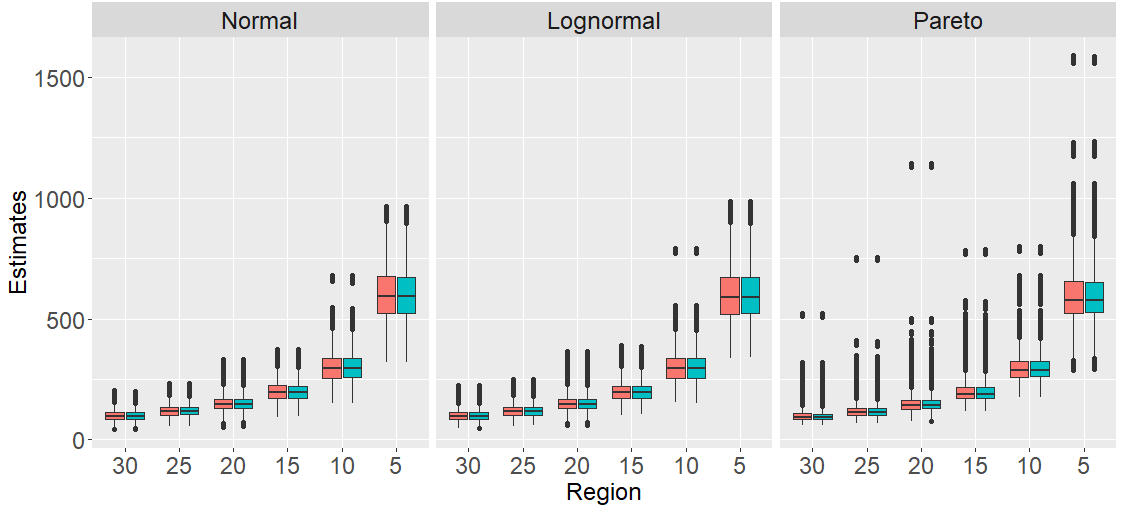}
        \caption{\small N = 3,000}
        \label{3000_box_high_probs}
    \end{subfigure}\hspace*{\fill}

    \begin{subfigure}[b]{.5\linewidth}
         \includegraphics[width=11cm,height=6cm]{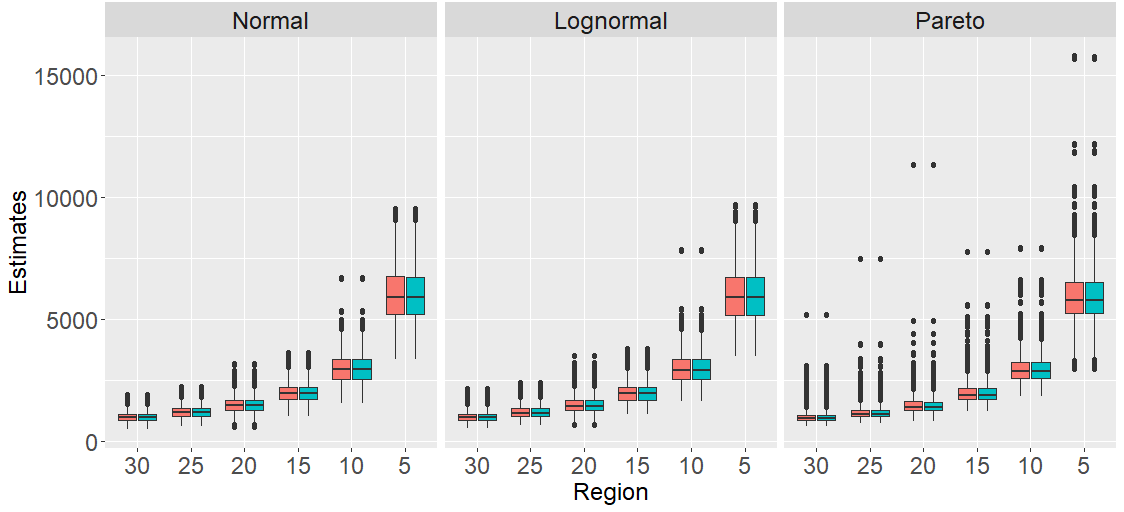}
         \caption{\small N = 30,000}
        \label{30000_box_high_probs}
    \end{subfigure}\hspace*{\fill}
    \begin{subfigure}[b]{.5\linewidth}
         \includegraphics[width=11cm,height=6cm]{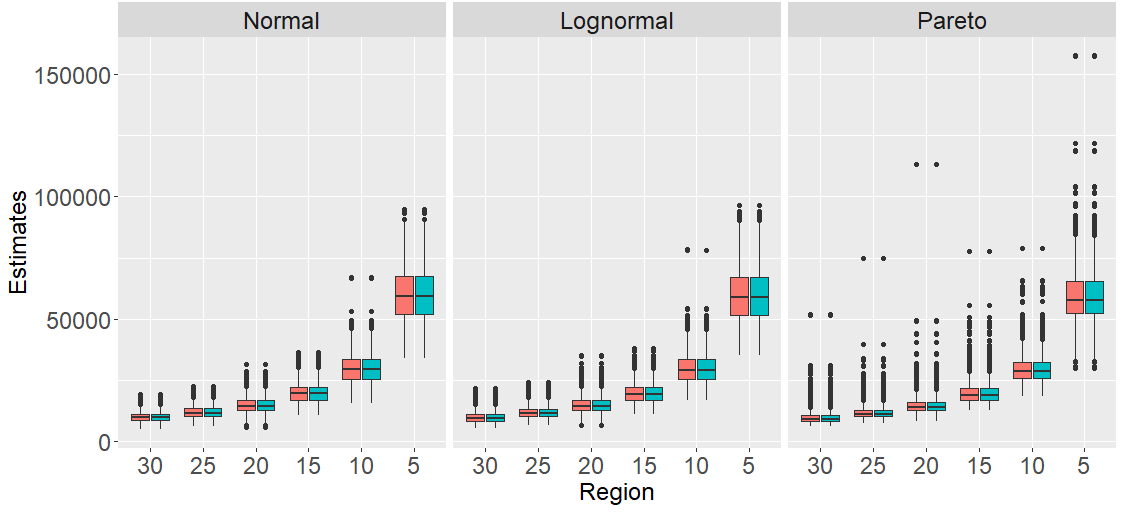}
         \caption{\small N = 300,000}
        \label{300000_box_high_probs}
    \end{subfigure}\hspace*{\fill}
    \end{minipage}
    \begin{minipage}[c]{0.1\textwidth}
        \includegraphics[width=3cm,height=2cm]{legend2.png}
    \end{minipage}

\caption{Regional level estimates where initially $\pi_A = 0.8$ and $\pi_B = 0.7$ and the random effects follow a normal, lognormal\\ or Pareto distribution. Note that the y-axis scale varies across $N$.}

\end{sidewaysfigure}

\begin{sidewaysfigure}[p]
\begin{minipage}{0.9\textwidth}
\centering
\begin{subfigure}[b]{.5\linewidth}
         \includegraphics[width=11cm,height=6cm]{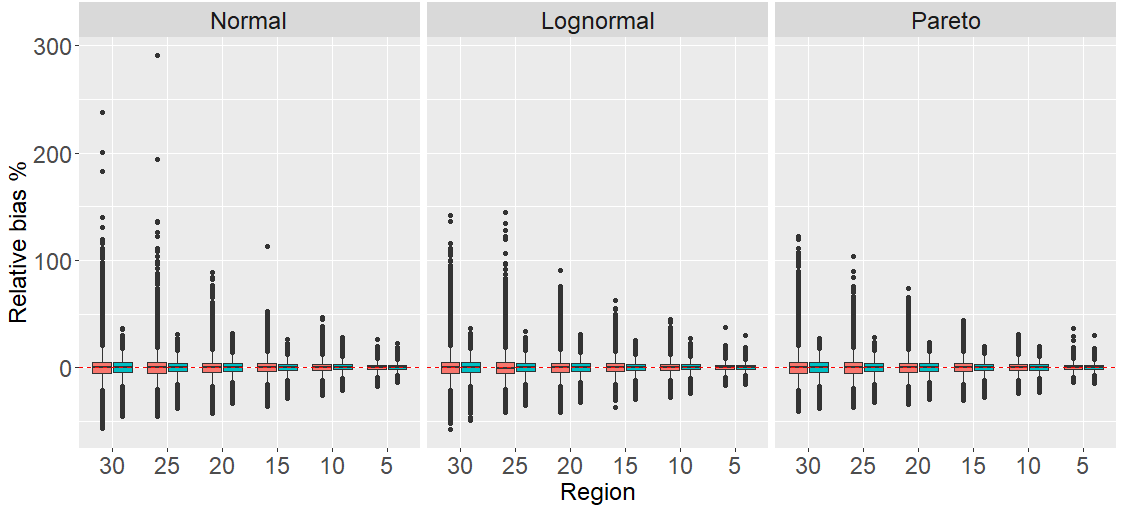}
         \caption{\small N = 500}
        \label{rb_500_box_plot_high_probs}
    \end{subfigure}\hspace*{\fill}
    \begin{subfigure}[b]{.5\linewidth}
         \includegraphics[width=11cm,height=6cm]{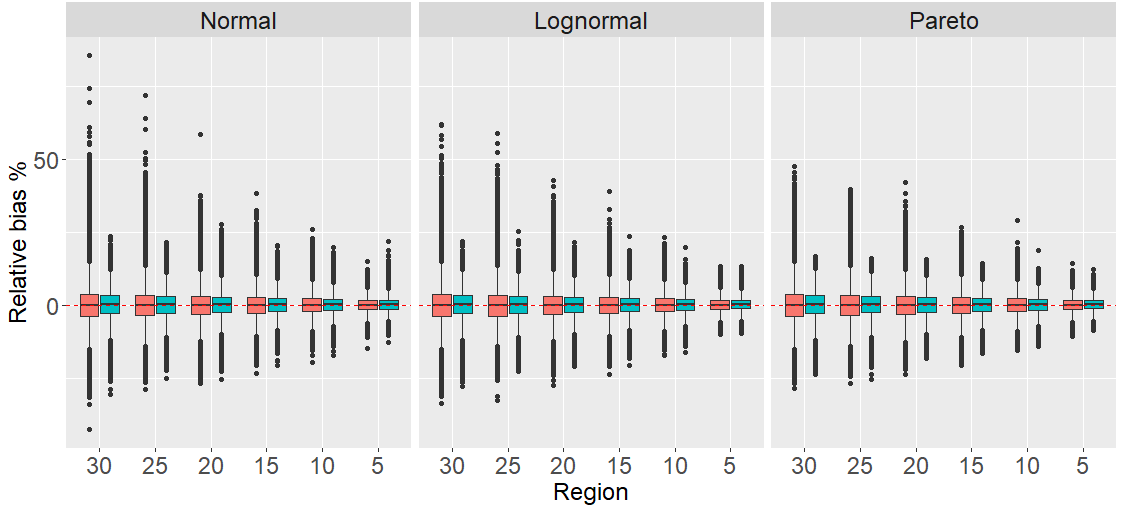}
        \caption{\small N = 1,000}
        \label{rb_1000_box_plot_high_probs}
    \end{subfigure}\hspace*{\fill}

    \begin{subfigure}[b]{.5\linewidth}
         \includegraphics[width=11cm,height=6cm]{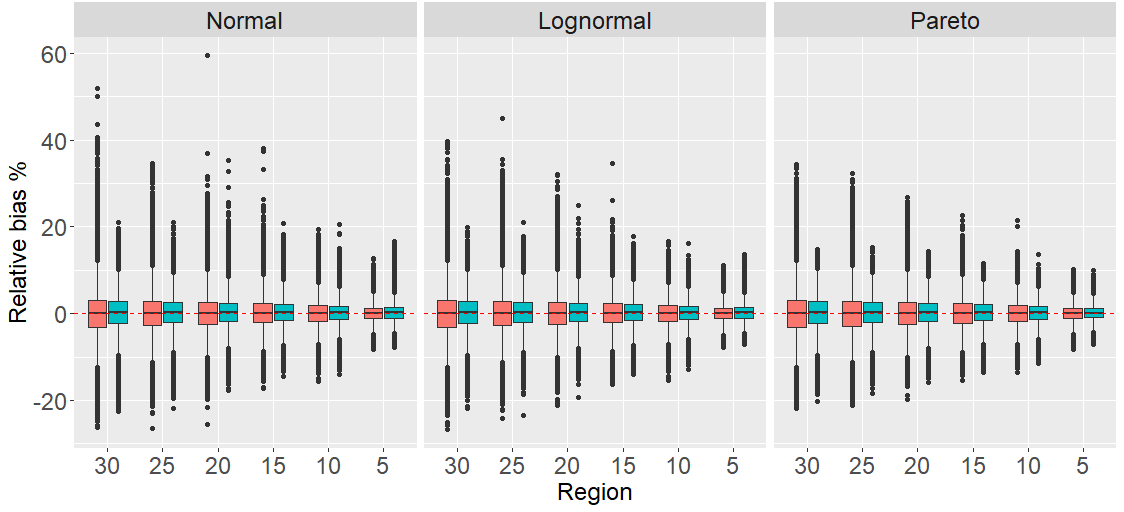}
         \caption{\small N = 1,500}
        \label{rb_1500_box_plot_high_probs}
    \end{subfigure}\hspace*{\fill}
    \begin{subfigure}[b]{.5\linewidth}
         \includegraphics[width=11cm,height=6cm]{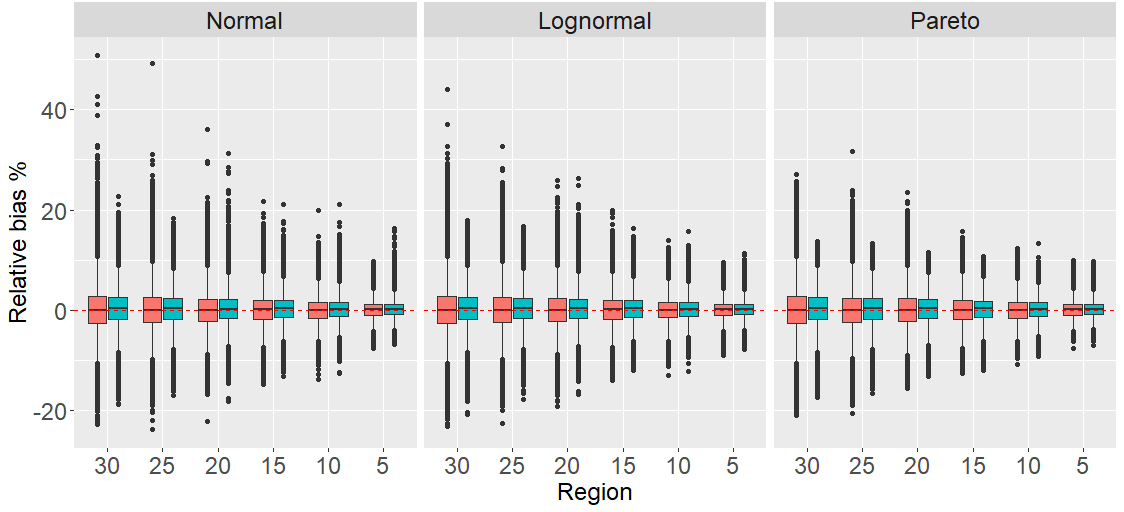}
         \caption{\small N = 2,000}
        \label{rb_2000_box_plot_high_probs}
    \end{subfigure}\hspace*{\fill}
    \end{minipage}
    \begin{minipage}[c]{0.1\textwidth}
        \includegraphics[width=3cm,height=2cm]{legend2.png}
    \end{minipage}

\caption{Relative bias (\%) of regional level estimates where initially $\pi_A = 0.8$ and $\pi_B = 0.7$ and the random effects follow\\ a normal, lognormal or Pareto distribution. Note that the y-axis scale varies across $N$.}\label{rb_high_probs_region}

\end{sidewaysfigure}

\begin{sidewaysfigure}[p]\ContinuedFloat
\begin{minipage}{0.9\textwidth}
\centering
\begin{subfigure}[b]{.5\linewidth}
         \includegraphics[width=11cm,height=6cm]{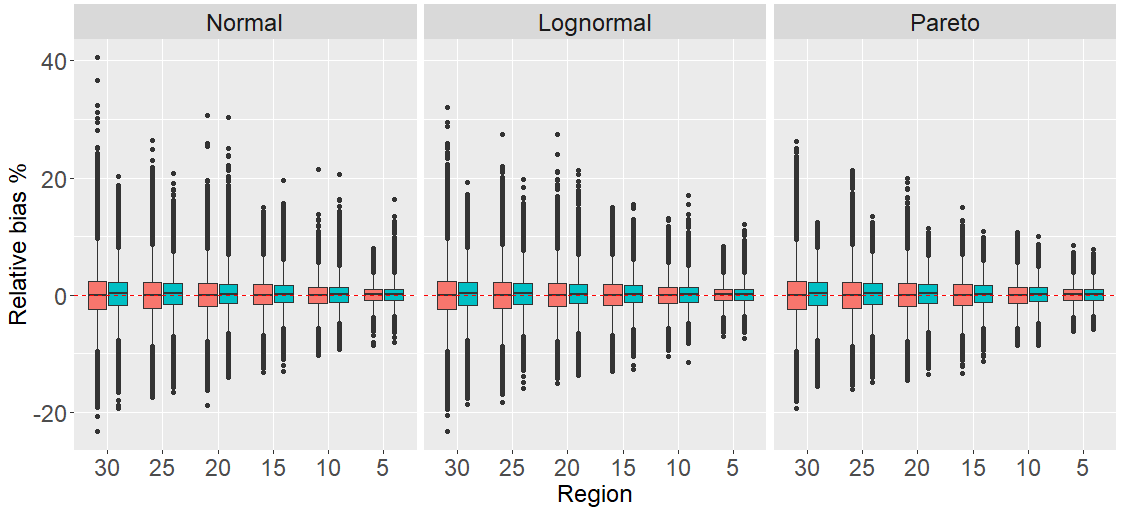}
         \caption{\small N = 2,500}
        \label{rb_2500_box_plot_high_probs}
    \end{subfigure}\hspace*{\fill}
    \begin{subfigure}[b]{.5\linewidth}
         \includegraphics[width=11cm,height=6cm]{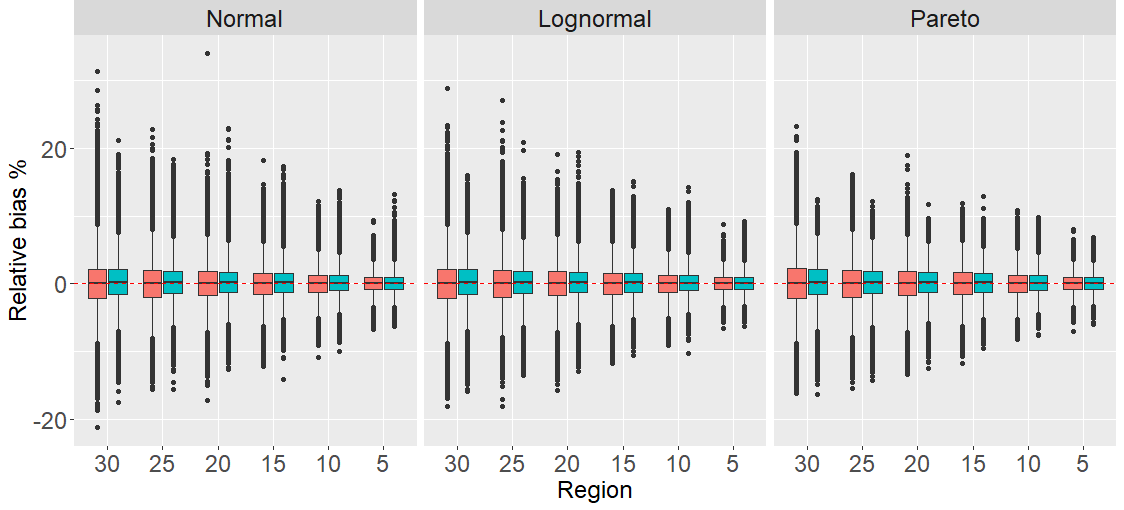}
        \caption{\small N = 3,000}
        \label{rb_3000_box_plot_high_probs}
    \end{subfigure}\hspace*{\fill}

    \begin{subfigure}[b]{.5\linewidth}
         \includegraphics[width=11cm,height=6cm]{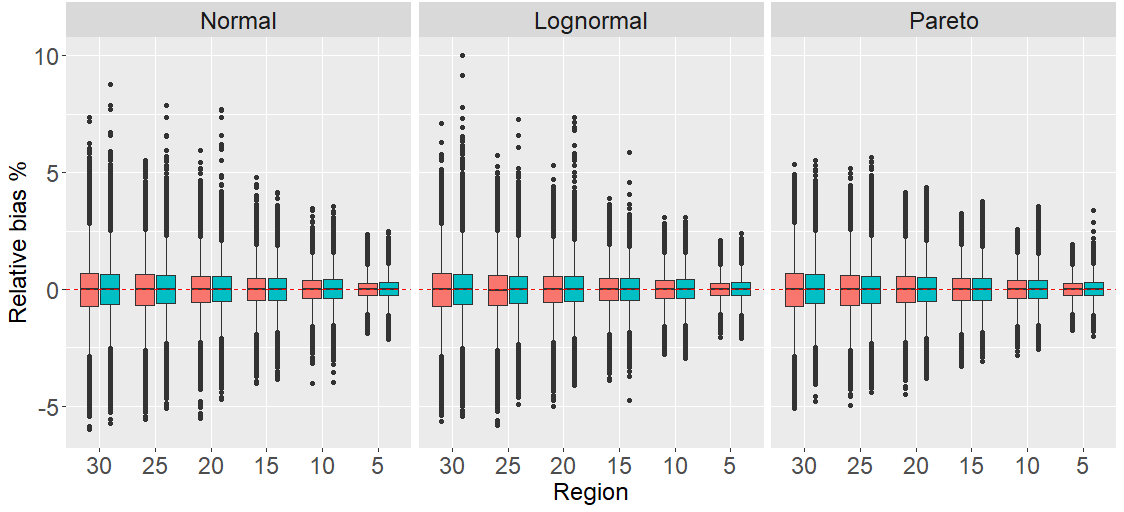}
         \caption{\small N = 30,000}
        \label{rb_30000_box_plot_high_probs}
    \end{subfigure}\hspace*{\fill}
    \begin{subfigure}[b]{.5\linewidth}
         \includegraphics[width=11cm,height=6cm]{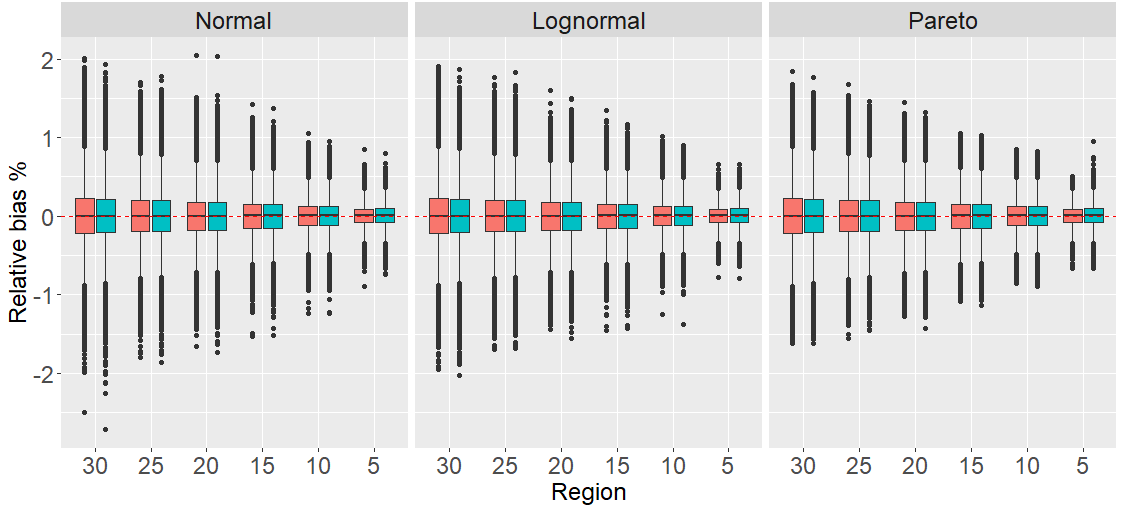}
         \caption{\small N = 300,000}
        \label{rb_300000_box_plot_high_probs}
    \end{subfigure}\hspace*{\fill}
    \end{minipage}
    \begin{minipage}[c]{0.1\textwidth}
        \includegraphics[width=3cm,height=2cm]{legend2.png}
    \end{minipage}

\caption{Relative bias (\%) of regional level estimates where initially $\pi_A = 0.8$ and $\pi_B = 0.7$ and the random effects follow\\ a normal, lognormal or Pareto distribution. Note that the y-axis scale varies across $N$.}

\end{sidewaysfigure}

\begin{table}[hbt!]\footnotesize
\centering
\captionsetup{width=5.6in}
\caption{\small Simulation results where $\pi_A = 0.4$ and $\pi_B = 0.2$, the random effects follow a normal distribution, with total population size N = 500, 1,000, 1,500, 2,000, 2,500, 3,000, 30,000 and 300,000, the mean relative bias (mrb)\%, the coefficient of variation (cv)\% and the mean squared error (mse) of the estimates for the fitted fixed effects (F), Chapman corrected fixed effects (CF) and mixed effects (M) models. $\infty$ indicates that, among the 10,000 samples drawn for each combination of $N$ and Number of regions, at least one estimated population size is infinite.}
\scalebox{0.9}{
\begin{tabular}{rrrrrrrrrrr*{1}{S[table-format=2]}} 
 \hline
\multirow{-2.5}{*}{}&\multicolumn{1}{c}{No.}&\multicolumn{1}{c}{F}&\multicolumn{1}{c}{CF}&\multicolumn{1}{c}{M}&\multicolumn{1}{c}{F}&\multicolumn{1}{c}{CF}&\multicolumn{1}{c}{M}&\multicolumn{1}{c}{F}&\multicolumn{1}{c}{CF}&\multicolumn{1}{c}{M}\\
\multicolumn{1}{c}{N}&\multicolumn{1}{c}{regions}&\multicolumn{1}{c}{mrb\%}&\multicolumn{1}{c}{mrb\%}&\multicolumn{1}{c}{mrb\%}&\multicolumn{1}{c}{cv\%}&\multicolumn{1}{c}{cv\%}&\multicolumn{1}{c}{cv\%}&\multicolumn{1}{c}{mse}&\multicolumn{1}{c}{mse}&\multicolumn{1}{c}{mse}\\
\hline
500&30&$\infty$&-13.577&3.681&$\infty$&48.217&20.016&$\infty$&53.6&12.0\\
500&25&$\infty$&-10.395&3.906&$\infty$&47.304&19.993&$\infty$&76.6&17.4\\
500&20&$\infty$&-7.350&3.815&$\infty$&45.657&19.263&$\infty$&115.8&25.2\\
500&15&$\infty$&-4.811&2.861&$\infty$&43.060&18.380&$\infty$&190.0&39.7\\
500&10&$\infty$&-2.891&2.571&$\infty$&35.968&17.530&$\infty$&307.8&81.0\\
500&5&$\infty$&-0.073&3.090&$\infty$&27.897&17.808&$\infty$&777.5&341.1\\
\hline
1000&30&$\infty$&-3.964&2.214&$\infty$&44.597&14.882&$\infty$&206.5&25.6\\
1000&25&$\infty$&-2.260&2.492&$\infty$&42.387&14.712&$\infty$&276.4&36.4\\
1000&20&$\infty$&-1.386&2.387&$\infty$&38.427&13.917&$\infty$&360.2&50.8\\
1000&15&$\infty$&-0.532&2.147&$\infty$&34.554&13.749&$\infty$&525.5&87.7\\
1000&10&$\infty$&-0.498&1.700&$\infty$&26.961&13.179&$\infty$&720.3&179.4\\
1000&5&4.561&0.752&2.597&20.455&18.650&12.964&1902.5&1413.5&720.8\\
\hline
1500&30&$\infty$&-1.211&1.800&$\infty$&39.742&12.409&$\infty$&386.5&39.7\\
1500&25&$\infty$&-0.232&2.050&$\infty$&37.210&12.319&$\infty$&496.7&56.9\\
1500&20&$\infty$&-0.484&1.482&$\infty$&32.989&11.876&$\infty$&606.6&81.2\\
1500&15&$\infty$&-0.367&1.132&$\infty$&29.276&11.667&$\infty$&851.5&138.2\\
1500&10&$\infty$&-0.255&1.453&$\infty$&20.995&11.213&$\infty$&987.1&290.9\\
1500&5&2.587&0.226&1.642&15.669&14.851&10.931&2378.2&1993.7&1118.3\\
\hline
2000&30&$\infty$&-0.352&1.611&$\infty$&35.041&11.067&$\infty$&542.3&55.9\\
2000&25&$\infty$&-0.509&1.192&$\infty$&31.753&10.832&$\infty$&639.1&76.2\\
2000&20&$\infty$&-0.415&0.985&$\infty$&28.326&10.822&$\infty$&796.1&118.4\\
2000&15&$\infty$&-0.077&1.134&$\infty$&25.419&10.524&$\infty$&1146.9&200.2\\
2000&10&3.425&-0.271&1.113&19.924&18.065&10.076&1736.9&1298.2&413.4\\
2000&5&1.835&0.117&1.234&13.373&12.870&9.815&3014.5&2656.2&1584.0\\
\hline
2500&30&$\infty$&-0.447&1.014&$\infty$&31.410&10.140&$\infty$&679.5&72.2\\
2500&25&$\infty$&-0.803&0.587&$\infty$&27.775&9.938&$\infty$&760.3&98.9\\
2500&20&$\infty$&-0.500&0.740&$\infty$&24.646&9.883&$\infty$&940.2&153.4\\
2500&15&$\infty$&-0.189&0.999&$\infty$&21.346&9.689&$\infty$&1261.0&264.3\\
2500&10&2.854&-0.023&1.162&17.200&16.051&9.396&1999.8&1609.3&563.9\\
2500&5&1.348&0.002&0.912&11.713&11.372&8.712&3563.5&3233.0&1933.2\\
\hline
3000&30&$\infty$&-0.445&0.794&$\infty$&28.342&9.386&$\infty$&796.7&88.6\\
3000&25&$\infty$&-0.804&0.399&$\infty$&25.358&9.243&$\infty$&912.2&122.8\\
3000&20&$\infty$&-0.485&0.661&$\infty$&22.077&9.355&$\infty$&1086.7&197.7\\
3000&15&$\infty$&-0.150&0.948&$\infty$&18.871&9.212&$\infty$&1420.2&343.9\\
3000&10&2.421&0.059&1.096&15.652&14.776&8.887&2357.4&1966.7&725.4\\
3000&5&1.030&-0.077&0.717&10.642&10.385&8.147&4194.4&3876.9&2420.9\\
\hline
30000&30&0.694&0.038&0.215&7.988&7.880&3.898&6512.1&6213.7&1520.2\\
30000&25&0.489&-0.053&0.114&7.253&7.172&3.779&7678.5&7398.9&2054.1\\
30000&20&0.480&0.047&0.186&6.497&6.439&3.642&9634.5&9338.2&2986.2\\
30000&15&0.352&0.028&0.150&5.575&5.539&3.431&12564.4&12277.4&4711.1\\
30000&10&0.178&-0.035&0.065&4.448&4.429&3.226&17897.6&17643.2&9358.3\\
30000&5&0.144&0.037&0.119&3.209&3.202&2.841&37244.0&36943.4&29100.3\\
\hline
300000&30&0.082&0.018&0.026&2.473&2.470&1.976&61309.8&61017.9&39032.1\\
300000&25&0.078&0.024&0.033&2.269&2.267&1.877&74353.1&74045.4&50705.3\\
300000&20&0.066&0.024&0.031&2.007&2.006&1.711&90862.3&90556.3&65841.4\\
300000&15&0.016&-0.016&-0.012&1.722&1.721&1.519&118619.0&118396.5&92316.4\\
300000&10&-0.028&-0.049&-0.046&1.413&1.413&1.311&179695.4&179602.5&154923.8\\
300000&5&0.024&0.014&0.018&1.010&1.010&0.994&367964.8&367586.5&356094.7\\
\hline
\end{tabular}}
\label{table:2a}
\end{table}

\clearpage

\begin{table}[hbt!]\footnotesize
\centering
\captionsetup{width=5.6in}
\caption{\small Simulation results where $\pi_A = 0.4$ and $\pi_B = 0.2$, the random effects follow a lognormal distribution, with total population size N = 500, 1,000, 1,500, 2,000, 2,500, 3,000, 30,000 and 300,000, mean relative bias (mrb)\%, the coefficient of variation (cv)\% and the mean squared error (mse) of the estimates for the fitted fixed effects (F), Chapman corrected fixed effects (CF) and mixed effects (M) models. $\infty$ indicates that, among the 10,000 samples drawn for each combination of $N$ and Number of regions, at least one estimated population size is infinite.}
\scalebox{0.9}{
\begin{tabular}{rrrrrrrrrrr*{1}{S[table-format=2]}} 
 \hline
\multirow{-2.5}{*}{}&\multicolumn{1}{c}{No.}&\multicolumn{1}{c}{F}&\multicolumn{1}{c}{CF}&\multicolumn{1}{c}{M}&\multicolumn{1}{c}{F}&\multicolumn{1}{c}{CF}&\multicolumn{1}{c}{M}&\multicolumn{1}{c}{F}&\multicolumn{1}{c}{CF}&\multicolumn{1}{c}{M}\\
\multicolumn{1}{c}{N}&\multicolumn{1}{c}{regions}&\multicolumn{1}{c}{mrb\%}&\multicolumn{1}{c}{mrb\%}&\multicolumn{1}{c}{mrb\%}&\multicolumn{1}{c}{cv\%}&\multicolumn{1}{c}{cv\%}&\multicolumn{1}{c}{cv\%}&\multicolumn{1}{c}{mse}&\multicolumn{1}{c}{mse}&\multicolumn{1}{c}{mse}\\
\hline
500&30&$\infty$&-13.634&3.637&$\infty$&48.242&20.149&$\infty$&53.6&12.2\\
500&25&$\infty$&-10.249&4.215&$\infty$&47.414&20.218&$\infty$&77.0&18.0\\
500&20&$\infty$&-7.405&3.738&$\infty$&45.336&19.079&$\infty$&114.1&24.7\\
500&15&$\infty$&-4.681&2.983&$\infty$&43.098&18.413&$\infty$&190.7&40.0\\
500&10&$\infty$&-2.978&2.588&$\infty$&35.923&17.717&$\infty$&306.7&82.7\\
500&5&$\infty$&0.184&3.273&$\infty$&28.368&17.719&$\infty$&808.0&340.0\\
\hline
1000&30&$\infty$&-3.856&2.317&$\infty$&44.641&14.910&$\infty$&207.3&25.8\\
1000&25&$\infty$&-2.431&2.377&$\infty$&42.065&14.624&$\infty$&271.3&35.8\\
1000&20&$\infty$&-1.541&2.261&$\infty$&38.360&14.099&$\infty$&358.0&52.0\\
1000&15&$\infty$&-0.541&2.167&$\infty$&34.505&13.788&$\infty$&524.0&88.3\\
1000&10&$\infty$&-0.367&1.816&$\infty$&27.189&13.294&$\infty$&734.3&183.1\\
1000&5&4.452&0.662&2.488&20.422&18.578&12.908&1889.6&1399.9&711.8\\
\hline
1500&30&$\infty$&-1.193&1.817&$\infty$&39.857&12.552&$\infty$&388.9&40.6\\
1500&25&$\infty$&-0.352&1.943&$\infty$&37.064&12.310&$\infty$&491.4&56.6\\
1500&20&$\infty$&-0.514&1.447&$\infty$&32.915&12.018&$\infty$&603.5&83.1\\
1500&15&$\infty$&-0.307&1.174&$\infty$&29.323&11.720&$\infty$&855.1&139.7\\
1500&10&5.072&-0.214&1.502&24.660&21.064&11.369&1559.6&994.5&299.5\\
1500&5&2.503&0.155&1.557&15.510&14.725&10.865&2325.6&1957.9&1102.1\\
\hline
2000&30&$\infty$&-0.359&1.556&$\infty$&35.175&11.103&$\infty$&546.3&56.2\\
2000&25&$\infty$&-0.572&1.124&$\infty$&31.637&10.884&$\infty$&633.8&76.8\\
2000&20&$\infty$&-0.517&0.902&$\infty$&28.256&10.908&$\infty$&790.8&120.0\\
2000&15&$\infty$&-0.078&1.131&$\infty$&25.214&10.481&$\infty$&1128.3&198.6\\
2000&10&3.342&-0.343&0.995&19.869&18.033&9.996&1722.5&1291.9&405.6\\
2000&5&1.819&0.107&1.213&13.397&12.903&9.860&3025.6&2670.1&1598.7\\
\hline
2500&30&$\infty$&-0.401&1.069&$\infty$&31.194&10.233&$\infty$&670.6&73.6\\
2500&25&$\infty$&-0.849&0.529&$\infty$&27.857&9.974&$\infty$&764.1&99.5\\
2500&20&$\infty$&-0.451&0.784&$\infty$&24.700&9.907&$\infty$&945.2&154.4\\
2500&15&$\infty$&-0.113&1.020&$\infty$&21.599&9.724&$\infty$&1293.0&266.5\\
2500&10&2.790&-0.076&1.107&17.140&16.033&9.382&1982.0&1604.2&561.5\\
2500&5&1.375&0.029&0.940&11.651&11.311&8.672&3530.6&3200.9&1918.1\\
\hline
3000&30&$\infty$&-0.406&0.854&$\infty$&28.444&9.467&$\infty$&803.1&90.3\\
3000&25&$\infty$&-0.809&0.391&$\infty$&25.347&9.304&$\infty$&911.6&124.4\\
3000&20&$\infty$&-0.428&0.707&$\infty$&21.907&9.259&$\infty$&1071.0&193.9\\
3000&15&$\infty$&-0.108&0.967&$\infty$&18.975&9.276&$\infty$&1436.8&349.0\\
3000&10&2.433&0.073&1.083&15.697&14.822&8.903&2373.3&1980.2&728.4\\
3000&5&1.058&-0.050&0.734&10.648&10.391&8.139&4204.7&3883.8&2418.4\\
\hline
30000&30&0.721&0.065&0.234&8.014&7.906&3.923&6561.3&6257.9&1540.2\\
30000&25&0.444&-0.096&0.068&7.259&7.179&3.813&7679.1&7407.5&2090.3\\
30000&20&0.471&0.039&0.174&6.497&6.439&3.647&9630.8&9337.0&2993.4\\
30000&15&0.347&0.023&0.140&5.573&5.537&3.438&12551.3&12267.4&4728.5\\
30000&10&0.162&-0.052&0.048&4.435&4.415&3.203&17775.8&17530.7&9228.2\\
30000&5&0.148&0.041&0.122&3.213&3.206&2.833&37336.6&37031.2&28957.2\\
\hline
300000&30&0.095&0.031&0.040&2.482&2.479&1.985&61816.9&61508.6&39400.0\\
300000&25&0.076&0.023&0.032&2.261&2.259&1.870&73811.7&73509.7&50339.3\\
300000&20&0.057&0.014&0.022&2.008&2.006&1.712&90892.3&90605.3&65948.7\\
300000&15&0.010&-0.022&-0.017&1.723&1.722&1.518&118843.6&118635.0&92198.8\\
300000&10&-0.029&-0.050&-0.046&1.414&1.413&1.310&179879.5&179791.1&154607.2\\
300000&5&0.026&0.015&0.020&1.015&1.015&0.997&371305.7&370913.0&357798.9\\
\hline
\end{tabular}}
\label{table:2c}
\end{table}

\clearpage

\begin{table}[hbt!]\footnotesize
\centering
\captionsetup{width=5.6in}
\caption{\small Simulation results where $\pi_A = 0.4$ and $\pi_B = 0.2$, the random effects follow a Pareto distribution, with total population size N = 500, 1,000, 1,500, 2,000, 2,500, 3,000, 30,000 and 300,000, the mean relative bias (mrb)\%, the coefficient of variation (cv)\% and the mean squared error (mse) of the estimates for the fitted fixed effects (F), Chapman corrected fixed effects (CF) and mixed effects (M) models. $\infty$ indicates that, among the 10,000 samples drawn for each combination of $N$ and Number of regions, at least one estimated population size is infinite.}
\scalebox{0.9}{
\begin{tabular}{rrrrrrrrrrr*{1}{S[table-format=2]}} 
 \hline
\multirow{-2.5}{*}{}&\multicolumn{1}{c}{No.}&\multicolumn{1}{c}{F}&\multicolumn{1}{c}{CF}&\multicolumn{1}{c}{M}&\multicolumn{1}{c}{F}&\multicolumn{1}{c}{CF}&\multicolumn{1}{c}{M}&\multicolumn{1}{c}{F}&\multicolumn{1}{c}{CF}&\multicolumn{1}{c}{M}\\
\multicolumn{1}{c}{N}&\multicolumn{1}{c}{regions}&\multicolumn{1}{c}{mrb\%}&\multicolumn{1}{c}{mrb\%}&\multicolumn{1}{c}{mrb\%}&\multicolumn{1}{c}{cv\%}&\multicolumn{1}{c}{cv\%}&\multicolumn{1}{c}{cv\%}&\multicolumn{1}{c}{mse}&\multicolumn{1}{c}{mse}&\multicolumn{1}{c}{mse}\\
\hline
500&30&$\infty$&-13.631&3.550&$\infty$&48.465&20.546&$\infty$&54.0&12.6\\
500&25&$\infty$&-10.154&4.369&$\infty$&47.288&20.674&$\infty$&76.7&18.9\\
500&20&$\infty$&-7.300&3.848&$\infty$&45.579&19.565&$\infty$&115.5&26.1\\
500&15&$\infty$&-4.636&2.856&$\infty$&42.986&19.125&$\infty$&189.7&43.1\\
500&10&$\infty$&-2.970&2.543&$\infty$&36.088&17.968&$\infty$&309.6&85.1\\
500&5&$\infty$&0.016&2.976&$\infty$&27.522&17.427&$\infty$&758.4&327.0\\
\hline
1000&30&$\infty$&-3.753&2.254&$\infty$&44.656&15.360&$\infty$&207.6&27.4\\
1000&25&$\infty$&-2.375&2.288&$\infty$&41.970&14.925&$\infty$&270.2&37.3\\
1000&20&$\infty$&-1.396&2.301&$\infty$&38.492&14.308&$\infty$&361.1&53.7\\
1000&15&$\infty$&-0.495&2.013&$\infty$&34.577&13.847&$\infty$&526.3&88.8\\
1000&10&$\infty$&-0.429&1.616&$\infty$&27.251&13.320&$\infty$&736.6&183.1\\
1000&5&4.560&0.800&2.394&20.279&18.571&12.483&1870.9&1403.2&666.6\\
\hline
1500&30&$\infty$&-1.217&1.692&$\infty$&39.499&12.951&$\infty$&381.6&43.2\\
1500&25&$\infty$&-0.328&1.817&$\infty$&37.350&12.591&$\infty$&499.3&59.1\\
1500&20&$\infty$&-0.471&1.444&$\infty$&32.676&12.160&$\infty$&595.3&85.3\\
1500&15&$\infty$&-0.253&0.973&$\infty$&29.572&11.680&$\infty$&870.4&138.3\\
1500&10&4.918&-0.299&1.268&24.022&20.956&11.188&1475.6&982.5&288.6\\
1500&5&2.480&0.152&1.334&15.478&14.714&10.341&2313.2&1954.4&994.1\\
\hline
2000&30&$\infty$&-0.248&1.514&$\infty$&35.220&11.342&$\infty$&548.8&58.7\\
2000&25&$\infty$&-0.702&0.867&$\infty$&31.747&11.071&$\infty$&636.5&79.1\\
2000&20&$\infty$&-0.362&0.936&$\infty$&28.088&11.071&$\infty$&783.5&124.1\\
2000&15&$\infty$&-0.085&0.940&$\infty$&25.156&10.433&$\infty$&1123.2&196.3\\
2000&10&3.467&-0.210&0.997&19.731&18.068&9.863&1707.2&1300.0&395.9\\
2000&5&1.743&0.041&0.957&13.450&12.947&9.384&3037.9&2683.8&1438.4\\
\hline
2500&30&$\infty$&-0.586&0.757&$\infty$&31.126&10.456&$\infty$&665.7&76.4\\
2500&25&$\infty$&-0.718&0.547&$\infty$&27.972&10.199&$\infty$&771.9&104.3\\
2500&20&$\infty$&-0.345&0.799&$\infty$&24.526&9.991&$\infty$&933.7&157.4\\
2500&15&$\infty$&-0.097&0.840&$\infty$&21.872&9.502&$\infty$&1326.1&253.8\\
2500&10&2.793&-0.072&0.951&17.135&16.060&9.077&1981.3&1609.6&524.4\\
2500&5&1.453&0.114&0.899&11.552&11.222&8.217&3481.0&3155.9&1723.8\\
\hline
3000&30&$\infty$&-0.478&0.671&$\infty$&28.212&9.701&$\infty$&789.0&94.6\\
3000&25&$\infty$&-0.575&0.518&$\infty$&25.256&9.520&$\infty$&908.6&130.8\\
3000&20&$\infty$&-0.307&0.675&$\infty$&22.362&9.343&$\infty$&1118.4&197.7\\
3000&15&$\infty$&-0.061&0.830&$\infty$&18.744&9.018&$\infty$&1403.3&329.5\\
3000&10&2.386&0.036&0.907&15.509&14.719&8.470&2313.2&1950.6&657.1\\
3000&5&1.107&0.003&0.692&10.590&10.340&7.798&4164.5&3848.4&2220.1\\
\hline
30000&30&0.676&0.026&0.172&7.964&7.859&3.896&6467.3&6178.6&1519.0\\
30000&25&0.485&-0.052&0.094&7.241&7.162&3.795&7652.4&7378.9&2072.5\\
30000&20&0.470&0.040&0.162&6.487&6.431&3.649&9602.7&9313.3&2998.3\\
30000&15&0.369&0.047&0.164&5.582&5.546&3.448&12605.2&12316.1&4764.1\\
30000&10&0.167&-0.047&0.063&4.454&4.434&3.123&17928.8&17682.6&8777.3\\
30000&5&0.152&0.047&0.130&3.205&3.198&2.740&37157.8&36855.2&27104.2\\
\hline
300000&30&0.093&0.030&0.041&2.472&2.469&1.879&61312.8&61012.5&35316.5\\
300000&25&0.087&0.034&0.045&2.254&2.252&1.780&73381.8&73066.6&45640.1\\
300000&20&0.067&0.025&0.034&2.007&2.006&1.640&90885.7&90579.7&60554.3\\
300000&15&0.012&-0.020&-0.013&1.727&1.726&1.454&119350.1&119136.8&84554.6\\
300000&10&-0.034&-0.055&-0.049&1.412&1.411&1.245&179419.9&179354.6&139717.2\\
300000&5&0.026&0.015&0.021&1.003&1.003&0.956&362553.8&362169.1&329486.9\\
\hline
\end{tabular}}
\label{table:2e}
\end{table}

\begin{sidewaysfigure}[p]
\begin{minipage}{0.9\textwidth}
\centering
\begin{subfigure}[b]{.5\linewidth}
         \includegraphics[width=11cm,height=6cm]{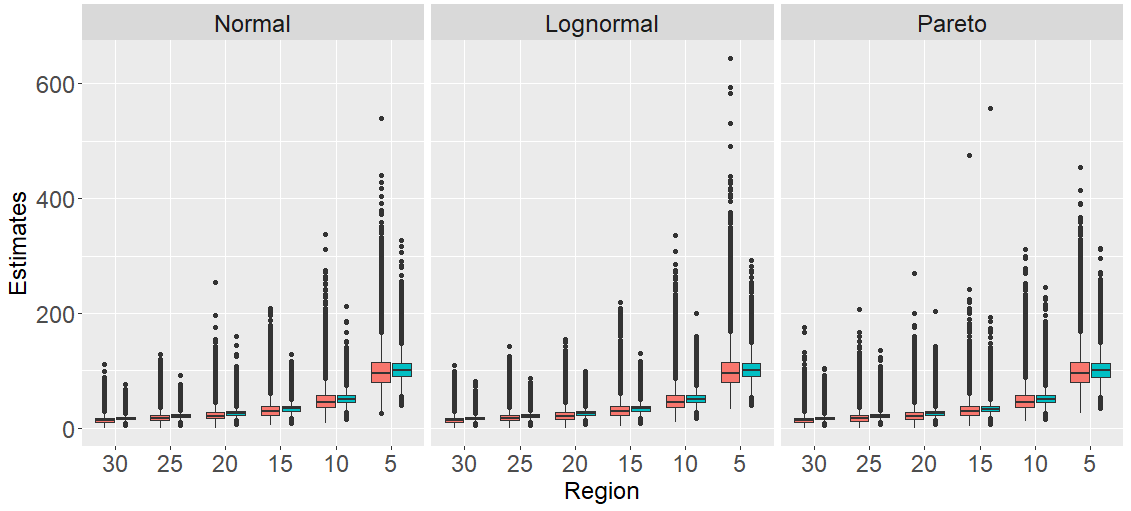}
         \caption{\small N = 500}
        \label{500_box_low_probs}
    \end{subfigure}\hspace*{\fill}
    \begin{subfigure}[b]{.5\linewidth}
         \includegraphics[width=11cm,height=6cm]{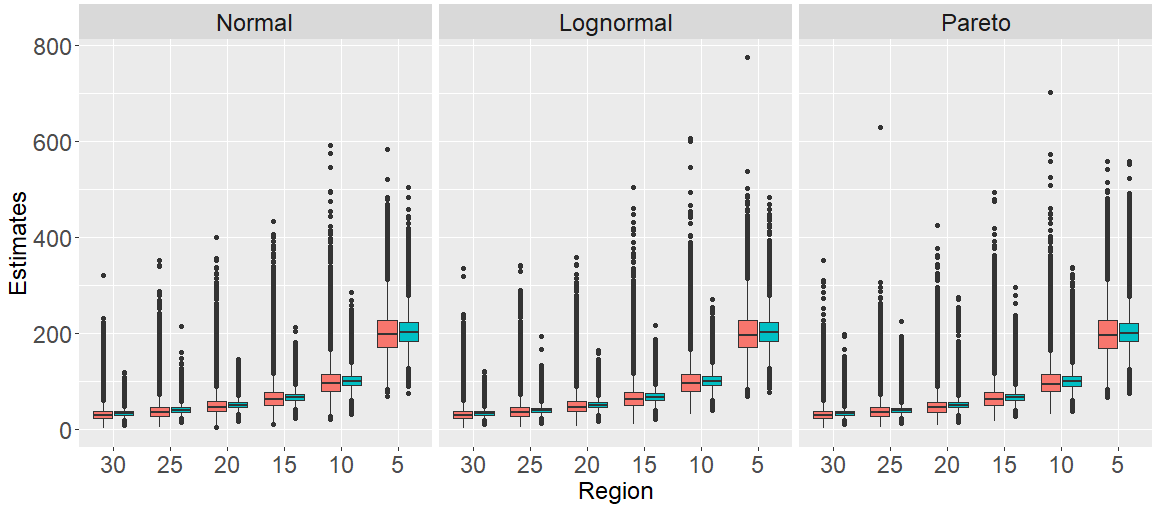}
        \caption{\small N = 1,000}
        \label{1000_box_low_probs}
    \end{subfigure}\hspace*{\fill}

    \begin{subfigure}[b]{.5\linewidth}
         \includegraphics[width=11cm,height=6cm]{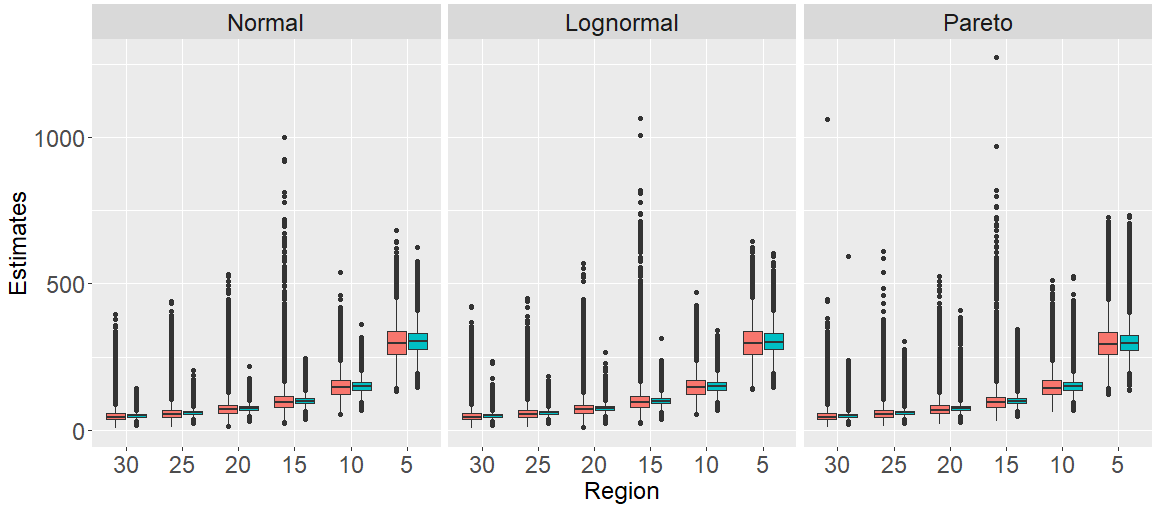}
         \caption{\small N = 1,500}
        \label{1500_box_low_probs}
    \end{subfigure}\hspace*{\fill}
    \begin{subfigure}[b]{.5\linewidth}
         \includegraphics[width=11cm,height=6cm]{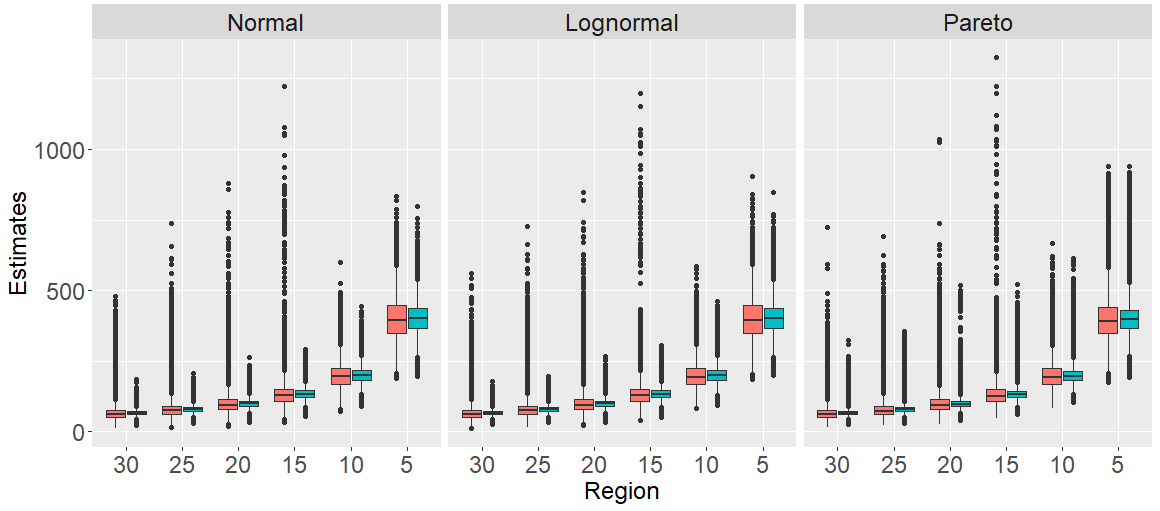}
         \caption{\small N = 2,000}
        \label{2000_box_low_probs}
    \end{subfigure}\hspace*{\fill}
    \end{minipage}
    \begin{minipage}[c]{0.1\textwidth}
        \includegraphics[width=3cm,height=2cm]{legend2.png}
    \end{minipage}
    
    \caption{Regional level estimates where initially $\pi_A = 0.4$ and $\pi_B = 0.2$ and the random effects follow a normal, lognormal\\ or Pareto distribution. Note that the y-axis scale varies across $N$.}
    \label{box_low_probs}
    
\end{sidewaysfigure}

\begin{sidewaysfigure}[p]\ContinuedFloat
\begin{minipage}{0.9\textwidth}
\centering
\begin{subfigure}[b]{.5\linewidth}
         \includegraphics[width=11cm,height=6cm]{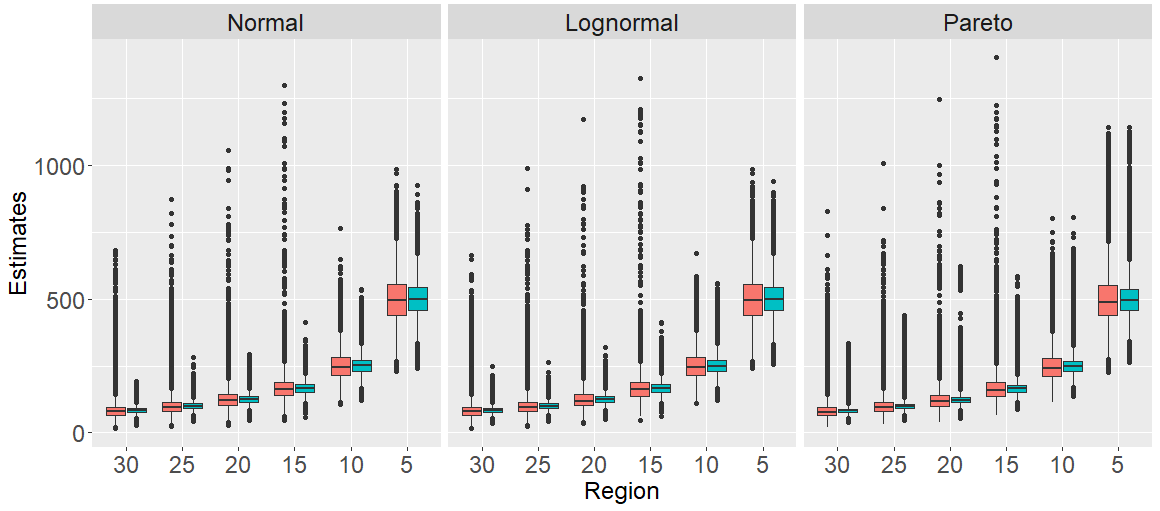}
         \caption{\small N = 2,500}
        \label{2500_box_plot_low_probs}
    \end{subfigure}\hspace*{\fill}
    \begin{subfigure}[b]{.5\linewidth}
         \includegraphics[width=11cm,height=6cm]{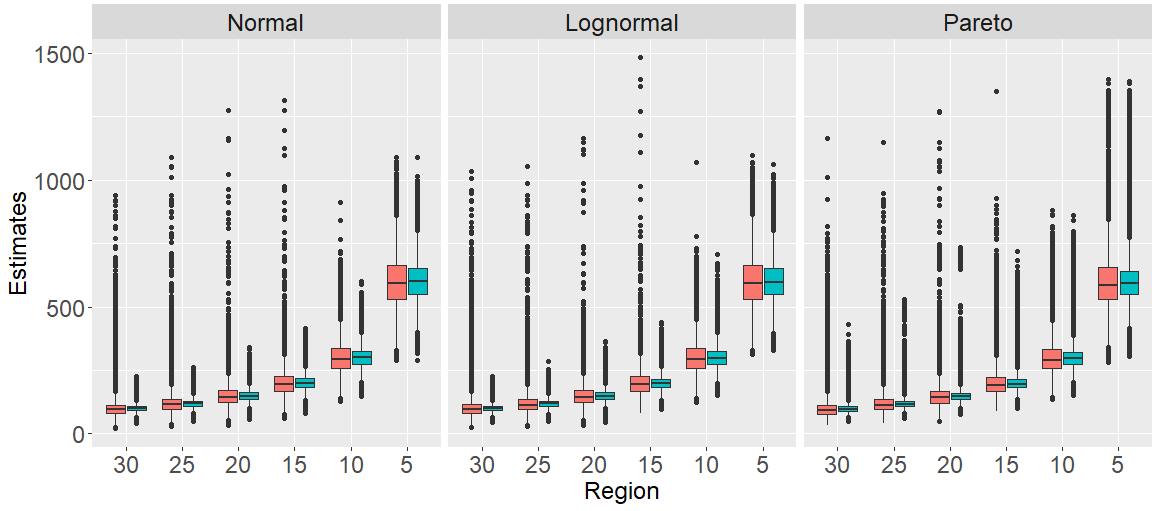}
        \caption{\small N = 3,000}
        \label{3000_box_plot_low_probs}
    \end{subfigure}\hspace*{\fill}

    \begin{subfigure}[b]{.5\linewidth}
         \includegraphics[width=11cm,height=6cm]{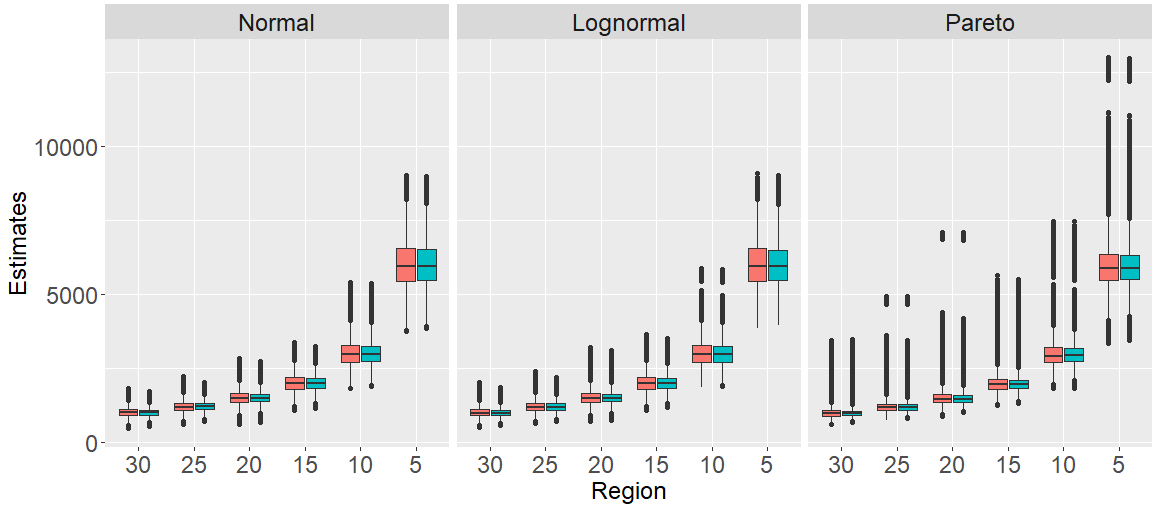}
         \caption{\small N = 30,000}
        \label{30000_box_plot_low_probs}
    \end{subfigure}\hspace*{\fill}
    \begin{subfigure}[b]{.5\linewidth}
         \includegraphics[width=11cm,height=6cm]{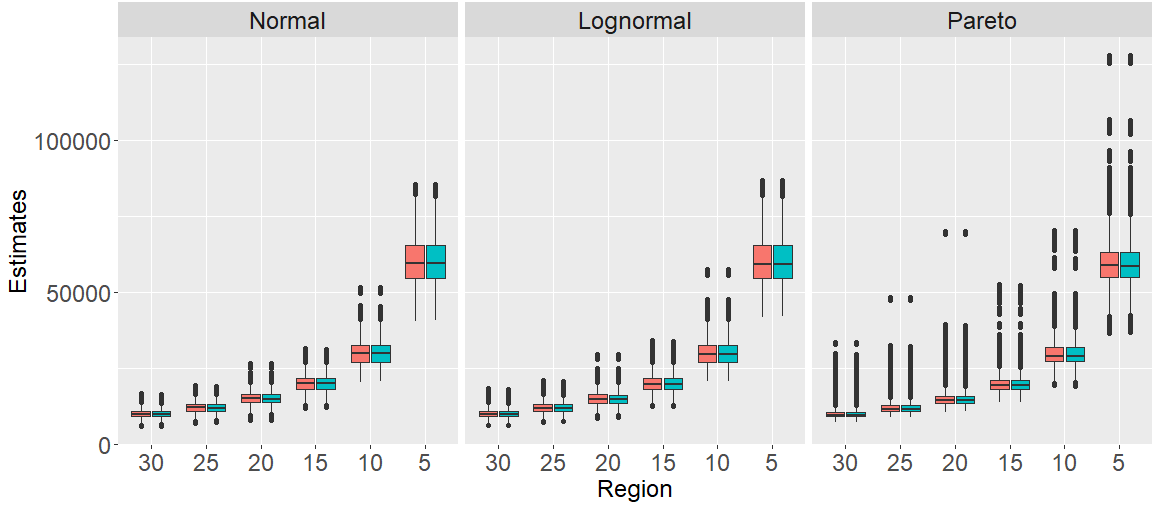}
         \caption{\small N = 300,000}
        \label{300000_box_plot_low_probs}
    \end{subfigure}\hspace*{\fill}
    \end{minipage}
    \begin{minipage}[c]{0.1\textwidth}
        \includegraphics[width=3cm,height=2cm]{legend2.png}
    \end{minipage}
    
    \caption{Regional level estimates where initially $\pi_A = 0.4$ and $\pi_B = 0.2$ and the random effects follow a normal, lognormal\\ or Pareto distribution. Note that the y-axis scale varies across $N$.}
\end{sidewaysfigure}

\begin{sidewaysfigure}[p]
\begin{minipage}{0.9\textwidth}
\centering
\begin{subfigure}[b]{.5\linewidth}
         \includegraphics[width=11cm,height=6cm]{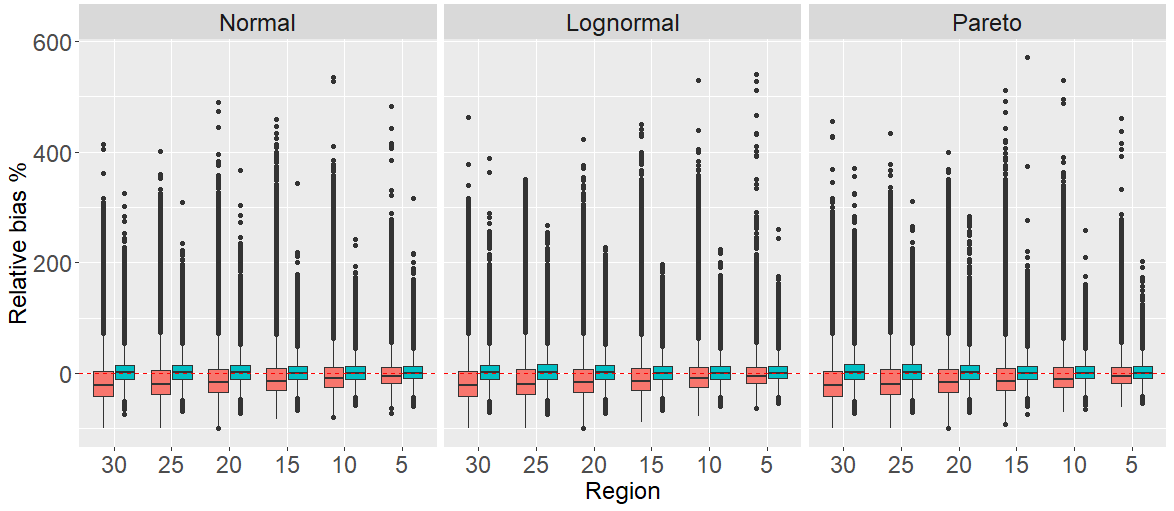}
         \caption{\small N = 500}
        \label{rb_500_box_plot_low_probs}
    \end{subfigure}\hspace*{\fill}
    \begin{subfigure}[b]{.5\linewidth}
         \includegraphics[width=11cm,height=6cm]{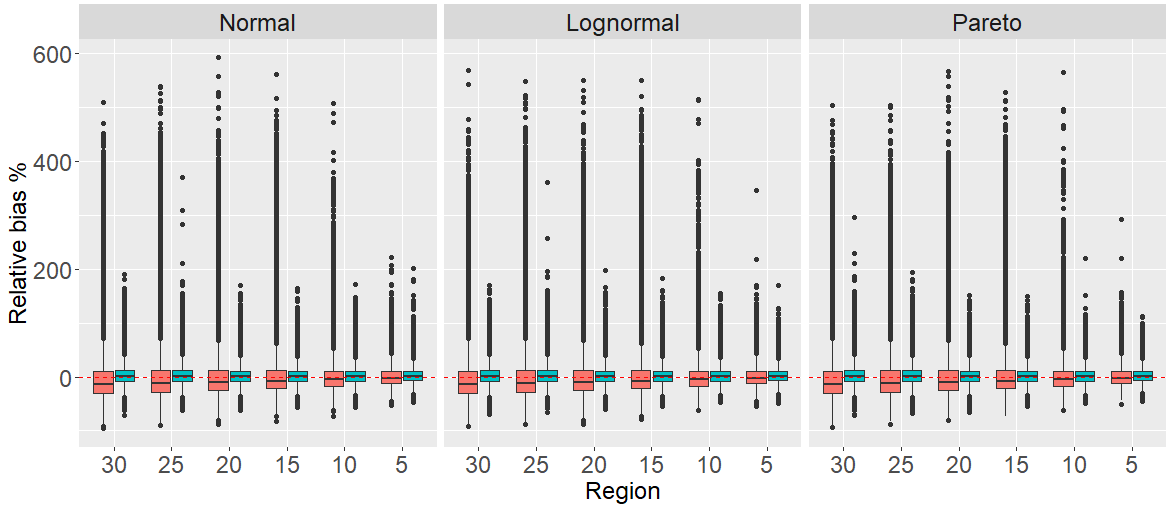}
        \caption{\small N = 1,000}
        \label{rb_1000_box_plot_low_probs}
    \end{subfigure}\hspace*{\fill}

    \begin{subfigure}[b]{.5\linewidth}
         \includegraphics[width=11cm,height=6cm]{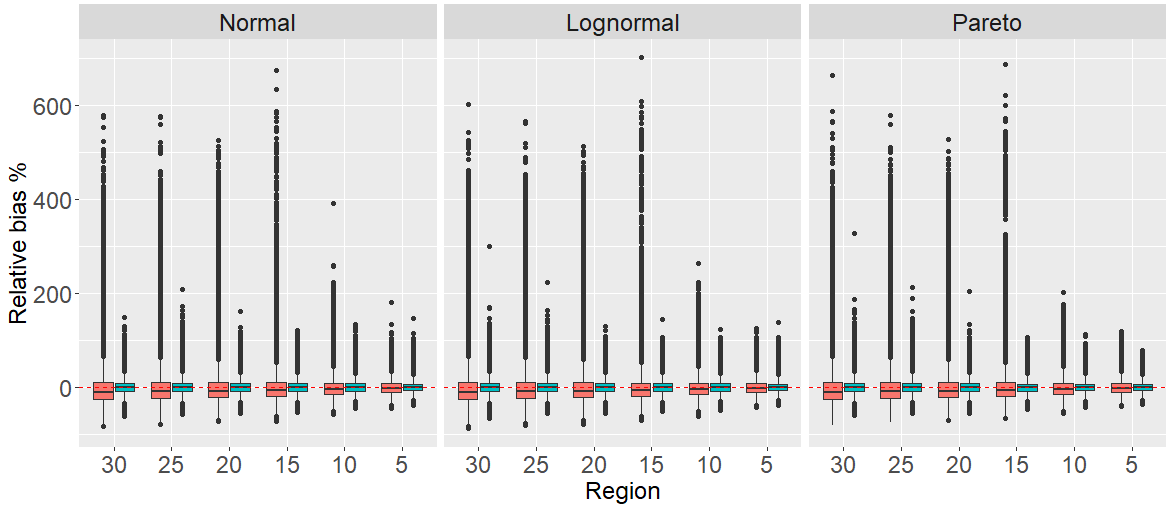}
         \caption{\small N = 1,500}
        \label{rb_1500_box_plot_low_probs}
    \end{subfigure}\hspace*{\fill}
    \begin{subfigure}[b]{.5\linewidth}
         \includegraphics[width=11cm,height=6cm]{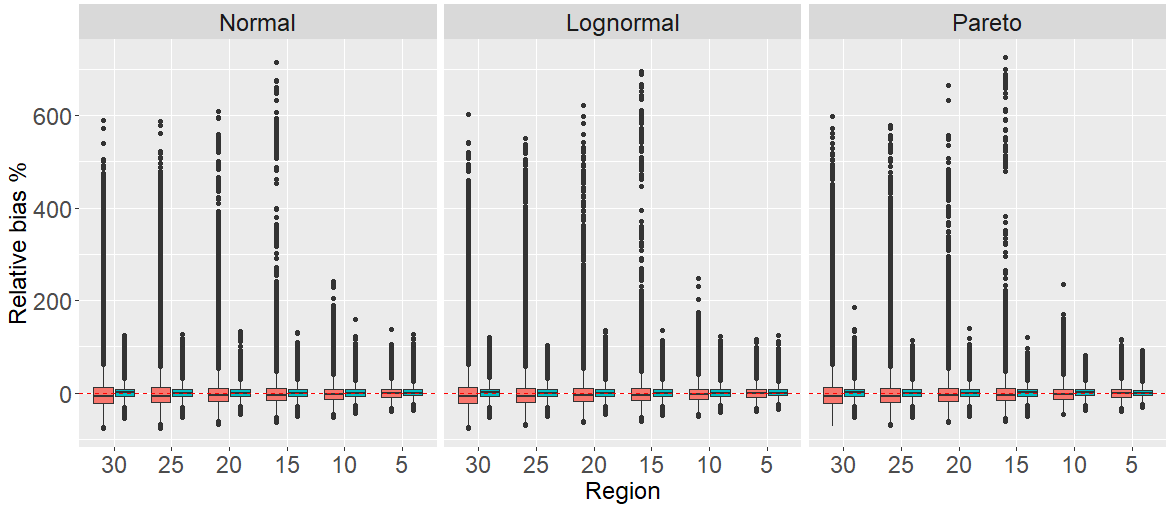}
         \caption{\small N = 2,000}
        \label{rb_2000_box_plot_low_probs}
    \end{subfigure}\hspace*{\fill}
    \end{minipage}
    \begin{minipage}[c]{0.1\textwidth}
        \includegraphics[width=3cm,height=2cm]{legend2.png}
    \end{minipage}

\caption{Relative bias (\%) of regional level estimates where initially $\pi_A = 0.4$ and $\pi_B = 0.2$ and the random effects follow\\ a normal, lognormal or Pareto distribution. Note that the y-axis scale varies across $N$.}\label{rb_low_probs_region}

\end{sidewaysfigure}

\begin{sidewaysfigure}[p]\ContinuedFloat
\begin{minipage}{0.9\textwidth}
\centering
\begin{subfigure}[b]{.5\linewidth}
         \includegraphics[width=11cm,height=6cm]{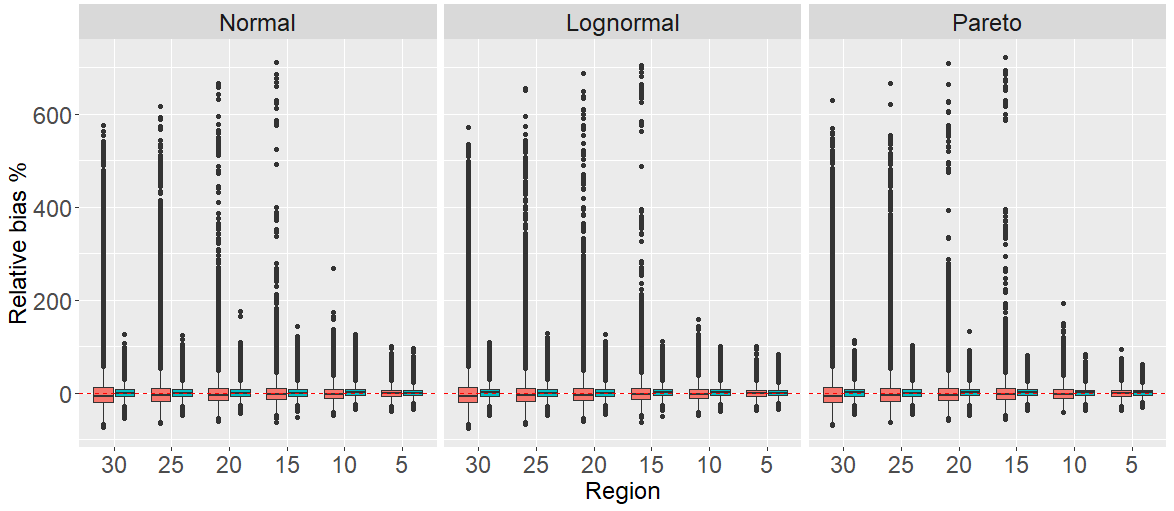}
         \caption{\small N = 2,500}
        \label{rb_2500_box_plot_low_probs}
    \end{subfigure}\hspace*{\fill}
    \begin{subfigure}[b]{.5\linewidth}
         \includegraphics[width=11cm,height=6cm]{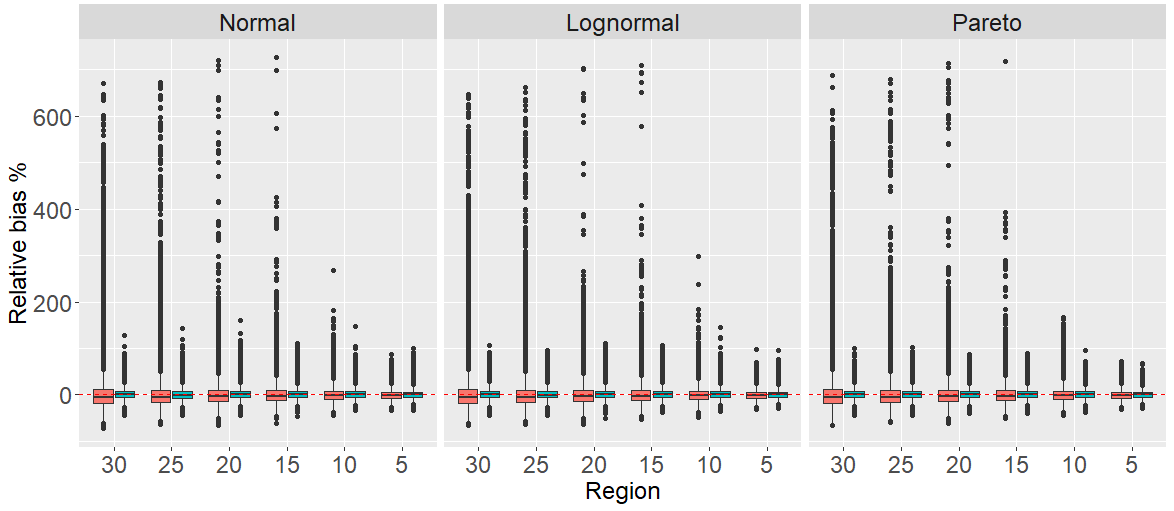}
        \caption{\small N = 3,000}
        \label{rb_3000_box_plot_low_probs}
    \end{subfigure}\hspace*{\fill}

    \begin{subfigure}[b]{.5\linewidth}
         \includegraphics[width=11cm,height=6cm]{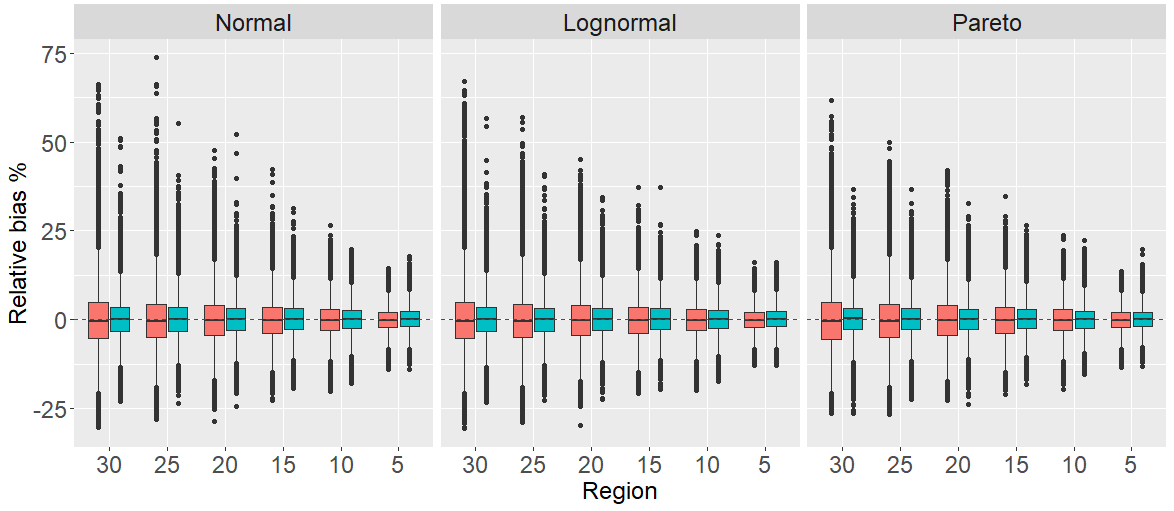}
         \caption{\small N = 30,000}
        \label{rb_30000_box_plot_low_probs}
    \end{subfigure}\hspace*{\fill}
    \begin{subfigure}[b]{.5\linewidth}
         \includegraphics[width=11cm,height=6cm]{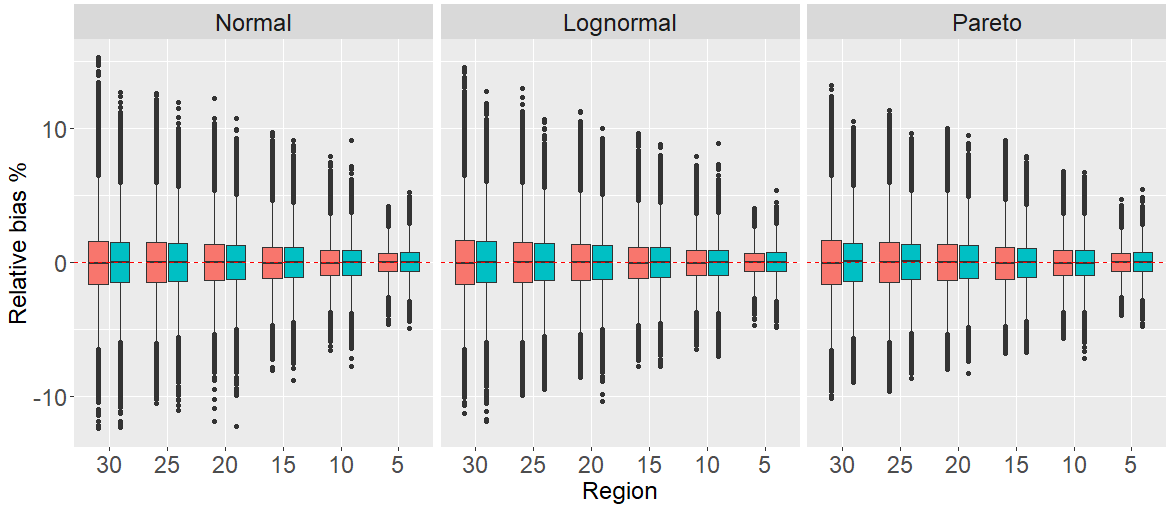}
         \caption{\small N = 300,000}
        \label{rb_300000_box_plot_low_probs}
    \end{subfigure}\hspace*{\fill}
    \end{minipage}
    \begin{minipage}[c]{0.1\textwidth}
        \includegraphics[width=3cm,height=2cm]{legend2.png}
    \end{minipage}

\caption{Relative bias (\%) of regional level estimates where initially $\pi_A = 0.4$ and $\pi_B = 0.2$ and the random effects follow\\ a normal, lognormal or Pareto distribution. Note that the y-axis scale varies across $N$.}

\end{sidewaysfigure}

\end{document}